\documentclass[12pt,preprint]{aastex}

\newcommand{\mm}{millimeter continuum}
\newcommand{\um}{\mbox{$\,\mu$m}}
\newcommand{\hii}{H{\sc ii}}
\begin{document}
\title{A Millimeter Continuum Survey for Massive Protoclusters in the
  Outer Galaxy}
\shorttitle{Millimeter Survey for Protoclusters}
\author{R. Klein\altaffilmark{1,3}, B. Posselt\altaffilmark{2,3},
  K. Schreyer\altaffilmark{2}, J. Forbrich\altaffilmark{4,2},
  Th. Henning\altaffilmark{5}}
\altaffiltext{1}{University of California at Berkeley, 366 Le Conte Hall, Berkeley, CA 94720-7300}
\email{r\_klein@calmail.berkeley.edu}
\altaffiltext{2}{Astrophysikalisches Institut und Universit\"ats-Sternwarte
  (AIU) Jena, Schillerg\"a{\ss}chen 2-3, D--07745 Jena, Germany}
\altaffiltext{3}{Max-Planck-Institut f\"ur extraterrestrische Physik (MPE),
  Giessenbachstr., D--85748 Garching, Germany}
\altaffiltext{4}{Max-Planck-Institut f\"ur Radioastronomie (MPIfR), Auf dem
  H\"ugel 69, D--53121 Bonn, Germany}
\altaffiltext{5}{Max-Planck-Institut f\"ur Astronomie (MPIA), K\"onigstuhl 17,
  D--69117 Heidelberg, Germany}
\begin{abstract}  
  Our search for the earliest stages of massive star formation turned
  up twelve massive pre-protocluster candidates plus a few
  protoclusters.  For this search, we selected 47 FIR-bright IRAS
  sources in the outer Galaxy. We mapped regions of several square
  arc\-minutes around the IRAS source in the millimeter continuum in
  order to find massive cold cloud cores possibly being in a very
  early stage of massive star formation.  Masses and densities are
  derived for the 128 molecular cloud cores found in the obtained
  maps.  We present these maps together with near-infrared,
  mid-infrared, and radio data collected from the 2MASS, MSX, and NVSS
  catalogs.  Further data from the literature on detections of
  high-density tracers, outflows, and masers are added.  The
  multi-wavelength datasets are used to characterize each observed
  region. The massive cloud cores ($M>100\,M_\odot$) are placed in a
  tentative evolutionary sequence depending on their emission at the
  investigated wavelengths.  Candidates for the youngest stages of
  massive star formation are identified by the lack of detections in
  the above-mentioned near-infrared, mid-infrared, and radio surveys.
  Twelve massive cores prominent in the millimeter continuum fulfill
  this requirement.  Since neither FIR nor radio emission have been
  detected from these cloud cores massive protostars must be very
  deeply embedded in these cores. Some of these objects may actually
  Pre-Proto-cluster cores: an up to now rare object class, where the
  initial conditions of massive star formation can be studied.
\end{abstract}

\section{INTRODUCTION}
\label{sec:intro}
\newcommand{\PPC}{PPclCs}

\subsection{Motivation}
\label{ssec:motiv}

Massive star formation (MSF) is far less understood than low-mass star
formation, despite of its high impact on the dynamics of the
interstellar medium and its importance for galactic evolution. There
are two crucial questions: What are the earliest stages of massive
star formation? Can we identify regions which are going to form
massive stars together with their low-mass companions? \citet{Evans02}
pointed out that there are no clear examples for Pre-Proto-cluster
Cores (\PPC), as they named the high-mass analogs of the low-mass
pre-stellar cores \citep{Ward-Thompson94}. {\PPC} may even not exist.
Low-mass stars may already form while the cloud core still gathers
mass to form massive stars. Then there would be no cold massive cloud
core without signs of star formation, but still, a massive cloud core
is needed to form a star cluster containing one or more massive stars.
\citet{Evans02} use a cloud of $4600\,M_\odot$ with an average density
of about $10^6\rm\,cm^{-3}$ and a temperature of 10\,K as a template
for such a PPclC -- a cold source prominent at millimeter wavelengths
($\sim$5\,Jy @ 1\,mm for a 5.5\,kpc distant core). When a PPclC starts
to collapse, certainly forming more than one protostar, a massive
protocluster is born deeply embedded in the core.  The appearance in
the {\mm} certainly does not change much in comparison with the
original cold massive core, although the central temperature has
increased.  Therefore, if a cloud core, prominent in the {\mm}, lacks
strong far-infrared emission, it is a candidate for a PPclC or a young
protocluster.  The main goal of our {\mm} survey has been to find
candidates for these earliest stages of massive star formation.

\subsection{The survey}
\label{ssec:survey}

But where to look for these early stages? Candidate cloud cores may be
found in the vicinity of regions with on-going MSF. On-going MSF
reveals itself by dust heated by the UV radiation of forming massive
stars. The dust re-emits the absorbed energy at far infrared (FIR) and
millimeter wavelengths.  Thus, evolved protoclusters and young massive
clusters can be revealed as luminous infrared point sources
\citep{Henning90,Henning00b,Osterloh97}.  Our strategy has been to use
the brightest IRAS sources as lighthouses to guide us to regions of
presumably young MSF sites (cf. \citealt{Henning92b}) and then have a
closer look at longer wavelengths in search for younger objects, i.e.
massive cloud cores off-set from the bright IRAS sources.  Comparisons
with observations at other wavelengths help to characterize the
detected cloud cores.

We selected bright FIR sources in the outer Galaxy (to avoid source
confusion) from the IRAS Point Source Catalog (PSC) to map them in the
{\mm}.  Large maps of 44 star-forming regions were obtained and are
presented here.  We compared the maps to publicly available surveys
performed from near infrared (NIR) to radio wavelengths in order to
characterize the objects found in our {\mm} maps. We used the
2MASS\footnote{The Two Micron All Sky Survey (2MASS) is a joint
  project of the University of Massachusetts and the Infrared
  Processing and Analysis Center/California Institute of Technology,
  funded by the National Aeronautics and Space Administration and the
  National Science Foundation ({\tt http://pegasus.phast.umass.edu}).}
data (K band sensitivity: 14.3\,mag) in the NIR.  The mid-infrared
(MIR) data comes from the MSX PSC\footnote{MSX -- Midcourse Space
  Experiment. We use the MSX6C galactic plane and high latitude
  catalogs available at {\tt
    http://vizier.u-strasbg.fr/viz-bin/Cat?V/114} \citep{MSXEgan2003}.
  Processing of the data was funded by the Ballistic Missile Defense
  Organization with additional support from NASA Office of Space
  Science.}  (4\um -- 21\um, $>$90\% complete down to 0.15 to 20\,Jy
depending on wavelength).  The 1.4\,GHz radio continuum data is taken
from the NVSS\footnote{\cite{NVSS}} PSC (completeness limit:
2.5\,mJy).  Further data from the literature on maser and outflow
detections are collected in this work, too.

Similar {\mm} surveys of selected MSF regions have been conducted by
other groups.  \citet{Hunter2000} acquired 350 $\mu $m continuum data
for 24 ultra-compact {\hii} (UC{\hii})-regions (radio- and
IRAS-colors-selected).  \citet{Mueller2002} observed 51 dense cores
associated with H$_{2}$O masers.  \citet{Beuther2002I} and
\citet{Beuther2002II} carried out an extensive survey of 69 objects in
the 3.6\,cm and 1.2\,mm continuum, and in spectral lines at millimeter
and radio wavelengths. In their search for young sources, they
targeted FIR bright IRAS-sources with CS detections and no 5\,GHz
radio detection.  Recently, \citet{Faundez04} completed a 1.2\,mm
continuum survey of 146 southern high-mass star forming regions
selected by their IRAS colors and CS detections.  The overlap between
the source lists of all these surveys and our targets is limited to a
few sources.

\subsection{Massive pre-protocluster candidates}
\label{ssec:cand}
The {\mm} maps present us with many cloud cores. Criteria are needed
to identify the earliest stages of massive star formation. A mass
limit can be established by estimating the minimum mass needed to form
a massive star.  The average star formation efficiency is only a few
percent \citep{Franco94,Rodriguez95,Diaz-Miller98}. However, taking
only cloud cores into account, ignoring the rest of the star-forming
molecular cloud, the star formation efficiency reaches probably values
of the order of 50\%. Such a value has been estimated for the Orion
cluster \citep{Hillenbrand98}. According to \citet{MillerScalo79},
about 12\% of the stellar mass is in massive stars ($M_*>8\,M_\odot$).
Thus, to build a $8\,M_\odot$ star and its low-mass companions, one
needs at least a $130\,M_\odot$ cloud core.  In this work we will use
a lower mass limit for massive cores of $100\,M_\odot$.

In order to characterize the massive cloud cores, which we find in our
{\mm} survey, we employ a phenomenalistic evolutionary model outlined
in Table~\ref{tab:stages}.  The model starts out with the hypothetical
{\em\PPC} (stage 0). The eventually collapsing center will start to
produce massive and low-mass protostars, however the outside
appearance of this stage 1, the {\em early protocluster}, will hardly
be different from a PPclC.  All radiation will be reprocessed and
emitted by the surrounding cloud core and the spectral energy
distribution (SED) resembles a single temperature (modified)
blackbody.  In stage 2, the {\em protocluster} heats the interior of
the cloud core and clears a cavity.  The massive stars develop UC{\hii}
regions as the accretion flow cannot quench them anymore. The object
can be detected at radio wavelengths.  Due to the heating, the SED
peak shifts to shorter wavelengths and resembles no longer a
(modified) blackbody.  When the rising MIR emission is strong enough
to be detected by the MSX satellite, we speak of stage 3, an {\em
  evolved protocluster}, within the scope of this work.  When the
cluster emerges from its parental cloud and gets detectable in the NIR
(2MASS), we have a {\em young cluster}, stage 4.  It continues to
disperse the surrounding cloud core remnants. A star {\em cluster} is
born, stage 5.

In this sequence, the luminosity in the MIR and then in the NIR rises
whereas the millimeter luminosity slowly decreases.  Therefore, our
model complies with the suggestion by \citet{Minier04} to use the
ratio of the sub-millimeter luminosity to the bolometric luminosity
($L_{\rm sub}/L_{\rm bol}$) as indicator of the evolutionary state as
\citet{Andre93} applied it in low-mass star formation. Unfortunately,
we cannot derive this quantity for our cloud cores, because of the
lack of reliable FIR data to derive $L_{\rm bol}$. Our maps show that
often the IRAS data cannot be assigned to a {\mm} core unambiguously.

The evolutionary sequences is broken down into stages in such a way,
that we can assign the stages according to detections in the used
surveys.  However, the first two stages (0 -- {\PPC} and 1 -- early
protocluster) cannot be distinguished within this work. The on-set of
star formation separates these two stages, but this event does not
change the appearance in the regarded survey of the cloud core
immediately.  We will call the cloud cores that fall into the first
two stages {\em massive pre-protocluster candidates}.  They have no
association at infrared and radio wavelengths (2MASS, MSX, NVSS).
Taking the spatial resolution of the surveys and our observations and
pointing accuracies into account, there should be no 2MASS, MSX, or
NVSS association within 10\arcsec{} of the {\mm} peak and the mass
must be 100\,$M_\odot$ or higher.

\section{OBSERVATIONS}
\label{sec:obs}
\subsection{The Source Sample}
\label{ssec:srcsel}

Our strategy to find the earliest stages of MSF is to search around
bright FIR sources. We use the sample of bright FIR sources that has
been compiled by \citet{Henning92b} for a maser search extending the
sample of \citet{Snell88,Snell90} who performed an outflow search.
The sample is created by applying the following criteria to the IRAS
PSC:
\begin{itemize} 
\item  S$_{100\,\mu\rm m}\geq 500$\,Jy with a flux quality $\geq  2$,
\item  Right ascension range from 0\,h to 12\,h,
\item  Declination range from $-30\degr$ to $90\degr$.
\end{itemize} 
The first criterion selects all bright FIR objects seen by IRAS. The
second criterion excludes the inner part of the Galaxy to avoid source
confusion, while the last criterion ensures observability from the
northern hemisphere. In Galactic coordinates, this selects longitudes
$l=120\degr$ to $250\degr$ along the galactic plane.  These criteria
are fulfilled by 67 young stellar objects excluding IRC +10216 and M82
\citep{Snell90,Henning92b}.

From December 1998 to March 2001, we searched this sample for massive
pre-protocluster candidates by mapping 47 of these sources in 44
regions in the {\mm}.  The remaining 20 sources were not mapped
because these IRAS sources are mainly located in well-studied MSF
regions such as Orion and W3.  The {\mm} maps are the focus of this
work.  The source list together with some basic properties of the
regions is presented in Table \ref{tab:srclst}.  The 40 {\mm} maps
with detections are presented in Figure~\ref{fig:maps} together with
2MASS, MSX PSC and NVSS PSC data.  In the bolometer maps we reached an average
sensitivity of 50\,mJy/beam but the sensitivity from map to map varies
strongly (3 to 700\,mJy/beam).  Additionally, the results of the above
mentioned maser search and molecular line studies \citep{Schreyer96}
together with the results of outflow and maser searches by other
groups are compiled for the 47 objects in Table~\ref{tab:Literatur}.
However, these searches targeted the respective IRAS PSC positions and
often the results cannot be assigned to a source in our maps,
because many cloud cores are displaced from the IRAS position.

\subsection{Telescopes}
\label{ssec:tel}

The {\mm} maps were collected at three observatories: the Heinrich
Hertz Submillimeter Telescope (SMT, Mt. Graham, AZ, USA), the
"Institut de Radioastronomie Millim\'etrique" (IRAM) 30\,m telescope
(Pico Veleta, Spain), and the James Clerk Maxwell Telescope (JCMT,
Mauna Kea, HI, USA). The sections below  describe the instruments
used at these three telescopes together with the respective observing
techniques.  The basic parameters of these observations are given in
Table~\ref{tab:beam}.

The general observing strategy is common to all telescopes and
instruments. To ensure good pointing and calibration, pointing, focus,
and sky-dip (to measure the atmospheric opacity) observations were
performed in appropriate intervals during the observing shifts.  At
least once during an observing shift a planet, when available, was
observed for the flux calibration. At the JCMT only secondary flux
calibrators could be observed. The reader will find a discussion of
the flux calibration in \S~\ref{ssec:cal}.

\paragraph{Heinrich Hertz Submillimeter Telescope}
We used the SMT with the 19-channel bolometer developed by the
Max-Planck-Institut f\"ur Radioastronomie (MPIfR), Bonn, Germany. It
operates at 870\um.  The applied mapping technique and subsequent data
reduction is the same for the 19-channel bolometer and MAMBO (see
below).  At the SMT, the scanning speed and steps in elevation were
8\arcsec\,s$^{-1}$ and 8\arcsec{}, respectively. We achieved a
resolution of {26\arcsec} estimated from planet maps.

\paragraph{IRAM 30\,m Telescope}
We used MAMBO, the 37-channel bolometer with a central wavelength of
1.3\,mm developed by the MPIfR \citep{bolos}, at the IRAM 30m
telescope.  The on-the-fly mapping technique was applied, where the
bolometers are read out while the telescope is moving over the area.
The area of interest is mapped in horizontal coordinates, i.e.\ the
telescope moves in azimuth while integrating and then steps in
elevation for the next scan.  We scanned in azimuth with a speed of
5\arcsec\,s$^{-1}$ and in steps of 5\arcsec{} in elevation. Due to the
sky rotation, the subscans in azimuth are not parallel in equatorial
coordinates, and the maps are more or less fan-shaped.  Scanning may
create artifacts along the scanning direction. To minimize these
artifacts and to increase the sensitivity, we mapped the sources more
than once at different hour angles.  The data reduction was performed
using the NIC software package \citep{nic}, which employs the EKH
algorithm \citep{Emerson79} to restore the on-the-fly maps.  The beam
size has been {14\arcsec} to 17\arcsec.

\paragraph{James Clerk Maxwell Telescope}
The bolometer array used at the JCMT at {850\um} is SCUBA
\citep{scuba}.  Its 37 channel long-wavelength array has the same
hexagonal geometry as the MPIfR bolometers.  The on-the-fly observing
technique with SCUBA differs substantially from the ``MPIfR method''.
The observation is performed in equatorial coordinates instead of in
horizontal coordinates.  Following the work of \citet{EmersonII}, a
number of different chop configurations is used, in order to sample as
many spatial frequencies as possible.  Scanning artifacts are largely
suppressed this way. We used the default setting of six different chop
configurations: Chop throws of 20, 30 and 65{\arcsec} each with chop
position angles of 0\degr{} (DEC chopping) and 90\degr{} (RA chopping)
in equatorial coordinates.  The software package SURF \citep{surf} was
used for reducing the SCUBA data and rendering the maps. The achieved
beam size has been about 15\arcsec.

\subsection{Calibration and Flux Density Measurements}
\label{ssec:cal}
Planets with their known fluxes are used as primary flux calibrators
for {\mm} observations. At the JCMT, the calibration has been achieved
using the secondary calibrators HL Tau and CRL\,618.  The fluxes of
the planets and of these secondary calibrators are known up to a few
percent. The calibration measurements have been executed to a
precision of a few percent, too. These measurements provide the
conversion factor from instrumental units to physical units like
mJy/beam with an accuracy of about 10\%.

Uranus and the secondary calibrators are sufficiently point-like to
measure the beam profile (results see Table~\ref{tab:beam}).
The knowledge of the beam area is not only important for source
size deconvolution and the spatial resolution but also for the
calibration. It is necessary to convert mJy/beam to intensities not
related to the beam size.  We derived the beam sizes with accuracies
of 1 to 2\%.

For each of the seven observing runs, we derived one flux conversion
factor and the full width half maximum (FWHM) of the telescope beam,
$\theta_{beam}$, from the calibration measurements.  The total flux
density of the calibrators has been determined by integrating the
respective sky-subtracted map inside a manually placed polygon close
to the $3\sigma$ contour, i.e.  three times the root mean square (rms)
of the map's background noise. This procedure has been repeated six
times to average out the different polygons.  The beam size is derived
assuming  a spherical Gaussian beam shape. The actual conversion
factors and an extensive description of the calibration procedure can
be found in \citet{PosseltDIP}. The main uncertainty in determining
the absolute flux level in the maps is due to the variability of the
atmosphere.  The atmospheric opacity is frequently (every one or two
hrs) monitored by ``sky-dips'', then interpolated for the observations
in between.  Depending on the stability of the atmosphere, the error
in the flux level can be a few percent up to 10\%. Altogether, the
absolute calibration of the maps has an accuracy between 10 and 20\%
including systematic and statistical errors.

Similarly to the calibration, the integrated flux densities for a
cloud core are based on the mean of the integrated flux in three
polygons around the source peak position.  Where possible the polygon
borders are close to the $5\sigma$ contour.  In the case of
neighboring sources the polygon border between them corresponds to the
local minimum identified by eye.  A peak in the {\mm} maps is listed
as individual cloud core, if it has at least a $4\sigma$ detection.
Exceptions are made if the peaks are probably spurious due to map
artifacts (e.g. IRAS 04324+5106).  Multiple sources were identified by
eye, having at least $5\sigma$ and usually at least two closed
$\sigma$ contours around each peak.  Exceptions are made if the cloud
core morphology as a whole encourages the interpretation as multiple
sources (e.g. IRAS 03211+5446).  To estimate roughly the
beam-deconvolved angular diameter of the source, $\theta_{Source}$, a
spherical source is assumed, having a map area equal to the mean area
$A^{int}$ of the three polygons.  Thus, $\theta_{Source}$ can be estimated
by $A^{int} \approx A_{sph}= \pi (\theta_{beam}^2 +
\theta^2_{Source})$, where $\theta_{beam}$ is the corresponding beam
size of the observation run as described above.

\section{RESULTS}
\label{sec:results}

We detected {\mm} emission at a $3\sigma$-level in almost all of our
44 targeted regions.  Only in one region (\object[IRAS 05281+3412]{IRAS 05281+3412}), we did not detect any emission despite
a high sensitivity ($\sigma=15\rm\,mJy/beam$).  In three other
regions, no emission was detected, however due to limited sensitivity
($\sigma\approx500\rm\,mJy/beam$) only strong emission can be ruled
out. Figure~\ref{fig:maps} displays the maps of the 40 regions with
detections.  The mean source area $A^{int}$ as well as
$\theta_{Source}$, flux densities and the rms of the observations are
given in Table~\ref{tab:Messergeb}.

\subsection{Physical Quantities}
\label{ssec:quant}

\paragraph{Masses:}
In order to derive gas masses $M_g$ for each millimeter condensation
from the observed millimeter continuum emission, the following common
formula has been applied:
\begin{equation}
M_{g}=\frac{F_{\nu}\,D^{2}}{\kappa_m^d(\lambda)\,B_{\nu}(T_{d})}\,
        \frac{M_{g}}{M_{d}}  
\end{equation}
where $D$ is the distance, $B_{\nu}(T_{d})$ the Planck function at the
dust temperature $T_{d}$ and $\kappa_m^d (\lambda)$ is the mass
absorption coefficient of dust at the wavelength $\lambda$. The
measured flux density $F_{\nu}$ is integrated over the map area
$A^{int}$ as decribed above and listed in Table~\ref{tab:Messergeb}.
This approach assumes optically thin dust emission which is justified
by the long wavelength. The assumed gas to dust ratio is
$M_g/M_d=150$.  The linearöy interpolated dust opacities $\kappa_m^d(\lambda)$ from
\citet{Ossenkopf94}\footnote{thin ice mantles, density $n=10^{6}$
  cm$^{-3}$} have beeen used.

Furthermore, we need to assume a dust temperature $T_{d}$. Individual
estimates of dust temperatures for each IRAS source derived from IRAS
$60 \mu $m and $100 \mu $m data are listed in \citet{Schreyer96} and
repeated in in Table \ref{tab:srclst}, though these color temperature
are not physical dust temperatures.  \citet{Mueller2002} obtain a mean
temperature of $29\pm9$\,K from their radiative transfer models of 31
objects. Similarly, \citet{Hunter2000} derive an average dust
temperature of 35\,K.  It is safe to assume temperatures between 20
and 50\,K. We list the masses for a temperature of $T_{d}=20$\,K. For
$T_{d}=50$\,K, the gas masses are about 30\% of the values in
Table~\ref{tab:massenetc}.  The distances $D$ are in most cases
kinematic distances which imply uncertainties of up to 50\%.  A small
source of systematic uncertainty is the contribution of CO line
emission within the observed continuum bands.  \citet{Sandell2000}
finds a line flux contribution of about 8\% at 850\um{} for NGC\,6334
by comparing 850\um{} broadband JCMT UKT14 photometry with
position-switched spectra for both sources.  \citet{McCutcheon2000}
adopt this value for their 1.3\,mm observations. NGC\,6334 is a MSF
region similar to the star-forming regions investigated here.  CO
measurements of some of our sources indicate a line flux contribution
of less than 2\% (e.g.\ \citealt{Snell88,Carpenter90}).

\paragraph{Column densities:}
The molecular hydrogen column densities N(H) through the core centers
are derived as follows:
\begin{equation}
N(H)=\frac{{F_{\nu}}^{peak}}{{\kappa_m}^{d}(\lambda)
\Omega_{mb}B_{\nu}(T_{d})m_{H}}\frac{M_{H}}{M_{d}},
\end{equation}
where $m_{H}$ is the mass of a hydrogen atom, and $\Omega_{mb}$ is the
solid angle of the beam. The flux density ${F_{\nu}}^{peak}$ is the peak
flux density of each individual core.

\paragraph{Core densities:}
Assuming that the source has the same extent along the line of sight
as in the plane of the sky, one can derive the volume-averaged
hydrogen number density by
\begin{equation}
{n ( H )}_C =\frac{N( H) }{\theta_{Source}D},
\end{equation}
where $\theta_{Source}$ is the estimated beam-deconvolved angular
diameter of the source as described in \S\,\ref{ssec:cal}.  For crowded
fields of ``overlapping'' sub-sources the considered map area is
sometimes slightly smaller than the beam area.  Therefore,
$\theta_{Source}$ cannot be determined by beam deconvolution.  In
these cases we use the angular diameter of the beam to get a safe
lower limit on the core density even if the peaks are closer together.
The actual source diameters are listed in column 8 of Table
\ref{tab:Messergeb} where ``B'' marks the cases where the beam size
limited the size determination.

The masses, column densities, and number densities calculated from the
observed values (Table~\ref{tab:Messergeb}) are listed in
Table~\ref{tab:massenetc}.

\subsection{The Millimeter Continuum Maps}
\label{ssec:maps}
We show the {\mm} maps as contour maps together with 2MASS K-band
images and mark the positions of MIR and radio sources taken from the
MSX and NVSS PSC catalogs, respectively.  Thus, each panel in Figure
\ref{fig:maps} features:
\begin{itemize}
\item A grey-scale image representing the K-band 2MASS image.
\item Contours representing the (sub-)millimeter map. The spacing of
  the contours is given below the panel in multiples of $\sigma$, the
  rms of the map's background noise (Table \ref{tab:Messergeb}).  The
  first level is always $3\sigma$. The spacing between levels is
  indicated by the increment $\Delta$. If the increment changes, a
  thick contour is drawn at the level indicated in bold font.
\item The IRAS sources plus positional error denoted by a diamond
  together with an error ellipse.
\item Entries in the MSX point source catalog denoted by a plus sign
  (resolution 18\arcsec, positional uncertainty 2\arcsec).
\item NVSS radio point sources indicated by triangles (resolution 45\arcsec,
  positional uncertainty $<7\arcsec$).
\end{itemize}
As we aim to find particularly early stages of massive star formation,
protoclusters or {\PPC}, the most interesting millimeter sources are
those which have no association at shorter wavelengths and in the
radio continuum. The detected twelve massive pre-protocluster
candidates are compiled in \S\,\ref{sec:pre}.

\subsection{Source descriptions}

Each short description of the observed regions and the {\mm} maps is
titled with the name of the IRAS source in it. Regions with
pre-protocluster candidates are marked. The telescope used for the
observation and the respective wavelength are noted, too. When the
description mentions of NIR, MIR, or radio observations, the 2MASS
K-band images, the MSX PSC, or the NVSS PSC are meant, if not noted
otherwise.  For the NIR sources also 2MASS J-band and H-band were
taken into account to identify embedded sources by their reddening.  A
classification of the massive cloud cores according to Table
\ref{tab:stages} is tried.

\label{ssec:src}
\newcommand{\ra}{{(IRAM 1.27\,mm)}}
\newcommand{\rb}{{(IRAM 1.27\,mm)}}
\newcommand{\rc}{{(JCMT 850\um)}}  
\newcommand{\rd}{{(JCMT 850\um)}}  
\newcommand{\re}{{(JCMT 850\um)}}  
\newcommand{\rf}{{(JCMT 850\um)}}  
\newcommand{\rg}{{(SMT 870\um)}}   

\newcommand{\Sa}[1]{{ core \##1: \em pre-protocluster candidate}}
\newcommand{\Sb}[1]{{}}
\newcommand{\Sc}[1]{{}}
\newcommand{\Sd}[1]{{}}
\newcommand{\Se}[1]{{}}

\paragraph{\protect\object[IRAS 01195+6136]{IRAS 01195+6136} \re:}
This region is also known as S\,187 \citep{Sharpless59}. No \mm{}
emission was detected in the immediate vicinity of \object[IRAS 01195+6136]{IRAS 01195+6136}.
Only a small clump was detected 4\arcmin\ southwest of \object[IRAS 01195+6136]{IRAS 01195+6136}
and {2\arcmin} south of a star cluster.  The panel in
Figure~\ref{fig:maps} only shows the extract of the {\mm} map with the
clump and the star cluster. This extract includes \object[IRAS 01198+6136]{IRAS 01198+6136}
close to the cluster.

\paragraph{\protect\object[IRAS 02244+6117]{IRAS 02244+6117} \rb:}

The {\mm} emission has the shape of a band extending from north to
south delineating the border of a large emission nebula to the east.
The Digitized Sky Survey (DSS) images suggest that we see the border
of a bubble created by W\,4 (associated with the open cluster IC 1805)
about {30\arcmin} to the east.  \citet{Kraemer03} published MSX images
of \object[IRAS 02244+6117]{IRAS 02244+6117}, showing an arc of sources parallel to the {\mm}
emission shifted to the north-west and \object[IRAS 02245+6115]{IRAS 02245+6115}, a compact
{\hii} region. The dust cloud causing the {\mm} emission can be seen
as an extinction band against the background in the MSX 8.3\um{}
image.

\paragraph{\protect\object[IRAS 02575+6017]{IRAS 02575+6017} \ra \Sd{1a}:}
\label{src02575}
This source is also known as AFGL\,4029. We classify the centrally
peaked massive core as a young cluster since it is associated with
strong NIR/MIR sources.  The envelope of the core looks compressed on
the western side, which can be explained by inspecting the respective
DSS plate, showing an emission arc at the border of the {\mm}
emission.  This arc is the edge of the extended {\hii} region S\,199,
which is excited by an O7 star (HD 18326). It is possible that we see
triggered star formation here.  The region of \object[IRAS 02575+6017]{IRAS 02575+6017} has been
studied at different wavelengths by \citet{Deharveng1997},
\citet{Zapata2001}, and \citet{Ogura2002}, however with different
results on the star formation history of this region.
To the south, a small patch of {\mm} emission close to another IRAS
source is found. The two IRAS sources may be related since the 
above-mentioned arc encompasses both of them.

\paragraph{\protect\object[IRAS 02593+6016]{IRAS 02593+6016} \ra \Sc{1}:}

This IRAS source is located in the {\hii} region S\,201. Due to its
MIR appearance it is also known as AFGL\,416. Despite being only
12\arcmin{} east of the border of the extended {\hii} region S\,199
(see \object[IRAS 02575+6017]{IRAS 02575+6017}) there is no apparent connection between these
two regions. The region itself is powered by an O9.5 star
\citep{Mampaso89}.

In comparison with the DSS plate showing a bipolar emission nebula,
the {\mm} structure intersects the optical emission nebula and, thus,
is responsible for the dark lane in the nebula. Three molecular cores
can be identified in the {\mm} emission. The highest peak is closest
to the IRAS position, but there is a clear separation. If the offset
to the MSX source close to the peak is real, this core would be a
candidate for a massive protocluster (\S\ \ref{sec:pre}). However with
the close MSX source and the {\mm} peak almost within the IRAS error
ellipse, we classify this massive core as an {evolved protocluster}.

\paragraph{\protect\object[IRAS 03064+5638]{IRAS 03064+5638} \rg \Sa{1a}\Sd{1b}:}

For this region, also known as AFGL 5090, we have only a shallow map.
We classify the massive core \#1b as a {young cluster}: a faint IR
cluster, an MSX detection, and radio emission are present.  The
prominent K-band source in the cluster is extremely red. The MSX
detection is likely to be the same object deeply embedded in cloud
core \#1b.  Core \#1a, however is a massive {pre-protocluster
  candidate}.  Our {\mm} observations and the CS ($J=2\to1$) detection
by \citet{Carpenter93} indicate that there is still a large amount of
gas associated with the cluster. In contrast to \citet{Carpenter93},
we use a distance of 4.1~kpc \citep{Henning92b}, because it
corresponds better to the Galactic velocity field \citep{Brand93}.
Note that the CS-traced gas mass estimated by \citet{Carpenter93} is a
factor of 5 lower than our derived gas mass estimate from {\mm}
emission using the same distance of 2.2kpc. This may indicate that
most of the gas is not concentrated in high-density cores traced by
CS. Other reasons can be a different dust opacity, CS abundance, or
dust-to-gas mass ratio.

\paragraph{\protect\object[IRAS 03211+5446]{IRAS 03211+5446} \rg \Sa{1}\Sd{2}:}

An emission nebula, NIR sources together with an MSX as well as a
radio point source are found at the position of massive cloud core
\#2, thus classified as a young cluster, while the IRAS source is
about 1\arcmin{} to the east. The massive core \#1 does not show any
association making it a {pre-protocluster candidate}. The relatively
shallow {\mm} emission map indicates a rather large gas mass still
remaining in this star-forming region. Targeting the IRAS source, no
other indications of on-going star formation except for an H$_2$O
maser are known (Table \ref{tab:Literatur}), though these
non-detections may have missed the double-peaked cloud core.

\paragraph{\protect\object[IRAS 03236+5836]{IRAS 03236+5836} \rc:}

The {\mm} map around \object[IRAS 03236+5836]{IRAS 03236+5836} (AFGL
490) reveals a ``bridge'' to \object[IRAS 03233+5833]{IRAS
  03233+5833}. Both IRAS sources coincide with prominent molecular
cores. The ``bridge'' hosts another two, however less prominent cores.
Both prominent cores have MSX detections as well.  The source AFGL 490
is identified as a deeply embedded intermediate-mass young stellar
object surrounded by a disk \citep{Schreyer2002}. It is of great
interest because it is in a transition stage to a Herbig\,Be star.  In
contrast to \object[IRAS 03236+5836]{IRAS 05281+3412}, the other IRAS source
(\object[IRAS 03233+5833]{03233+5833}) is hardly studied.

\paragraph{\protect\object[IRAS 03595+5110]{IRAS 03595+5110} \rg:}

This region is known as the {\hii} region S 206 (NGC 1491, AFGL 5111).
The ``fluffy'' structure of the {\mm} emission and the large extent in
the FIR (listed in the IRAS small structure catalog) suggest that this
is a developed region without ongoing star formation, especially without outflow activity
\citep{Mook99}. But still there are two molecular cores with masses
of almost $100\,M_\odot$ each. One of them is associated with an MSX
point source.

\paragraph{\protect\object[IRAS 04073+5102]{IRAS 04073+5102} \rg \Sa{1, 2, 6, 7}s\Sc{1, 3}\Sd{5}:}

The {\hii} region S 209 (AFGL 550) may be one of the most distant
galactic {\hii} regions known. It is associated with \object[IRAS 04073+5102]{IRAS 04073+5102} and
\object[IRAS 04072+5100]{IRAS 04072+5100}.  \citet{Brand93} used \object[IRAS 04073+5102]{IRAS 04073+5102} to derive their
velocity field of the outer galaxy.  Therefore, we adopted their
distance of 8.2\,kpc, whereas \citet{Caplan00} chose a distance of
9.8\,kpc. In contrast to these far distances, \citet{Bica03b} put the
NIR cluster at a distance of 4.9\,kpc.

The shallow {\mm} map shows a ring-like structure. A star cluster and
some associated radio sources are not exactly in the center of this
ring, but they possibly shaped the ring. Since the {\mm} emission
coincides with a dark region in the faint optical emission nebula, the
molecular cores are located in front of the {\hii} region. The most
prominent core (\#1) with $4100\,M_\odot$ is close to \object[IRAS 04073+5102]{IRAS 04073+5102}
and an MSX source but clearly separated more than {10\arcsec}, thus it
is classified as a {pre-protocluster candidate} like three other cores
in the ring (\#2, \#6, \#7), which do not have any associations with
sources at other wavelengths. The cores \#3 and \#4 have NIR (core \#3
also MIR) detections which put them into stage 4 -- young clusters.
\object[IRAS 04072+5100]{IRAS 04072+5100} is close to cloud core \#5 and associated with an MSX
source and faint NIR emission -- another {young cluster}.

\paragraph{\protect\object[IRAS 04269+3510]{IRAS 04269+3510} \rg:}

This IRAS source is a Herbig Be star also known as LkH$\alpha$ 101.
This B0 - 0.5 star has a massive disk \citep{Tuthill02}.
\citet{Barsony91} identified a surrounding young star cluster (age
$\sim10^6$\,yr) with more than 100 members including low-mass objects
and brown dwarf candidates. The Herbig Be star illuminates the
reflection nebula NGC\,1579. The dark regions of this nebula match the
two condensations seen in the {\mm} emission. The 10\,$M_\odot$ cores
have no counterparts at other wavelengths.  Note that the distance to
this object is uncertain.  \citet{Stine98} and \citet{Tuthill02} put the
region at distances of 150\,pc and 340\,pc, respectively. We adopt the
original distance of 800\,pc determined by \citet{Herbig71}, because
\citet{Tuthill02} cannot rule out neither the 150\,pc nor the 800\,pc.

\paragraph{\protect\object[IRAS 04324+5106]{IRAS 04324+5106} \ra \Sd{1}:}
This region is also known as AFGL\,5124.  The many strongly peaked
patches around two major components of the {\mm} emission are scanning
artifacts. However, the observing procedure (chopping while scanning)
should preserve the fluxes and, thus, the detected halo around the two
major components should be real although it is probably not that
clumpy.  The two major components are associated with NIR emission.
Additionally, the massive core \#1, classified as {young cluster}, has
been detected by the MSX and NVSS surveys. The coordinates of the
radio source F3R 4467 \citep{Furst90} match the coordinates of core
\#2 also showing NIR emission. However, this is likely to be a strange
coincidence since the resolution of the observations by
\citet{Furst90} is too low to detect a different point source than the
NVSS point source.

\paragraph{\protect\object[IRAS 04329+5047]{IRAS 04329+5047} \rg \Sc{1}:}
AFGL\,5125 or S\,211 seems to be an older star-forming region than
AFGL\,5124 (above).  It belongs to the same cloud complex, but looks
much more evolved. A star cluster has appeared. In the shallow {\mm}
emission map, we detected remnants of the parental molecular cloud on
the outskirts of the cluster, though the remnants are still massive
and may form massive stars. The northern edge is associated with
\object[IRAS 04329+5047]{IRAS 04329+5047} from our sample, while the
southern edge is associated with \object[IRAS 04329+5045]{IRAS
  04329+5045}. Core \#1 is associated with an MSX source, therefore
the massive cloud fragment is classified as {evolved protoclusters}.

\paragraph{\protect\object[IRAS 05100+3723]{IRAS 05100+3723} \rc \Sd{1a}:}

This IRAS source is associated with LBN\,784, also known as AFGL\,5137
or S\,228. The dust emission has a compact, almost circular shape but
features two peaks.  The IRAS source coincides with the southern peak
(core \#1a).  Dense gas is detected, too (Table \ref{tab:Literatur}).
An NIR source along with a nebula \citep{Carpenter93} lead to the
classification as a young cluster. The most prominent K-band object on
the 2MASS image just the north-east of the IRAS PSC coordinates shows
red NIR colors. It is quite deeply embedded in the cloud. The northern
core shows no sign of ongoing star formation.

\paragraph{\protect\object[IRAS 05197+3355]{IRAS 05197+3355} \rg \Sb{1}:}

An NIR cluster of reddened stars is associated with \object[IRAS 05197+3355]{IRAS 05197+3355}.
However, the area of the cloud core is free of NIR sources. Only the
IRAS position is close to the peak. The MSX source is more than
{10\arcsec} away. The core could host a {protocluster}. The
envelope of the cloud core is relatively fragmented.

\paragraph{\protect\object[IRAS 05281+3412]{IRAS 05281+3412} \re:}
No {\mm} emission was detected towards this IRAS source.  Thus, no map
is shown in Figure~\ref{fig:maps}. Other efforts to detect tracers of
molecular gas failed as well (Table \ref{tab:Literatur}).
\object[IRAS 05281+3412]{IRAS 05281+3412} is associated with the open cluster
NGC\,1931.  The molecular gas apparently was dispersed by the stellar
cluster.

\paragraph{\protect\object[IRAS 05327-0457]{IRAS 05327-0457} \rb:}
This region is in Orion about 30{\arcmin} north of the Orion Nebula.
The dust emission seems to trace the northern outskirts of the Orion A
molecular cloud. \citet{Mookerjea2000} observed \object[IRAS 05327-0457]{IRAS 05327-0457} in the
FIR. Their radiative transfer models predict a flux of about 10\,Jy at
1\,mm. However, they have no {\mm} data to constrain the dust
emissivity, making their extrapolation difficult. We have not detected
any dust emission from the IRAS source, neither do CO observations
show the source.  As in our measurements, the molecular cloud has a
border at the IRAS position in the CO line maps by \citet{Chini97}. It
is puzzling why the prominent FIR source reported by
\citet{Mookerjea2000} has no {\mm} counterpart.

\paragraph{\protect\object[IRAS 05341-0530]{IRAS 05341-0530} \rg:}
This IRAS source is located on the eastern border of the Orion Nebula.
A shallow observation of this area revealed no {\mm} detection.
However, other tracers of dense molecular gas have been detected
(Table \ref{tab:Literatur}).

\paragraph{\protect\object[IRAS 05345+3157]{IRAS 05345+3157} \re:}
The IRAS source is associated with a star cluster and a nebulosity.
However, the star cluster is not associated with any {\mm} emission.
The cluster has apparently ``burned'' a hole into the molecular cloud
leaving an arc of dust emission to the north and east. Two prominent
condensations are detected in the arc {1\arcmin} north-east of the
IRAS source which may be forming stars of intermediate mass. At least
core \#1 is associated with an MSX source.  Interferometric HCO$^+$
observations by \citet{Molinari02} do not show these two prominent
cores, but a clumpy cloud.

\paragraph{\protect\object[IRAS 05355+3039]{IRAS 05355+3039} \re:}
A small dust cloud is surrounding \object[IRAS 05355+3039]{IRAS 05355+3039}. The IRAS source is
associated with a nebulosity (actually about 7.5\arcsec{} south).
\citet{Ishii02} find that NIR emission is not only scattered light,
but also PAH emission. Additionally, there is an MSX source at the
position of the nebulosity. Since the dust emission is surrounding the
nebulosity, the embedded source probably cleared a cavity in its
parental could.  The part of the ellipse in the north-western corner
of the map is the error ellipse of \object[IRAS 05354+3041]{IRAS 05354+3041}.

\paragraph{\protect\object[IRAS 05375+3540]{IRAS 05375+3540} \rf \Sd{1a}:}
This and the following IRAS source (\object[IRAS 05377+3548]{05377+3548}) 
are the FIR-brightest IRAS sources in the molecular cloud S\,235. This
molecular cloud was studied by \citet{Evans81a} and \cite{Evans81b} in
the NIR and MIR as well as with molecular line emission. They
distinguish two velocity components of this cloud. The dust emission
recorded in this map belongs to the $v_{LSR}=-17$\,km/s component.

\object[IRAS 05375+3540]{IRAS 05375+3540} is associated with the
optical nebulosity S\,235\,A.  The K-band image shows two bright
objects in this nebulosity. An MSX source and a radio source are also
located at the IRAS source position. Already \citet{Evans81a}
identified an infrared source there (EB IRS 3). However the peak of
core \#1a (230\,M$_\odot$) is about 20\arcsec{} south of the the IRAS
source. NIR sources at the core's center lead to the classification as
a {young cluster}. The core has two smaller extensions to the north
and to the south.  Another bright K-band source is 20\arcsec{} south
of the maximum of the dust emission. It is associated with S\,235\,B
and the infrared source EB IRS 4. The massive cloud core hosts H$_2$O
masers as signs of deeply embedded on-going star formation, as also
pointed out by \citet{Felli97}.

The map targeting \object[IRAS 05375+3540]{IRAS 05375+3540}
encompasses the IRAS source \object[IRAS 05375+3536]{05375+3536} which
is close to S\,235\,C.  A second, smaller condensation has been
detected to the south slightly east of \object[IRAS 05375+3536]{IRAS
  05375+3536}.

\paragraph{\protect\object[IRAS 05377+3548]{IRAS 05377+3548} \rf:}
The second velocity component of the molecular cloud S\,235 is at
$v_{LSR}=-20$\,km/s (see \object[IRAS 05375+3540]{IRAS 05375+3540}). It hosts the bright source
\object[IRAS 05377+3548]{IRAS 05377+3548}, but also \object[IRAS 05374+3549]{IRAS 05374+3549}. The {\mm} emission from
this cloud part is most intriguing.  It features many separated
prominent cores and various elongated structures.  The western area is
populated with NVSS radio sources. A large emission nebula can be
found in the DSS in the same area. MSX sources with apparent
counterparts are associated with three cores (\#2a, \#4 and \#7), with
the star BD+35\,1201, an O9.5V star, and a red NIR object (MSX6C:
G173.6098+02.8183). The cores 2a and 4 show K-band counterparts.
These objects were also detected by \citet{Evans81a} as EB IRS 1 and
2. The components 2b and 2c are elongated and pointing towards
component 2a.  We interpret these structures as pillars eroded by the
radiation of the forming stars in core 2a.  The cores 1 and 3 are
similar in mass (50-60\,M$_\odot$) and shape to the cores 2a and 4.
The lack of NIR and MIR emission suggests that these cores are younger
and less evolved than the cores 2a and 4.  They may be
intermediate-mass pre-stellar cores.  The two patches of dust emission
to the east (components 7, 8, and 9) are approximately at the eastern
edge of the molecular could mapped by \citet{Evans81a}. Core \#9 is
massive with MIR and NIR emission, thus classified as a young cluster.

\paragraph{\protect\object[IRAS 05480+2544]{IRAS 05480+2544} and \protect\object[IRAS 05480+2545]{IRAS 05480+2545} \rb:}

Our map covers two bright IRAS sources.  \object[IRAS 05480+2545]{IRAS 05480+2545} is associated
with {\mm} emission and an MSX source while this is not the case for
the other IRAS source. At the north-eastern edge of the cloud core,
where the MSX source is located, the 2MASS K-band image shows
extended emission and several reddened sources. An explanation could
be that embedded young stars (the MSX source) are emerging at this
side of the cloud core.  The IRAS source 05480+2544 is associated with
the {\hii} region BFS\,48 \citep{Blitz82}.

\paragraph{\protect\object[IRAS 06006+3015]{IRAS 06006+3015} \rb \Sd{1}:} 

This source is associated with AFGL\,5176 in S\,241.  We detected a
massive core with an envelope. The peak coincides with the IRAS
position. \citet{Mueller2002} found a weak source in their 350~{\um}
continuum map.  The mass derived from our data agrees with the mass
obtained by \citet{Mueller2002}.  No MSX point sources or NVSS radio
point sources are present at the dust core.  However, there is a
moderately bright and red NIR source at the IRAS position leading to a
classification as a {young cluster}.

\paragraph{\protect\object[IRAS 06013+3030]{IRAS 06013+3030} \rg:}

A shallow observation of this area revealed no {\mm} detection. Other
tracers of dense molecular gas have not been detected either (Table
\ref{tab:Literatur}). No molecular gas remained in the reflection
nebula which is associated with this IRAS source.  Only the recently
reported star cluster remains as 
 sign of past star formation
activities \citep{Bica03b}.

\paragraph{\protect\object[IRAS 06055+2039]{IRAS 06055+2039} \rb \re:}
\object[IRAS 06055+2039]{IRAS 06055+2039} is associated with
S\,252\,A, which was described in detail by \citet{Koempe89}. While
\citet{Shepherd96} found that the main peak in $^{12}$CO $(J=2-1)$ is
located slightly east of the IRAS source, the strongest CS(J=2-1)
emission was reported north of \object[IRAS 06055+2039]{IRAS
  06055+2039} by \citet{Carpenter95b}. \citet{Mueller2002} detected a
350~$\um$ source at the IRAS position.

There are two maps for \object[IRAS 06055+2039]{IRAS 06055+2039}: one obtained at the JCMT at a
wavelength of 850{\um} (JCMT) and one obtained at the IRAM 30m
telescope at 1300{\um} (IRAM).  In both maps there is a strong, nearly
spherically shaped millimeter peak about $30\arcsec$ east of the IRAS
source. MSX point sources and a NVSS radio point source are associated
with the IRAS source, but not with the {\mm} peak. In the 2MASS K-band
image, an extended star cluster around the IRAS source can be seen.
Dense gas and masers but no outflow were detected towards this source
(Table~\ref{tab:Literatur}). It would be an excellent candidate for a
massive pre-protocluster but it has ``only'' a mass of 97\,$M_\odot$
(cf.~\S~\ref{ssec:cand}).  At $850\mu $m, the source is slightly
elongated, and in the direction of its elongation, one can find
further small dust clumps.  They may represent remnants of a
fragmentation process.

\paragraph{\protect\object[IRAS 06056+2131]{IRAS 06056+2131} \ra \rc \Sd{1, 2} and 06058+2138 \rb \rc \Sa{1}:}

This region, adjacent to the {\hii} region S\,247, is intriguing not
only because of its two bright IRAS sources, but even more because
these IRAS sources are connected by a bridge of material seen at
{850\um} (JCMT).  Our maps obtained at the IRAM 30m telescope were not
large (and sensitive) enough to detect this bridge.  This bridge may
be caused either by tidal forces between the two IRAS sources or may
be a remnant of the fragmentation process. This filament is located
close to the ionization front limiting the visible {\hii} region
S\,247 \citep{Koempe89}. (The NVSS sources are inside S\,247.)
Massive cloud cores reside at each endpoint of the bridge which are
associated with the targeted IRAS sources.

After the multi-line study by \citet{Koempe89},
\citet{Carpenter95a,Carpenter95b} made a large-scale molecular line
and NIR study of the Gemini OB association which includes these two
IRAS sources.  A detailed FIR study was conducted by \citet{Ghosh00}.
All these investigations characterize this region as an active
star-forming region. The detected {\mm} cores are associated with
embedded NIR star clusters and MIR sources.  However, core \#1 of
\object[IRAS 06058+2138]{IRAS 06058+2138} is separated by more than
{10\arcsec} from MSX and notable NIR sources. Thus, this is a
{pre-protocluster candidate}. The other cores are classified as young
clusters. Masers are found in the region.  Obviously the {\hii} region
S\,247 is more evolved than the cloud cores and may now compress the
molecular cores to the east.  \citet{Ghosh00} suggests that
\object[IRAS 06058+2138]{IRAS 06058+2138} is the youngest of the three
bright IRAS sources 06056+2131, \object[IRAS 06058+2138]{IRAS
  06058+2138}, and \object[IRAS 06061+2151]{IRAS 06061+2151} (further
up to the north-east, see next paragraph).

Cloud core \#1 of \object[IRAS 06058+2138]{IRAS 06058+2138} is rather
spherical. We derive a radial density profile for this core in
Appendix \ref{sec:radial}.  The problems and limitations of such an
analysis are discussed there, too.

\paragraph{\protect\object[IRAS 06061+2151]{IRAS 06061+2151} \rb:}
This is the third bright IRAS source in the molecular cloud
surrounding S\,247. This source was also included in the above
mentioned studies of the other two IRAS sources (
\object[IRAS 06056+2131]{IRAS 06056+2131}, \object[IRAS 06058+2138]{IRAS
  06061+2151}).  Recently, \cite{Anandarao04} have found H$_2$
emission knots indicating a jet originating from the deeply embedded
star cluster associated with \object[IRAS 06061+2151]{IRAS 06061+2151}
($A_{\rm V} \leq $ 30\,mag).  Based on their NIR study of the cluster,
they identified several Class\,I and II sources and a massive
protostar. This protostar is close to the position of the MSX source.
However, the peak of the {\mm} emission is 10\arcsec{} west of the
cluster and the MSX source.  Thus, there might be a deeply embedded
protostar of intermediate mass in the center of the cloud core.

\paragraph{\protect\object[IRAS 06063+2040]{IRAS 06063+2040} \rf:}
This object is associated with AFGL 5183 and the ultra-compact {\hii}
region S\,252\,C. \object[IRAS 06063+2040]{IRAS 06063+2040} is a part of the western molecular
cloud fragment of S\,252, though it is not in the dense ridge which
delineates the visible {\hii} region on the western side \citep{Felli77}.

We detect extended millimeter emission covering a rich NIR cluster.
The emission has a cometary structure extending from a prominent core
to the north towards the IRAS source. A bright NIR object and an NVSS
source coincide exactly with the IRAS source.  On the western slope
close to the peak (7\arcsec), there are an MSX point source and 2MASS
K-band sources.  The rich near-infrared cluster has a gas reservoir to
form more stars.

\paragraph{\protect\object[IRAS 06068+2030]{IRAS 06068+2030} \rf:}
 
\object[IRAS 06068+2030]{IRAS 06068+2030} belongs to the eastern molecular cloud fragment of
S252 and is associated with the compact {\hii} region S\,252\,E. The
region was investigated by, e.g., \citet{Felli77}, \citet{Koempe89},
and \citet{Carpenter95a,Carpenter95b}. A near-infrared cluster is also
present at this position \citep{Bica03a}.

The millimeter emission reveals the presence of two dust cores and an
extended envelope. It surrounds an NIR cluster associated with
\object[IRAS 06068+2030]{IRAS 06068+2030}, an MIR, and radio source
with an opening to the east. Core \#1 is located south of the cluster,
core \#2 is located north-west of it. The envelope is extending from
core \#2 to the north-east showing a radio source and two MSX sources
on the cluster's side. The cluster may have shaped the opening in its
parental cloud.

\paragraph{\protect\object[IRAS 06073+1249]{IRAS 06073+1249} \rg \Sd{1}\Sa{2}:}

The {\hii} region S270 is associated with this IRAS source.  Neither
\citet{Carpenter90} nor \citet{Fich93} found structures in the radio
continuum emission. \citet{Carpenter93} mapped the region in J, H, K
and CS($J=2-1$). There is a star cluster in the NIR at the IRAS
position. In CS, \citet{Carpenter93} found two maxima, one at the IRAS
position, but the stronger second peak is situated around 70$\arcsec$
east of \object[IRAS 06073+1249]{IRAS 06073+1249}.  The {\mm} emission
also has two peaks, however the stronger peak is coincident with the
IRAS source, while the second weak millimeter peak lies around
50$\arcsec$ east of \object[IRAS 06073+1249]{IRAS 06073+1249}. The
main {\mm} peak (core \#1) is coincident with an MSX point source, an
NVSS radio point source and a bright 2MASS K-band source -- {a young
  cluster}.  The core \#2 is a massive {pre-protocluster candidate},
since it is not detected in any of these other surveys.

\paragraph{\protect\object[IRAS 06099+1800]{IRAS 06099+1800} \rg \Sb{1}\Sd{2}:}
\object[IRAS 06099+1800]{IRAS 06099+1800} is located between two large {\hii} regions S\,255 and
S\,257, it is associated with the star cluster S\,255-2.
\citet{Mezger88} observed the region at 350~$\mu $m and 1300~$\mu $m
and identified three components, FIR1 - FIR3. The star cluster
S\,255-2 was investigated by \citet{Howard97} and \citet{Itoh2001}.
Both concluded that S255-2 is a very young stellar cluster with
massive stars in different evolutionary stages.

The {\mm} emission has the shape of a bar extending in north-south
direction and separates the {\hii} regions (note the NVSS sources east
and west).  The bar breaks up in several individual cloud cores.
Apart from the sources already found by \citet{Mezger88} -- FIR1 (core
\#1), FIR2 (core \#2), and FIR3 (core \#4a) -- three more, weaker
sub-sources are apparent. Where available in \citet{Mezger88}, the
masses agree with the masses derived here. The dust cores are massive
and very dense.  The cloud core \#2 is associated with the IRAS source
and NIR sources, thus a {young cluster}.  \citet{Crowther03} published
the MSX map of the cluster. The northern cloud core (\# 1) hosts an
MSX source and an UC{\hii} region which shows two peaks separated by
$\sim2\arcsec$ \citep{Kurtz94}.  This UC{\hii} region is also present
as radio source in the NVSS catalog.  While the star cluster that
formed in the southern core already emerged from its parental cloud,
the forming massive star(s) still have an UC{\hii} region. However,
the core gives already rise to an MSX detection leading to a
classification as an evolved protocluster.

\paragraph{\protect\object[IRAS 06105+1756]{IRAS 06105+1756} \rg \Sa{1a, 1b, 2, 3}:}
This object is associated with S\,258, and there are not many
observations reported in the literature (Table \ref{tab:Literatur}).
The {\mm} map displays a cloud core with a fragmented envelope.  The
cloud core is located east of \object[IRAS 06105+1756]{IRAS 06105+1756}, however, it is still
within the error ellipse of the IRAS position.  Since an NVSS radio
point source and an MSX point source coincide much better with the
IRAS source and a small cluster of NIR sources, we attribute the IRAS
source rather to the cluster than to core \#1a, which is then a
{pre-protocluster candidate}. There is another {\mm} source about
4$\arcmin$ east of \object[IRAS 06105+1756]{IRAS 06105+1756}. We notice no sign of star
formation associated with any of the cloud fragments. Since the
fragments have enough mass to form massive stars, they are classifiead
as {pre-protocluster candidates}, too.

\paragraph{\protect\object[IRAS 06114+1745]{IRAS 06114+1745} \re:}
There is not much known about the nature of \object[IRAS 06114+1745]{IRAS 06114+1745},
associated with AFGL 5188.  \citet{Carpenter95b} found an emission
maximum in CS ($J=2-1$) about 1$\arcmin$ south of the IRAS source.
This would correspond to the least massive (6$M_{\odot}$) core \#3.
The IRAS source itself is located on the southern border of the most
massive core \#1a (17$M_{\odot}$).  Furthermore, it corresponds
to an MSX point source, an NVSS radio point source and a bright object
in the star cluster seen in 2MASS K-band emission. The 2MASS K-band
also shows an agglomeration of sources around the southern core \#3.
It seems that star formation has lead to the disruption of the
molecular cloud with some intermediate-mass cores still surviving.

\paragraph{\protect\object[IRAS 06117+1350]{IRAS 06117+1350} \rb \Sd{1}:}
This source, associated with AFGL 902, is located in the western part
of the H{\sc ii} region S\,269. Using near-infrared observations,
\citet{Eiroa94} identified \object[IRAS 06117+1350]{IRAS 06117+1350} as a double source, IRS2e
and IRS2w, separated by roughly 4$\arcsec$. \citet{Jiang2003} studied
the rich embedded cluster associated with S\,269 in the near infrared.
From line ratios, an H$_2$ molecular jet, strong Br$\gamma$ emission,
and an IR excess they inferred that both IRS2e and IRS2w are massive
young stellar objects in an evolutionary stage comparable to low-mass
class I sources. We classify the massive core as a young cluster,
because of the bright NIR source.

The {\mm} emission extends from its peak associated with the young
stellar objects, IRS2e and IRS2w, to the south-west. The emission
extends to the locations of the more deeply embedded objects
(subregion 2 in \citealt{Jiang2003}). The millimeter peak is coincident
with an MSX point source, too. A few NVSS radio sources can be found
in the optical H{\sc ii} region. This emission nebula is probably located in
front of the molecular cloud. Otherwise one would expect dark areas in
the optical nebula associated with the millimeter structure which is
not the case.

\paragraph{\protect\object[IRAS 06155+2319]{IRAS 06155+2319} \rb:}
\object[IRAS 06155+2319]{IRAS 06155+2319} is associated with the {\hii} region BFS 51. The
embedded cluster around the IRAS source was investigated by
\citet{Carpenter90,Carpenter93}. The north-south-elongated {\mm}
emission is located north-east of \object[IRAS 06155+2319]{IRAS 06155+2319}, similar to the CS
emission detected by \citet{Carpenter93}. Both the CS and the {\mm}
emission morphology show a slightly steeper gradient to the east
compared to the opposite side.

The {\mm} emission peak is associated with a deeply embedded NIR
source. Other bright NIR sources are associated with the IRAS source
and a NVSS radio source at the western side of the molecular cloud.
Much fainter {\mm} emission is found close to \object[IRAS 06156+2321]{IRAS 06156+2321}. The
bright K-band object within the millimeter emission structure is
highly reddened. It is not present in the 2MASS J-band image.

\paragraph{\protect\object[IRAS 06308+0402]{IRAS 06308+0402} \rd:}
This IRAS source, located about 1$^{\circ}$ south of the Rosette
Nebula, is embedded in very extended dust emission with many
intermediate-mass dust cores. The most massive one (core \#1a) has a
mass of 12~M$_{\odot}$. An NIR star cluster is embedded in this
molecular cloud. While some of the stars in the apparent center of the
cluster are not deeply embedded in the cloud, most of the NIR objects
show high reddening.  The IRAS source is located close to the cluster
center.

A molecular outflow and H$_2$O masers have been detected towards
\object[IRAS 06308+0402]{IRAS 06308+0402} further indicating on-going
star formation.  This region and its molecular cloud Monoceros OB2
have been studied e.g.\ by \citet{Cox90} and \citet{Phelps97}, who
both suggested sequential star formation triggered by the nearby
stellar association NGC 2244.

\paragraph{\protect\object[IRAS 06319+0415]{IRAS 06319+0415} \rg \Sd{1}:}
This source is also called AFGL~961 and it is a well studied object
in the Rosette Nebula Giant Molecular Cloud. This pre-main sequence
binary consists of early B-type stars, the western part powering an
extended outflow while bow shocks indicate past outflow activity of
both components \citep{Aspin98}.

The strong millimeter peak has nearly spherical morphology and
corresponds to the IRAS source, an MSX point source and a bright
object in the 2MASS K-band emission within positional uncertainties.
The high mass of 550\,M$_{\odot}$ and the relatively high number
density at the peak position ($5.0\cdot10^5\rm\,cm^{-3}$) mark
\object[IRAS 06319+0415]{IRAS 06308+0402} as one of the most dense
massive cores.

\citet{Phelps97} investigated the NIR sources around core \#1. The
masive core is classified as young cluster since the NIR sources are
apparent in the 2MASS K-band image. One of the theories of
\citet{Phelps97} for the region is sequential star formation,
triggered by the nearby OB-association NGC~2244, but they also took
spontaneous star formation into consideration. Recent X-ray
observations with {\sl Chandra} by \citet{Townsley2003} provide
observational evidence for strong wind shocks in the Rosette Nebula.
This could support the sequential star formation theory.

We note that a distance of 1.4 kpc has been adopted by
\citet{Townsley2003} for NGC~2244 and the Rosette Nebula, while we use
1.6 kpc for the dust emission around \object[IRAS 06319+0415]{IRAS 06319+0415}.

\paragraph{\protect\object[IRAS 06380+0949]{IRAS 06380+0949} \rg:}

A shallow observation of this part of NGC 2264 revealed no {\mm}
detection.

\paragraph{\protect\object[IRAS 06384+0932]{IRAS 06384+0932} \rc \Sd{1a, 1b}:}
\object[IRAS 06384+0932]{IRAS 06384+0932} is located on the southern edge of the NGC\,2264
molecular cloud. It is associated with the prominent NIR source known
as Allen's source \citep{Allen72} or NGC\,2264\,IRS1. A
multi-wavelength study by \citet{Schreyer2003a} suggests that
NGC\,2264\,IRS1 is a young B-type star with low-mass companions
located in a low-density cavity surrounded by a clumpy, shell-like,
and dense cloud remnant. Several outflow systems and their driving
sources were identified using interferometer data. The associated core
\#1a is thus classified as a {young cluster}. The other massive
core (\#1b), also a {young cluster}, is associated with a very red
NIR source, only detected by 2MASS in the K-band, also listed by
\citet{Bica03a}.

The single-dish map of the {\mm} emission surrounding NGC\,2264\,IRS1
shows a multiply-peaked cloud. \citet{Wardthomson} have already
observed this part of the NGC\,2264 molecular cloud at 1300{\um} ,
800{\um}, 450{\um} and 300{\um} and identified five sources: MMS1 to
MMS5, also marked on our map. The two sources MMS5 and MMS4 cannot be
resolved in our map (850\um). \citet{Wardthomson} could separate these
two sources only at shorter wavelengths. The masses we derived differ
by a factor of 2 from their masses for MMS2 and MMS4 combined together
with MMS5 and by a factor of 5 from the mean mass of MMS3.  While also
the column densities correspond roughly to each other, their volume
densities are up to two orders of magnitudes larger because
\citet{Wardthomson} averaged over a smaller source area (the source's
FWHM) and thus a smaller extension along the line of sight to convert
the column density to a volume density.

North of \object[IRAS 06384+0932]{IRAS 06384+0932}, we detected further patches of {\mm}
emission closer to \object[IRAS 06382+0939]{IRAS 06382+0939}\footnote{This source almost passed
  our selection criteria: $S_\nu(100\um)=499.3$\,Jy.}. A deeper map
was obtained by \citet{Wolf2003}, who investigated this cloud part at
far-infrared, submillimeter and millimeter wavelengths. They also
detected the cloud core \#2, which is close to the origin of the
massive molecular outflow NGC\,2264\,D \citep{Margulis88} and might
harbor the driving source.

\paragraph{\protect\object[IRAS 06412-0105]{IRAS 06412-0105} \re:}
We detected a relatively low-mass dust core in this region. It is
coincident with a bright object in the 2MASS K-band, an MSX point
source and an NVSS radio point source. All these emissions line up
very well, but are separated by 1$\arcmin$ from the IRAS source in the
west.  To reconcile this discrepancy we inspected the IRAS ISSA
images. These images show a source elongated in east-west direction
with the IRAS PSC position ($06^h41^m12\fs5$,
$-01\degr05\arcmin02\arcsec$) located in the western part of the
emission and not in the center. The cataloged position does not
accurately describe the peak and center of the IR emission. We fitted
Gaussians to the 4 IRAS maps and found as center coordinates
($06^h41^m16\fs3\pm0\fs7$, $-01\degr05\arcmin14\arcsec\pm2\arcsec$)
without any trend with the wavelength. This position aligns much
better with the NIR, MIR and {\mm} emission (the latter one being
$06^h41^m15\fs8$, $-01\degr05\arcmin11\arcsec$). We conclude that IRAS
measurements, e.g.  the luminosity of 2000\,L$_\odot$, can be
attributed to the cloud core, however observations with better
resolution pointing to the IRAS position might have missed the core
and resulted in the non-detections as the observations listed in
Table~\ref{tab:Literatur}. This example demonstrates very clearly that
newer FIR data is needed.

\paragraph{\protect\object[IRAS 06567-0355]{IRAS 06567-0355} \rd \Sd{1}:}
This IRAS source is associated with the bipolar nebula NS\,14
\citep{Neckel84}, also known as H{\sc ii} region BFS 57. Detailed
observations of this cloud core, associated with an MSX and a NVSS
source, were performed by \citet{Neckel89} and \citet{Howard98}. Both
conclude that a trapezium of stars (spectral types B0 to A5) power a
small H{\sc ii} region. The millimeter continuum map by
\citet{Neckel89} corresponds to our findings. The mass estimate for
the cloud core \#1, {a young cluster}, of 200\,M$_\odot$ matches
with their estimate.  They fitted the SED with three dust components
and concluded a gas mass of 355\,M$_\odot$ for the coldest component
(10\,K).

We detected {\mm} emission from another IRAS source (
\object[IRAS 06567-0350]{06567-0350}) {5\arcmin} to the north. A star
cluster is embedded in this cloud fragment \citep{Lada03}, also giving
rise to MSX detections.

\paragraph{\protect\object[IRAS 06581-0846]{IRAS 06581-0846} and \protect\object[IRAS 06581-0848]{IRAS 06581-0848} \re:}
No dust emission was detected from either IRAS source. Extended
diffuse {\mm} emission can be found in the western neighborhood of the
IRAS positions. The young star cluster in this region was reported by
\citet{Ivanov02} as a new discovery. This region is also known as
the H{\sc ii} region BSF\,64. In the mapped region, the NVSS catalog
lists two sources and so does the MSX catalog. The cloud remanent is
still massive and for the NIR sources classified as young cluster.

\paragraph{\protect\object[IRAS 07029-1215]{IRAS 07029-1215} \rd:}
This IRAS source is located within the {\hii} region S 297 and is also
associated with the reflection nebula vdB 94 \citep{Bergh1966}. The
{\hii} region is powered by the B1II/III star HD 53623
\citep{Houk1988}.  West of the IRAS source, a dark cloud can be found.
Little is known about \object[IRAS 07029-1215]{IRAS 07029-1215}.  The {\mm} maps show emission
from three distinct filaments, the eastern one containing the IRAS
source, many MSX and NVSS radio point sources. There is only weak dust
emission at the IRAS position.  The central filament contains the
strongest emission peak (core \#1a) and is known as UYSO\,1
\citep{Forbrich04}. UYSO\,1 seems to be a very young intermediate-mass
protostellar object with neither 2MASS K-band, MSX nor NVSS point
source counterparts.  The western filament contains a weak emission
peak within the dark cloud.  This source (core \#3) also has no
counterparts at the other wavelengths considered.

\paragraph{\protect\object[IRAS 07299-1651]{IRAS 07299-1651} \rb:}
This region has designations as a reflection nebula (DG 121,
\citealt{Dorschner63}) and as an H$\alpha$ emission nebula (RCW 7,
\citealt{Rodgers60}).  The IRAS source is coincident with a single
{\mm} peak with a small, rugged envelope.  The 28\,M$_\odot$ core has
an NIR counterpart and an MSX detection.  \citet{Walsh99,Walsh2001}
study the ultra-compact H{\sc ii} region and the methanol maser
embedded in the detected cloud core.
Despite a similar sensitivity of the IRAM and the JCMT maps of this
source, we did not detect the cloud core in the latter.

\section{DISCUSSION}
\label{sec:discussion}

\subsection{Masses and Densities}
\label{ssec:othersurv}

Distances, dust temperatures and dust opacities are the main
uncertainties for the mass estimates. Accurate distances are crucial
since masses etc.\ depend quadratically on the distance, but the
distances are difficult to obtain. Most distances are kinematic
distances already compiled by \citet{Henning92b} and, if necessary,
corrected using new molecular line data and the velocity field of the
outer galaxy by \citet{Brand93}.  Kinematic distances have fairly
large uncertainties (up to 50\%), especially in the distant outer
galaxy where the rotation curve is uncertain.  The mass estimate
depends also on the assumed effective temperature, as mentioned in \S
\ref{ssec:quant}. And still the dust opacities are uncertain within a
factor of 2.

Given these difficulties, a comparison of the mass estimates with
\citet{Mueller2002} is in order for those source we have in common.
Considering the combination of our larger number of cloud components,
the masses agree within 40~\% for IRAS 06006+3015, 20~\% for IRAS
03236+5836 and IRAS 06055+2039, and 10~\% for IRAS 05377+3548, which
is reasonable.  \citet{Mueller2002} also used the dust opacities by
\citet{Ossenkopf94}. The temperature and distances were adjusted for
the comparison.

The total gas masses derived in the described way range from 1 to
5000\,$M_\odot$ for all measured objects, including all the faint and
small sources.  Similar values are reported by \citet{Beuther2002II},
\citet{Hunter2000}, and \citet{Faundez04} ($10^2\,M_\odot -
10^4\,M_\odot$, 260\,$M_{\odot} - 10^5\,M_\odot$, and 6\,$M_{\odot} -
10^4\,M_\odot$, respectively). Our total range extends to smaller
values and does not reach these high masses, because we are interested
in individual cores and break down the total masses into sub-sources
for each peak.

Column densities estimated for the core centers are of the order of
$10^{22}$~cm$^{-2}$ to $10^{23}$~cm$^{-2}$. These values are
comparable to those of other surveys (e.g. \citealt{Beuther2002II}).
\citet{Faundez04} stated column densities considerably smaller than
$10^{26}$~cm$^{-2}$, but did not report individual values (average
density $5\cdot10^{23}\rm\,cm^{-2}$, \citealt{GaraySicily05}).

Number densities vary from $10^{3}$ to $10^{5}$~cm$^{-3}$ for most
sources.  Only in the case of IRAS
  06099+1800 (components \#1 and \#2) and IRAS
  04269+3510 (both components) the number densities reach
$10^{6}$~cm$^{-3}$.  For IRAS 06099+1800 an
underestimation of the source size due to sub-source crowding is a
likely explanation.  IRAS 04269+3510 is
located relatively nearby at a distance of 0.8 kpc.  The source size
equals nearly the beam size. Therefore, the deconvolved source size
and subsequently the core density have high uncertainties.

\subsection{MIR sources and massive molecular cloud cores}
\label{ssec:multi}
With the large maps (on average larger than $20^{\square\arcmin}$), we
identified a large amount of cores and clumps with a variety of
morphologies, rarely spherically shaped and often overlapping. Single
sources were detected in only 23\% of the 40 mapped regions, a
percentage similar to the findings of \citet{Faundez04} (27\%).  We
detected many cores not associated with IRAS sources.  Only 50\% of
the bright IRAS sources are themselves associated with a {\mm} peak.
Thus, data collected in previous studies only on the IRAS sources have
only limited value for the interpretation of the early phases of
star-forming cores.  It seems to be useful to check for associations
to other infrared catalogs.  In the following, we analyze the
association of our cloud components to the MSX point source catalog.

\paragraph{Mass vs Distance:}
The mass distribution of the cloud components, we identified, is
displayed in Figure~\ref{fig:MassvsDist} giving the source counts for
logarithmic mass intervals.  Each of the 128 cloud components is
considered separately. We note that we take into account only the JCMT
observations for IRAS 06055+2039 and only the IRAM observations for
IRAS 06056+2131 and IRAS 06058+2138 to avoid double-counting.  The
turn-over occurs at masses of 100\,$M_\odot$.
Figure~\ref{fig:MassvsDist} also includes the information at which
distances the cores are found. It is conspicuous that the massive
cores are predominantly found at large distances. This is a result of
several selection effects. Massive cores are rare and, therefore,
unlikely to be found close to the sun.  On the other hand, the minimum
detectable mass rises with the distance, which explains the lack of
small masses at large distances.  For example with the average
sensitivity we reached, it is not possible to detect a 30\,$M_\odot$
core more distant than 6\,kpc.  Furthermore, a cluster of small cores
may not be resolved at large distances and thus looks like one massive
core. The distances, especially in the outer Galaxy, have severe
uncertainties which affect the masses and the mass distribution.
However, an 50\% error in the distance would lead to a mass wrong by a
factor between 0.25 and 2.25.  Concerning the mass bins (width 0.5 in
$log(M/M_\odot)$), this leads to a an uncertainty of only one bin.
Because of these selection effects one would need to restrict
distances to a small range effects in order to derive a realistic mass
function of the detected cores. Then, however, the numbers get too low
to establish significant statistical results.

\paragraph{Source associations:}
We investigate the correlation of {\mm} peaks (cloud cores) and MSX
PSC entries (\S\,\ref{ssec:survey}).  The regions of IRAS 04269+3510
($b=-9.01^o$) cannot be considered, because they were not observed by
the MSX satellite The remaining 126 components (36 massive, i.e.\ 
$>100\,M_\odot$, and 90 less massive cores) are broken down into three
classes:
\begin{enumerate}
\item The MSX point source lies within 10$\arcsec$ of the {\mm} peak.
  In this case, both sources represent probably the same object as the
  positional error of MSX is about 4$\arcsec$ to 6$\arcsec$ ($3
  \sigma$) \citep{MSXEgan2003}, similar to that of the
  millimeter observations.
\item In the cases where the MSX point source is separated from the
  millimeter peak by 10$\arcsec$ to 1$\arcmin$, one can exclude that
  the MSX source represents the core center, it is a different nearby source.
\item The last group includes those millimeter sources without any MSX
  point source in their vicinities up to 1$\arcmin$.
\end{enumerate}
We use angular distances and not linear distances because of the large
distance range of our sources. A criterion using linear distances
would be too coarse for nearby sources or beyond the resolution for
distant sources.

The analysis shows that there are more MSX point sources in the
vicinities of massive cloud cores than near less massive cores
(Figures~\ref{fig:MSXvsMass} and \ref{fig:MSXvsMassNear}).
Qualitatively, the black bars (cores with MSX sources within
10\arcsec) in the Figures \ref{fig:MSXvsMass} and
\ref{fig:MSXvsMassNear} do not drop as fast as the total hight of the
bars (total number of core per mass bin) after the turn over at
$100\,M_\odot$. In numbers, we found that 44\% of the massive cloud
cores have an MSX source within {10\arcsec} (16 of 36), whereas less
than 6\% of them have no MSX source within {1\arcmin} (2 of 36).  For
the less massive cores, there are only 26\% with a nearby MSX source
(23 of 90) and even 23\% of them have no MSX source within {1\arcmin}
(21 of 90). This might be a projection effect since the high-mass
cores are predominantly located at large distances
(Figure~\ref{fig:MassvsDist}) and a selection effect since only the
brightest MIR sources are detected by MSX in these distances. However,
analyzing only the 84 components closer than 2.5\,kpc yields a similar
result (Figure~\ref{fig:MSXvsMassNear}). In the ``near'' sample, 50\%
of the massive cores have close MSX sources (5 of 10) and all 10 cores
have a MSX source within 1\arcmin, whereas only 26\% of the other
cores have a close MSX source and 26\% have no MSX source within
1\arcmin. Of course we are dealing with low-number statistics here.
There are only 10 massive cores closer than 2.5\,kpc in our survey.
One could assume that the high fraction of the massive cores with
nearby MSX sources might be attributed solely to projection effects
among the large number of the far massive cores. Can the high fraction
of near massive cores with an nearby MSX source be explained just by
coincidence?  If we assume that the probability for an MSX source
being nearby a molecular cloud core is about 26\% independent of the
core mass, then the probability to get five out of ten massive cores
with nearby MSX source is only 7\%.  These 7\% are an upper limit for
the probability of a spurious result, but it only takes into account
the sources closer than 2.5\,kpc. It is difficult to draw quantitative
conclusions from such low-number statistics, but qualitatively we can
say that there is a higher fraction of MSX point source associations
to massive millimeter peaks than to less massive ones.  Thus, these
associations apparently live longer to be present more frequently.
Possibly massive cores can sustain star formation longer than low-mass
cores despite the faster formation of massive stars.

\subsection{Massive Pre-Protocluster Candidates and \PPC}
\label{sec:pre}

A very intriguing question concerns the initial conditions for massive
star formation. {\PPC} to study these conditions are rarely found.
Unfortunately, {\PPC} and protoclusters in an early stage cannot be
told apart within the scope of this work.  Both types of objects will
appear as strong {\mm} emission peaks without any other association in
the collected data.  Whether a massive protostar is going to form or
is already present in the center of a massive core can only be decided
when sensitive FIR observations with a better spatial resolution
become available.

As laid out in \S~\ref{ssec:cand}, we call those cloud cores {\em
  massive pre-protocluster candidates}, which have no association at
infrared and radio wavelengths (2MASS, MSX, NVSS) within 10\arcsec{}
of the peak and have a mass higher than 100\,$M_\odot$. An association
to sources at MIR/NIR wavelengths hints to a later stage of massive
star formation.  We were able to identify twelve massive pre-protocluster
candidates in our survey:
\begin{description}
\item[IRAS 03064+5638 \#1a] Quiescent part of a double-peaked cloud core.
\item[IRAS 03211+5446 \#1] Another quiescent part of a double-peaked
  cloud core.
\item[IRAS 04073+5102 \#1, \#2, \#6, \#7] A ring of cloud cores around
  a star cluster. Triggered star formation may take place here.
\item[IRAS 06058+2138 \#1] A single-peaked cloud core, but the MSX and
  IRAS sources are offset by 20\arcsec.
\item[IRAS 06073+1249 \#2] A relatively small cloud core compared to
  the main component, but still massive.
\item[IRAS 06105+1756 \#1a, \#1b, \#2, \#3] The cloud core \#1a has an
  IRAS and an MSX source on its flanks, but more than 10\arcsec{}
  away. The other components show further fragmentation, but even the
  fragments are more massive than 100\,$M_\odot$.
\end{description}
These massive pre-protocluster candidates are presumably the earliest
stages of massive star formation in our sample or even {\PPC}, where
initial conditions of massive star formation may be studied.
Observations in the FIR with higher sensitivity and spatial
resolution, e.g., by Spitzer, SOFIA, or Herschel are needed to
characterize the spectral energy distribution of the core centers and
to reveal possible deeply embedded sources. High-resolution
interferometric observations (Plateau de Bure, ALMA) looking for
outflows can help to confirm the evolutionary stage of the cores.
Interferometric observations are also needed to check whether these
cores are really single-peaked or whether they fragment and only form
low-mass stars.

Some more cores should be mentioned as pre-protocluster candidates
though they do not meet all of the above-mentioned criteria: The cloud
core close to IRAS 06055+2039 is almost an massive pre-protocluster
candidate. Our mass estimate just falls short of the somewhat
arbitrary 100\,$M_\odot$ limit. The mass estimated for
IRAS\,06055+2039 by \citet{Mueller2002} is above 100\,$M_\odot$.
IRAS\,02593+6061\,\#1 is clearly offset from the IRAS position and the
MSX source is only on the slope of the core, slightly closer than
10\arcsec{}.  The cloud cores \#1 and \#3 of IRAS\,05377+3548 are
rather early pre-protostar than pre-protocluster candidates with
intermediate masses (57 and 59\,$M_\odot$ respectively). A similar
case is core \#1a of IRAS 07029-1215 (UYSO\,1), but \citet{Forbrich04}
detected an outflow confirming the presence of a very young protostar.

The mass range of the pre-protocluster candidates goes up to
2900\,$M_\odot$, but this source is located at a distance of 8.2\,kpc
and individual cloud cores almost certainly are not resolved. The next
massive not so distant protocluster candidate is 03064+5638 \#1a with
1400\,$M_\odot$ at 4.1\,kpc. 
The column densities are relatively high ($6\cdot10^{22}\rm\,cm^{-2}$
to $2.5\cdot10^{23}\rm\,cm^{-2}$) compared to the other cloud cores.
The core densities (mean density $2\cdot10^4\rm\,cm^{-3}$) however,
are rather moderate when compared to the other cores in this survey .

\section{SUMMARY}
\label{sec:sum}

We used the 47 of the FIR-brightest IRAS sources in the outer galaxy (which
are in 44 fields) to search for early stages of star formation, ie.\ 
{\mm} emission from molecular cloud cores.  We presented relatively
large {\mm} maps of the 40 regions with detections.  For these
regions, we compiled NIR, MIR, radio, and molecular line data
including information on outflow and maser activities in these regions
in addition to the maps. This collection of data allows to
characterize the regions and to estimate the evolutionary stages of
the cloud cores.  The IRAS sources themselves are not the youngest,
most deeply embedded sources, but guided us to sites of on-going star
formation. Caution has to be exercised with observational data from the
literature obtained by observations pointed to IRAS sources alone. It
may not be clear to which source the measured flux has to be
attributed.  Only 50\% of the bright IRAS sources are themselves
associated with a {\mm} peak.  This study can provide a good starting
point for subsequent detailed investigations of especially intriguing
sources, e.g.\ the candidates for massive pre-protoclusters.

The masses and core densities have been estimated for all detected
cloud components. They compare well to the findings of similar
surveys, except that our mass range starts at lower masses.  The
reason is our attempt to separate the cloud components. A look at the
high-mass cores and their associations with MIR sources shows that
these associations are more frequent than for lower-mass cores.  We
suspect that massive cores can sustain star formation over a longer
time despite the fast evolution of massive stars.  When we focus on
massive protocluster candidates, they only have moderate densities
compared to the range of core densities estimated here.

The strategy of taking the FIR-brightest IRAS sources as lighthouses
guiding us to young, massive star-forming regions, proved successful.
We identified twelve massive pre-protocluster candidates.

\begin{acknowledgements}
  We thank the staff of the IRAM 30m telescope, the JCMT, and the SMT
  for their help with the observations and data reduction.
  Furthermore, we thank I. Zinchenko for valuable discussions and the
  referee for very helpful comments.
  
  The authors acknowledges support from the DFG through grants
  He\,1935/15-1,2, Sch 665/1, and Kl\,1330/1-1.  This research has
  made use of the SIMBAD database, operated at CDS, Strasbourg,
  France, of the NASA/IPAC Infrared Science Archive, which is operated
  by the Jet Propulsion Laboratory, California Institute of
  Technology, under contract with the National Aeronautics and Space
  Administration, and of Aladin \citep{Aladin}.  The Digitized Sky
  Survey was produced at the Space Telescope Science Institute under
  U.S. Government grant NAG W-2166. The images of these surveys are
  based on photographic data obtained using the Oschin Schmidt
  Telescope on Palomar Mountain and the UK Schmidt Telescope.
\end{acknowledgements}

\appendix

\section{An exemplary Radial Density Profile from single dish observations}


\label{sec:radial}

In this appendix, we present the radial intensity and the resulting
radial density profile of the best-suited source of our survey.
Pre-requisites as spherical symmetry of the cloud core envelope and
the restriction to radii larger than the beamwidth show the limitation
of the method if applied to single-dish data.

\smallskip

Density profiles of molecular cores are often used to discriminate
between different models of molecular cloud collapse. In the case of
clustered low-mass star formation it appears that cloud core envelopes
resemble rather finite-sized Bonnor-Ebert spheres \citep{Bonnor56}
than singular isothermal spheres used in the 'standard' isolated star
formation model by \citet{Shu} (e.g.
\citealt{Motte2001,Bacmann2000,Henriksen97,Bonnor56}). Concerning
massive-star formation, there is only a poor database of density
distributions.  Recent papers reported the following values for the
exponent $p$ in the radial density profile $\rho (r)\propto r^{-p}$:
$1\leq p\leq 1.5$ with exceptions showing $p=2$ \citep{Tak2000},
$p=1.5$ \citep{Hatchell2000}, $\left\langle p\right\rangle =1.8\pm
0.4$ \citep{Mueller2002}, and $\left\langle p\right\rangle =1.6\pm
0.5$ (for $r\leq 32\arcsec$, \citet{Beuther2002II}). We mention here
that using their own method \citet{Beuther2002II} found $p=1.9$ for
the submillimeter maps of \citet{Hatchell2000}. The difference to the
original value derived by another method indicates the current
difficulties in acquiring density profiles of cloud cores in
massive-star-forming regions.

As the beam profile of the each observation is not sufficiently known
for a proper deconvolution, we use the asymptotic expansion for an
observed monochromatic luminosity by \citet{Adams91}. A presumed
radial power-law density profile $\rho (r)\propto r^{-p}$ can be
derived by direct measurements of the radial beam-folded intensity
profile $I\propto r^{-\alpha }$ for radii greater than one full
beamwidth $\theta _{beam}$ in the case of spherical symmetry.  The
exponents $p$ and $\alpha $ are related by $\alpha =1-(p+Qq+\epsilon
_{proj})$, where $q$ refers to the temperature profile exponent,
$T\propto r^{-q}$.  The temperature profile exponent $q$ lies between
$0.33$ and $0.4$ \citep{Chandler2000}. We choose $q=0.4$, following
\citet{Beuther2002II}.  The quantity $Q$ is a correction factor
depending only on frequency $\nu $ and temperature $T$ in the form of
$Q=(x \cdot \exp ^{x})/(\exp ^{x}-1)$ with $x=h\nu /kT$
\citep{Adams91}. For comparison with \citet{Beuther2002II} we choose
$T=30$ K.  The correction factors for the IRAM 1300 $\mu $m and the
JCMT 850 $\mu $m maps are $Q=1.2$ and $Q=1.3$ respectively.  The
correction term $\epsilon _{proj}$ is due to deprojection effects
caused by the dual-beam mapping technique.  In our case $\epsilon
_{proj}$ is below $0.1$ \citep{Motte2001} and is only considered in
the error estimate.

To derive a radial density profile using the asymptotic expansion by
\citet{Adams91}, an (almost) spherical symmetric cloud core is needed
that is considerable larger than the beamwidth. The more distant the
considered source is and the lower the linear resolution, the more
unlikely it is that the interesting inner part of the molecular cloud
core is covered, needed e.g. to identify a Bonnor-Ebert sphere when
using this method.

As shown in Table~\ref{tab:beam}, the beamwidths in our survey range
from 14.1\arcsec{} to 26\arcsec{} in the worst case.  Furthermore, we
deal with large distances and many multiple sources.  After checking
for the largest and most 'spherical' dust emission cores at moderate
distances, we consider the one around IRAS 06058+2138 as the most
suitable and useful source to get an exemplary density distribution.
IRAS 06058+2138 also has the advantage of having maps at two
wavelengths, an IRAM 1300 $\mu $m map and a JCMT 850 $\mu $m map. We
provide the analysis of the intensity profile as a comparison to the
values derived by the other surveys. However, one should keep in mind
that only projection effects may render the cloud core spherical. A
thorough discussion of the problems with 1D-fitting can be found in
\citet{Steinacker04}.

\subsection{Density Profile of Core IRAS 06058+2138 \#1}
To get the radial intensity profile, we average the two-dimensional
intensity distribution in the map in circular annuli of 1.5\arcsec{}
around the peak position for $r>\theta _{beam}$. For the IRAM map, the
values are plotted against the average radius of the corresponding
ring in Figure~\ref{fig:Iprofil}. The data are fitted using the
Levenberg-Marquardt method \citep{Numericalrecipes86} with one power law
in the radial range from the beamwidth to 60\arcsec. The possibility of a
better fit with two power laws was also investigated, using the same
break value at 32\arcsec{} as used by \citet{Beuther2002II} for
comparison.

The fitted curves are shown in Figure~\ref{fig:Iprofil}. The resulting
values for $\alpha $ and $\chi ^{2}$, $p$ as well as the mean results
of the comparable surveys of \citet{Beuther2002II} and
\citet{Mueller2002} are summarized in Table~\ref{alphaundp}. The error
of $p$ is approximately 0.3, taking into account an error of $\sim0.1$
for $\alpha $ concerning ring widths and the peak position, an error
of the temperature distribution of $\sim0.1$ for $q$ and $\sim0.1$ as
the largest possible projection error $\epsilon _{proj}$
\citep{Beuther2002II,Motte2001}.

The radial density profiles of the IRAM and the JCMT map are
comparable, but $\chi ^{2}$ is about an order of magnitude better for
the IRAM data. There is good agreement with the mean density profile
exponent of \citet{Mueller2002} for the one power law case. Within the
error bars there is also an agreement with the surveys of
\citet{Tak2000} and \citet{Hatchell2000}. In the case with two power
laws the $p$ values agree very well with those of
\citet{Beuther2002II}. A flatter inner part and a steeper outer part
in the density profile is recognizable. A comparable behavior in the
case of low-mass star formation is regarded as agreement with a
theoretical description of the cloud core as a Bonnor-Ebert sphere
\citep{Motte2001,Bacmann2000,AndreB99}.

{\it Facilities:} \facility{HHT (19-channel bolometer)}, 
\facility{IRAM:30m (MAMBO)}, \facility{JCMT (SCUBA)}

\clearpage
\begin{figure}[htbp]
  \figurenum{1}
  \caption{Millimeter Continuum Maps --- see \S\,\ref{ssec:maps}}
  \label{fig:maps}
  \vspace{1ex}
  \includegraphics[bb=60 30 415 575, width= 70mm]{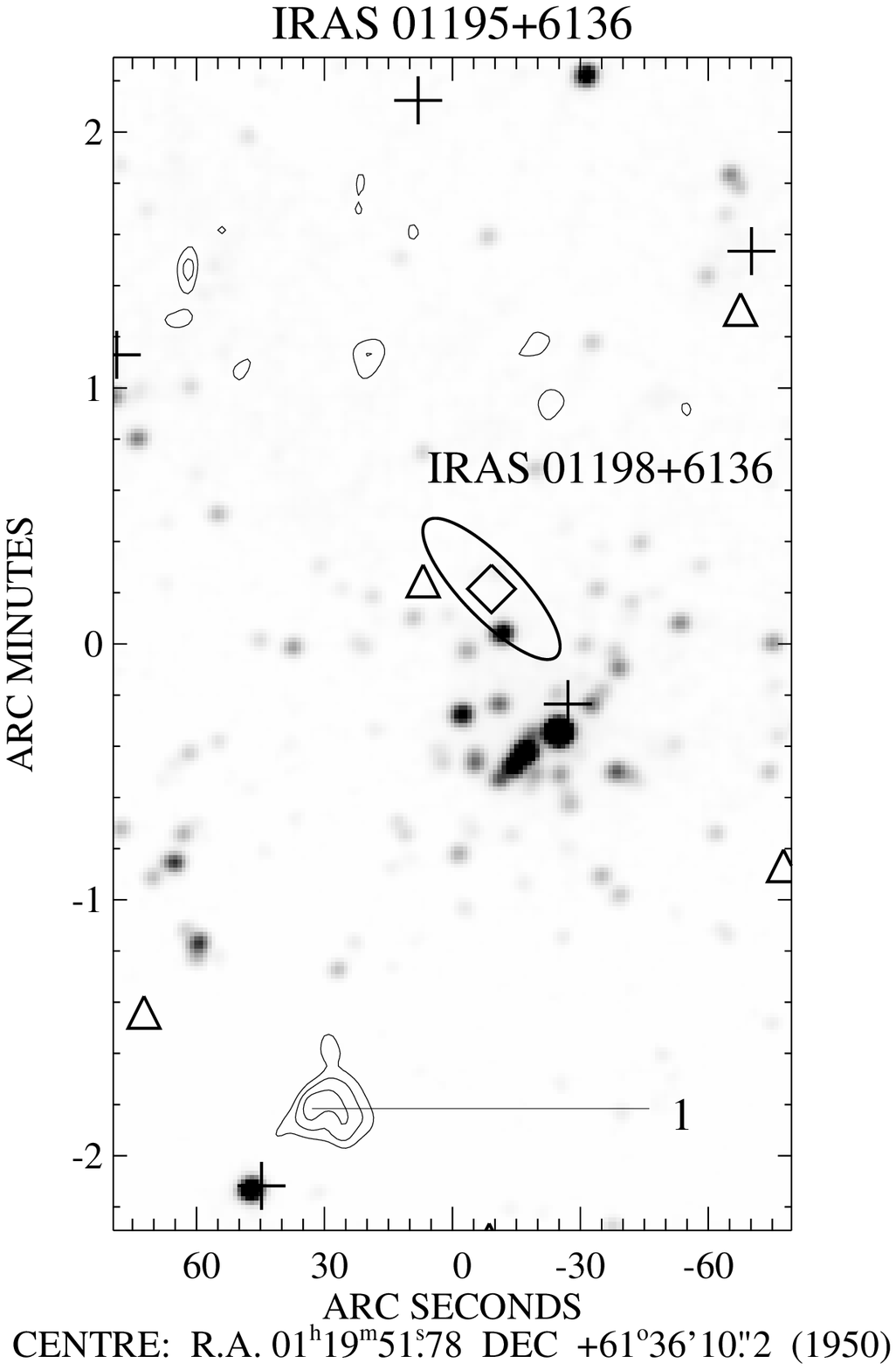}\hfill
  \includegraphics[bb=50 30 405 575, width= 70mm]{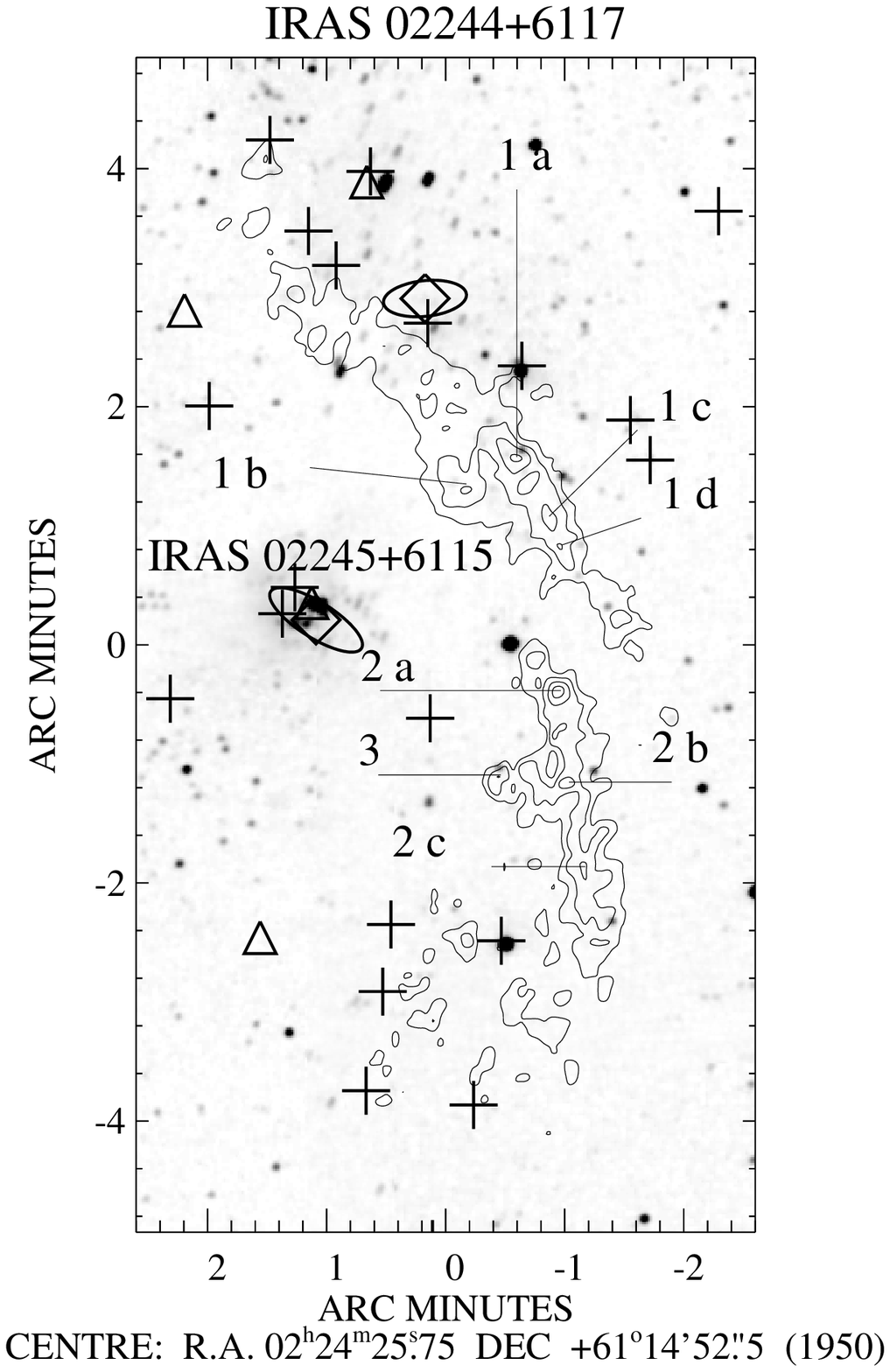}\\
  \hspace*{\fill}contours: (3$\sigma$, $\Delta=1\sigma$)\hfill 
  \hfill contours: (3$\sigma$, $\Delta=2\sigma$)\hspace*{\fill}\hspace*{\fill}\\[3ex]
  \includegraphics[bb=45 25 470 580, width= 60mm]{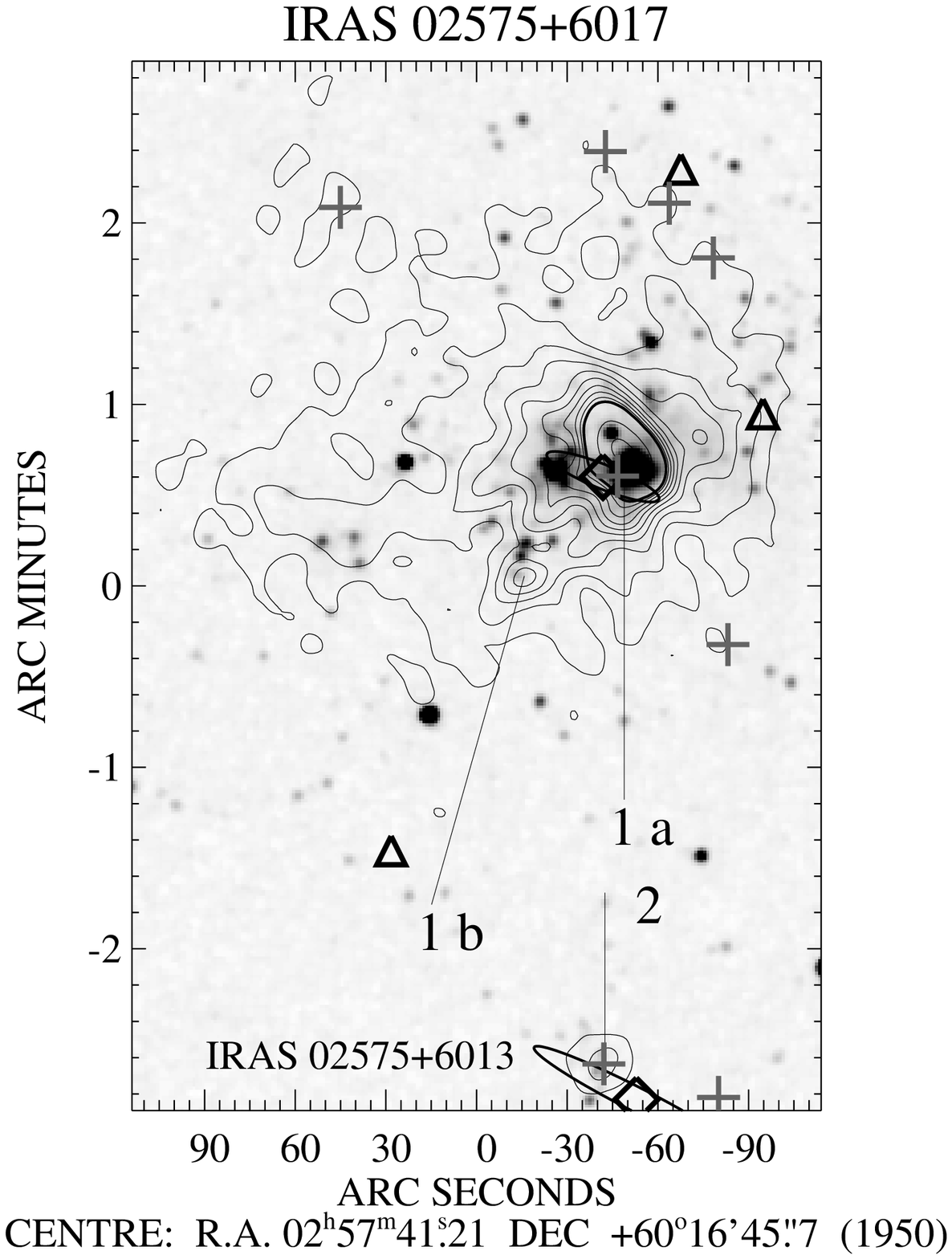}\hfill
  \includegraphics[bb=45 25 665 435, width= 100mm]{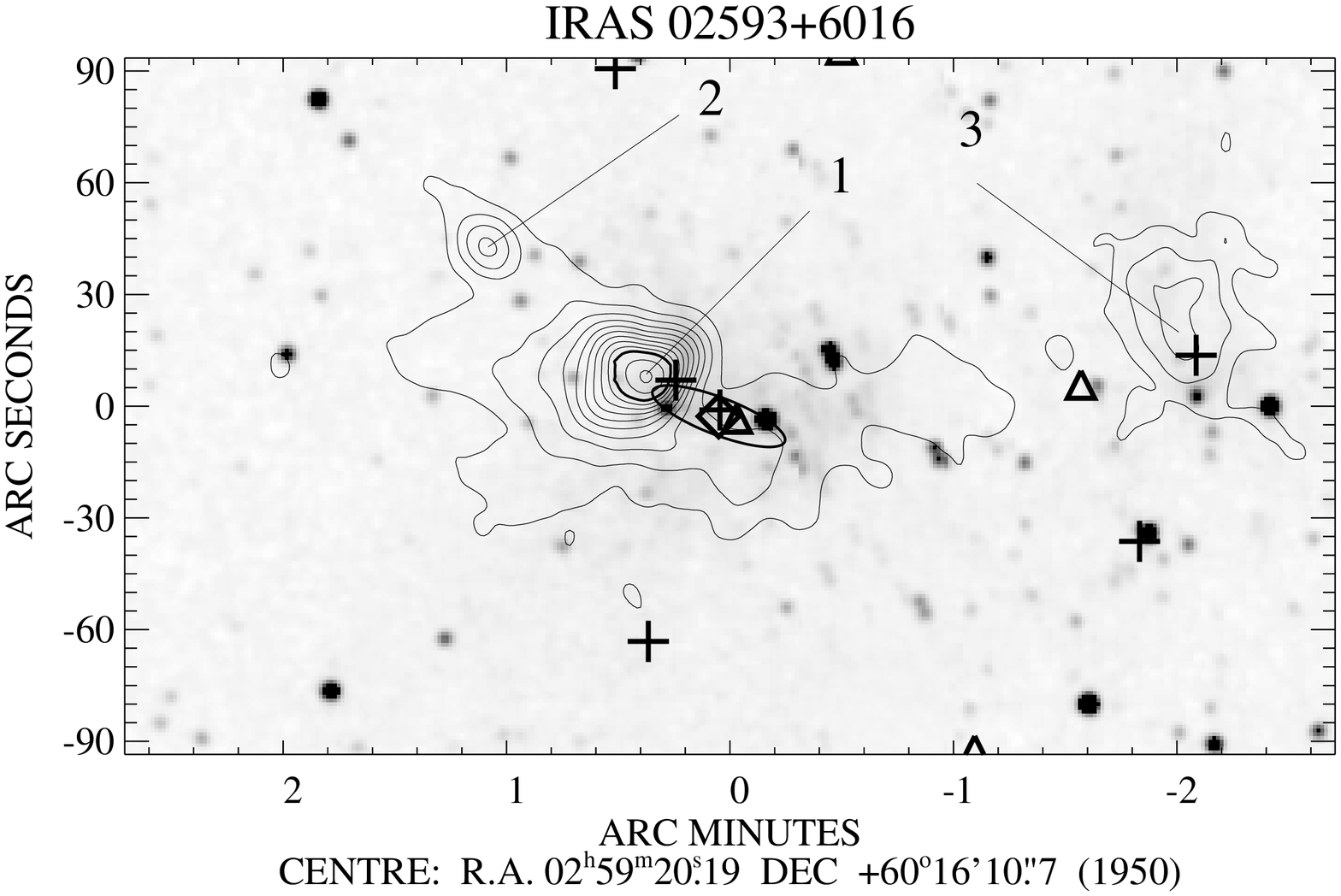}\\
  contours: (3$\sigma$, $\Delta=3\sigma$, {\boldmath$30\sigma$}, $\Delta=20\sigma$)
  \hfill\hfill
  contours: (3$\sigma$, $\Delta=3\sigma$, {\boldmath$30\sigma$}, $\Delta=5\sigma$)
  \hspace*{\fill}\hspace*{\fill}\\[6ex]
\end{figure}
\begin{figure}[htbp]
  \figurenum{1}
  \caption{Continued}
  \includegraphics[bb=45 15 580 580, width= 80mm]{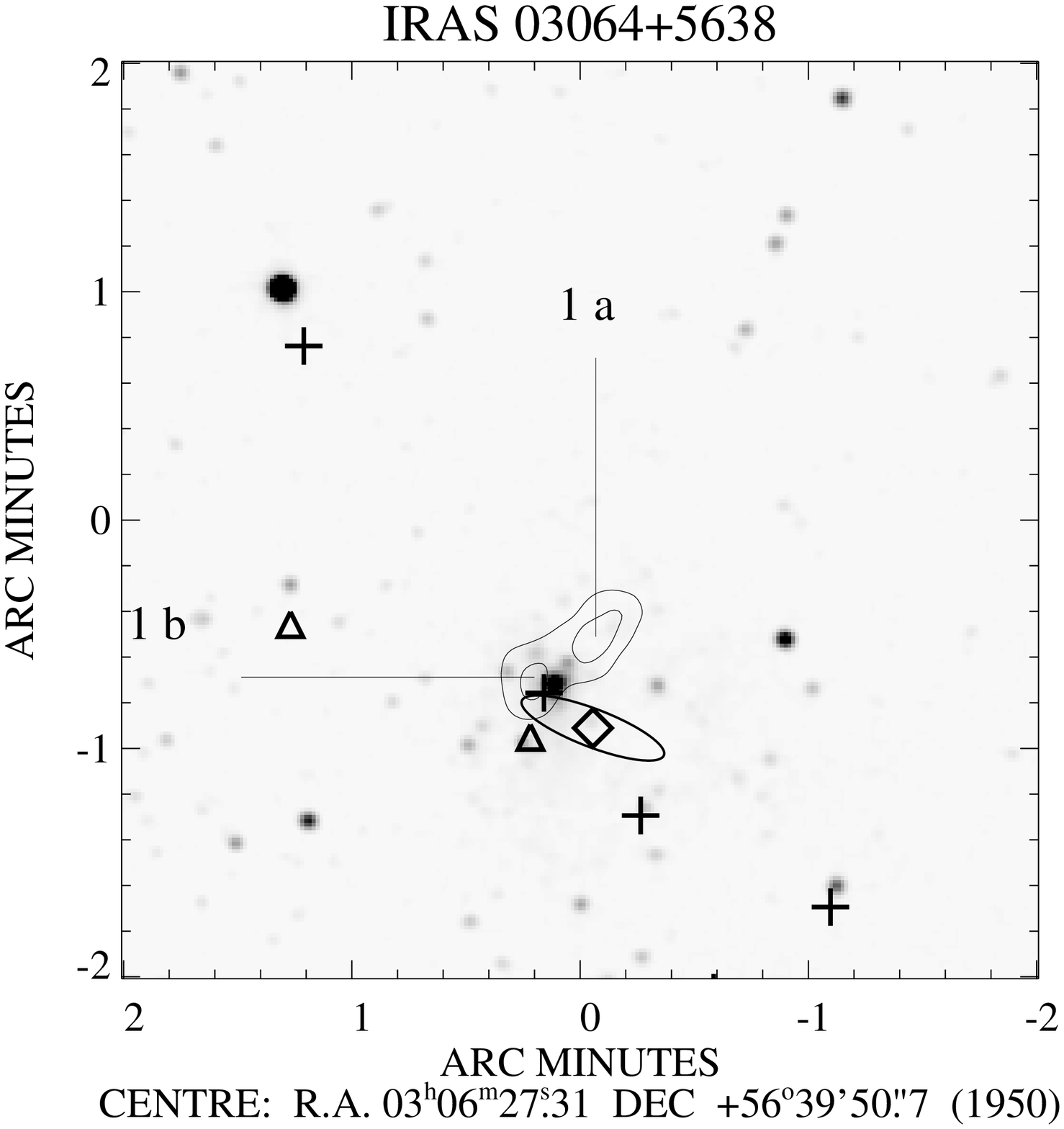}\hfill
  \includegraphics[bb=45 15 580 580, width= 80mm]{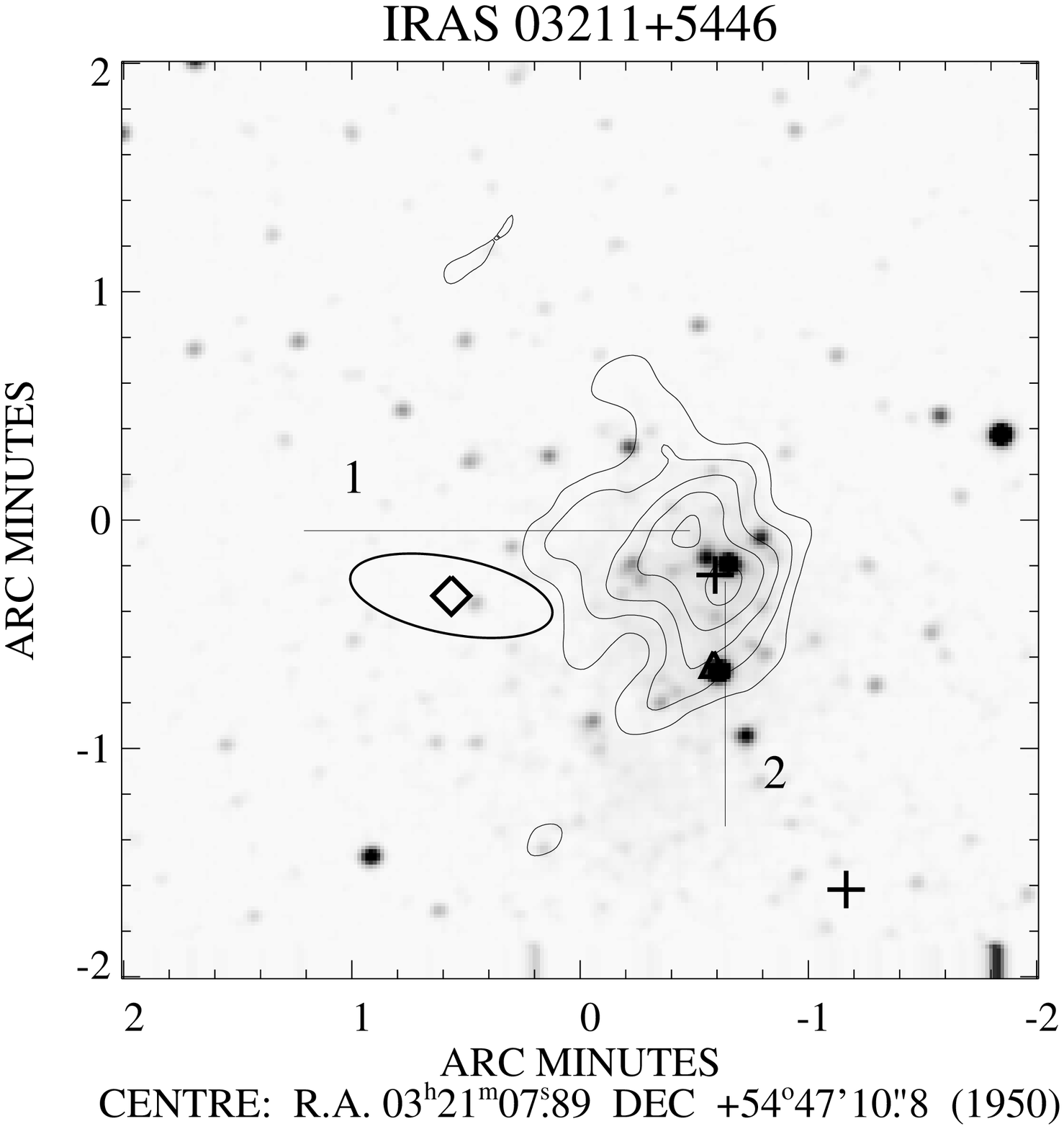}\\
  \hspace*{\fill}
  contours: (3$\sigma$, 4$\sigma$)\hfill\hfill
  contours: (3$\sigma$, $\Delta=1\sigma$)\hspace*{\fill}\\[3ex]
  \includegraphics[bb=50 25 510 580, width= 80mm]{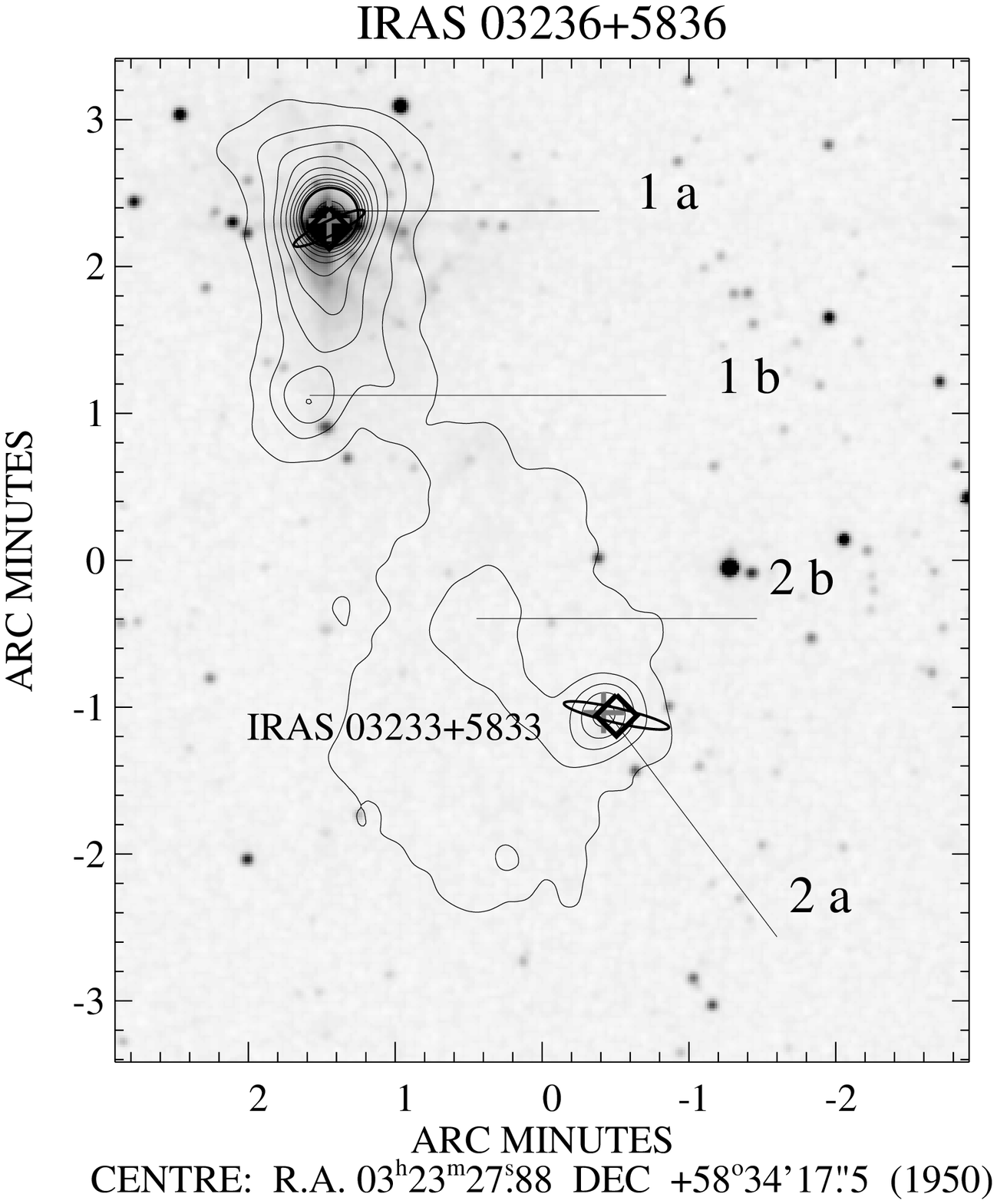}
  \includegraphics[bb=45 15 580 580, width= 80mm]{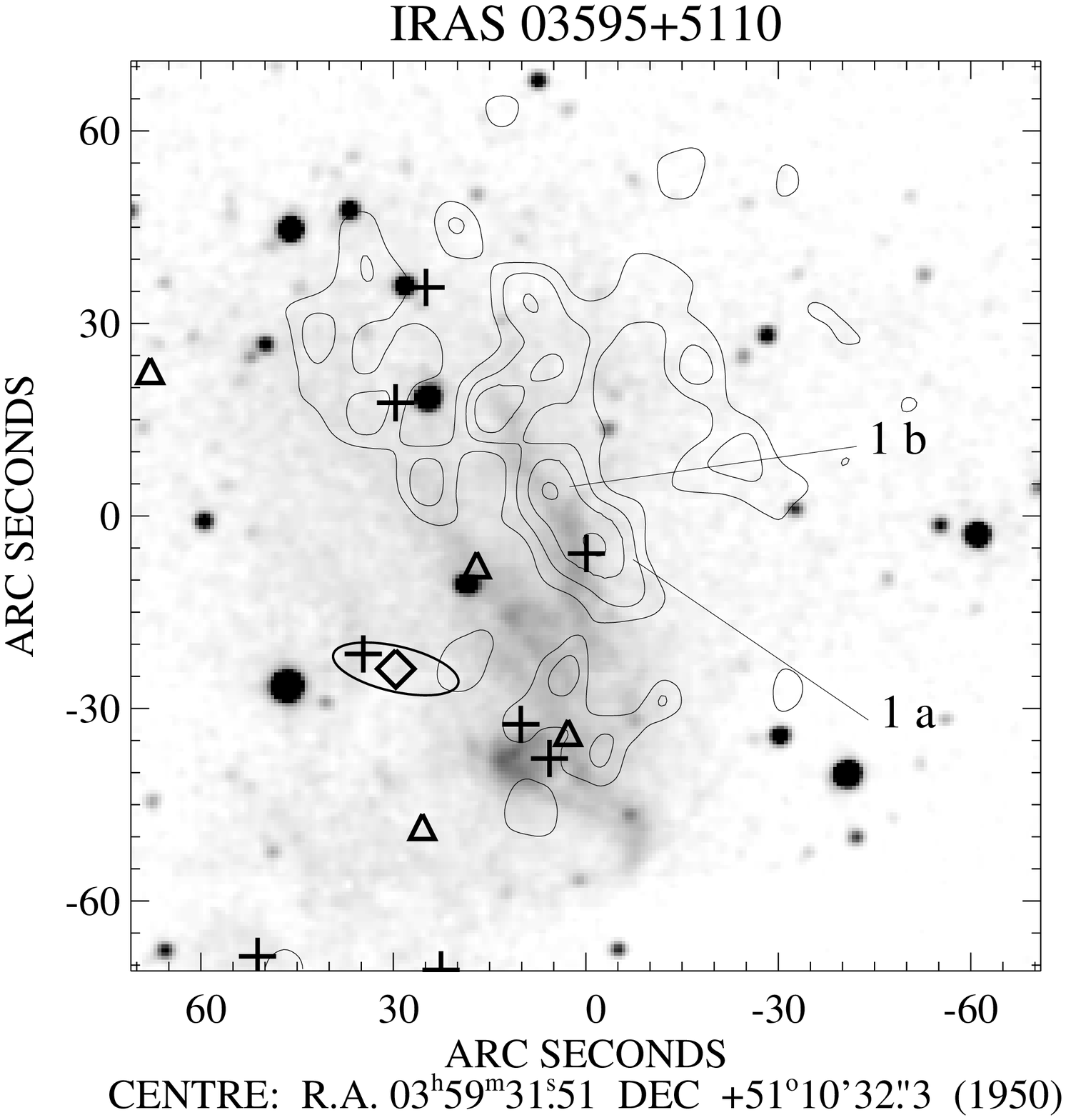}\\
  \hspace*{\fill}contours: (3$\sigma$, $\Delta=3\sigma$, {\boldmath$30\sigma$}, $\Delta=20\sigma$)
  \hfill\hfill
  contours: (3$\sigma$, $\Delta=1\sigma$)
  \hspace*{\fill}\hspace*{\fill}
\end{figure}
\begin{figure}[htbp]
  \figurenum{1}
  \caption{Continued}
  \includegraphics[bb=45 15 580 580, width= 80mm]{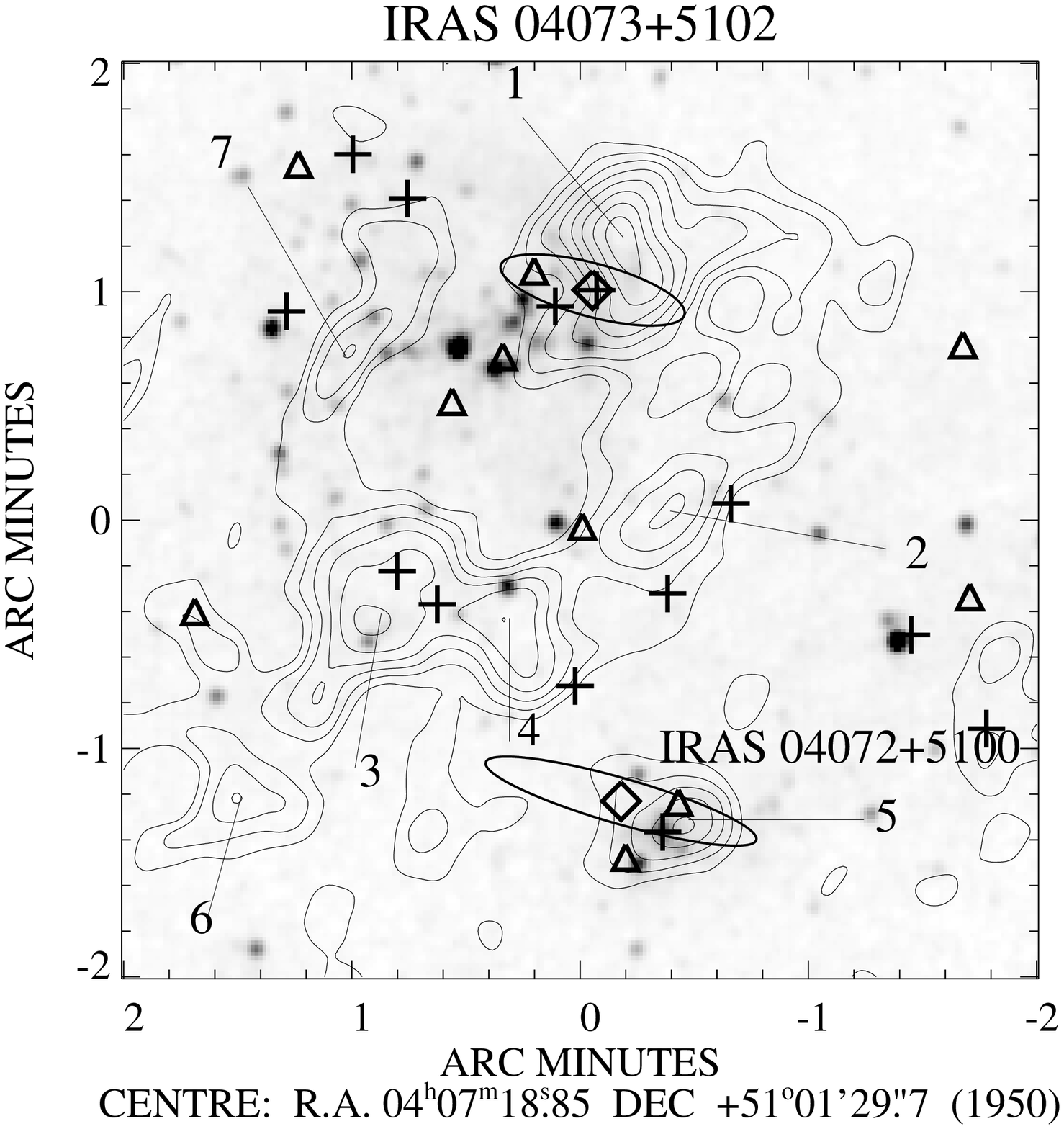}\hfill
  \includegraphics[bb=45 15 580 580, width= 80mm]{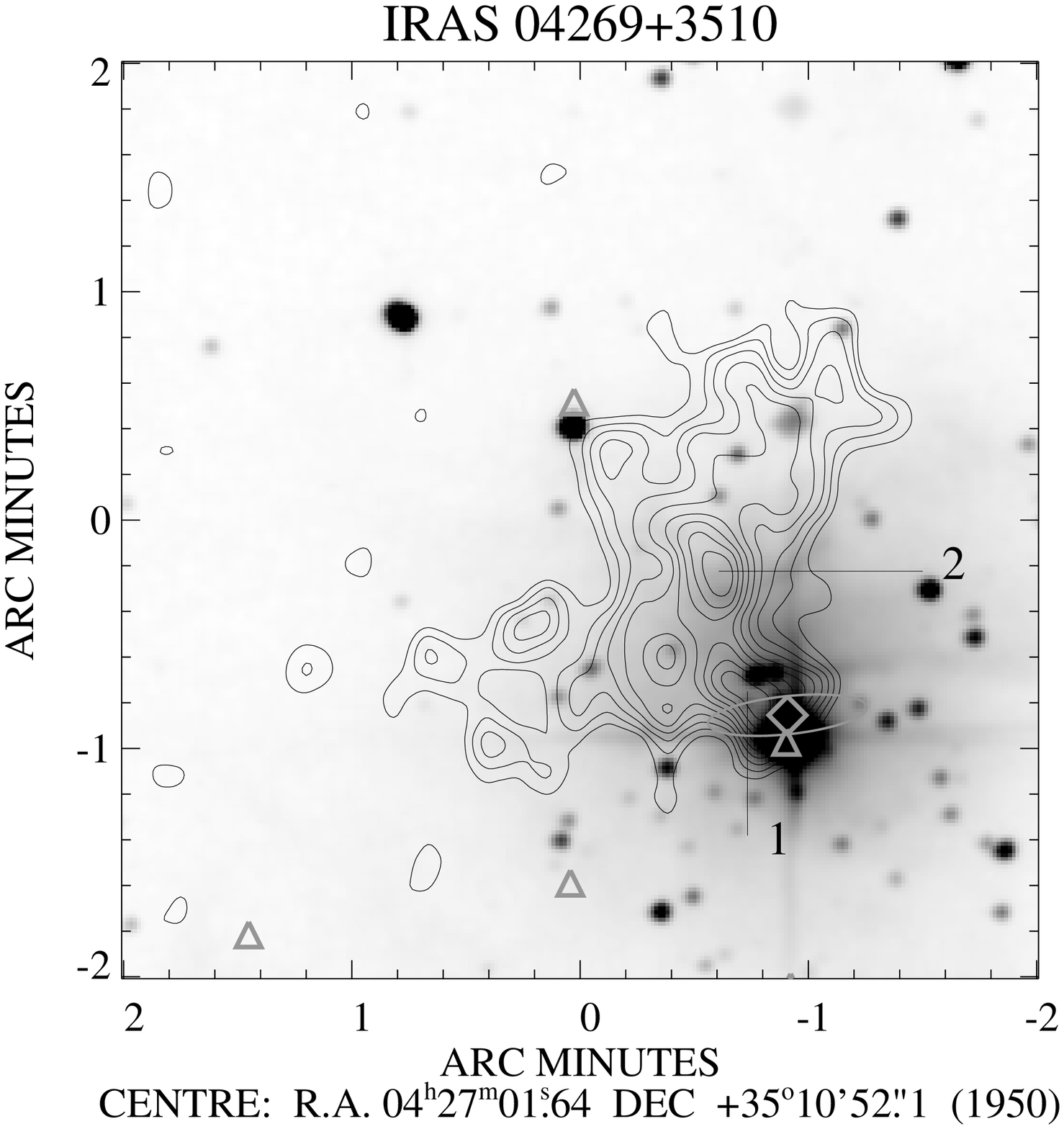}\\
  \hspace*{\fill}contours: (3$\sigma$, $\Delta=1\sigma$)\hfill\hfill
  contours: (3$\sigma$, $\Delta=1\sigma$)\hspace*{\fill}\\[3ex]
  \includegraphics[bb=45 15 580 580, width= 80mm]{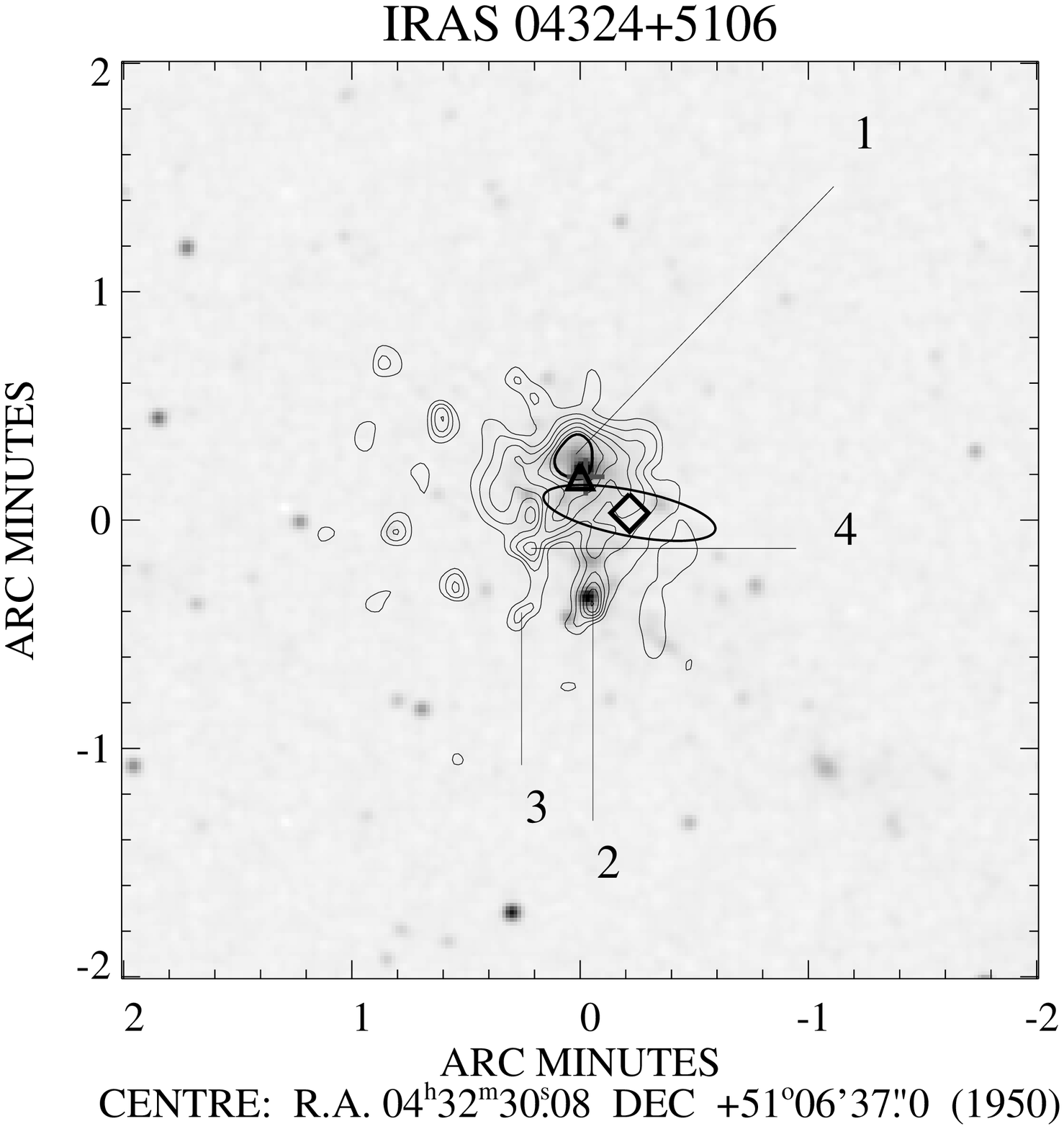}\hfill
  \includegraphics[bb=45 15 580 580, width= 80mm]{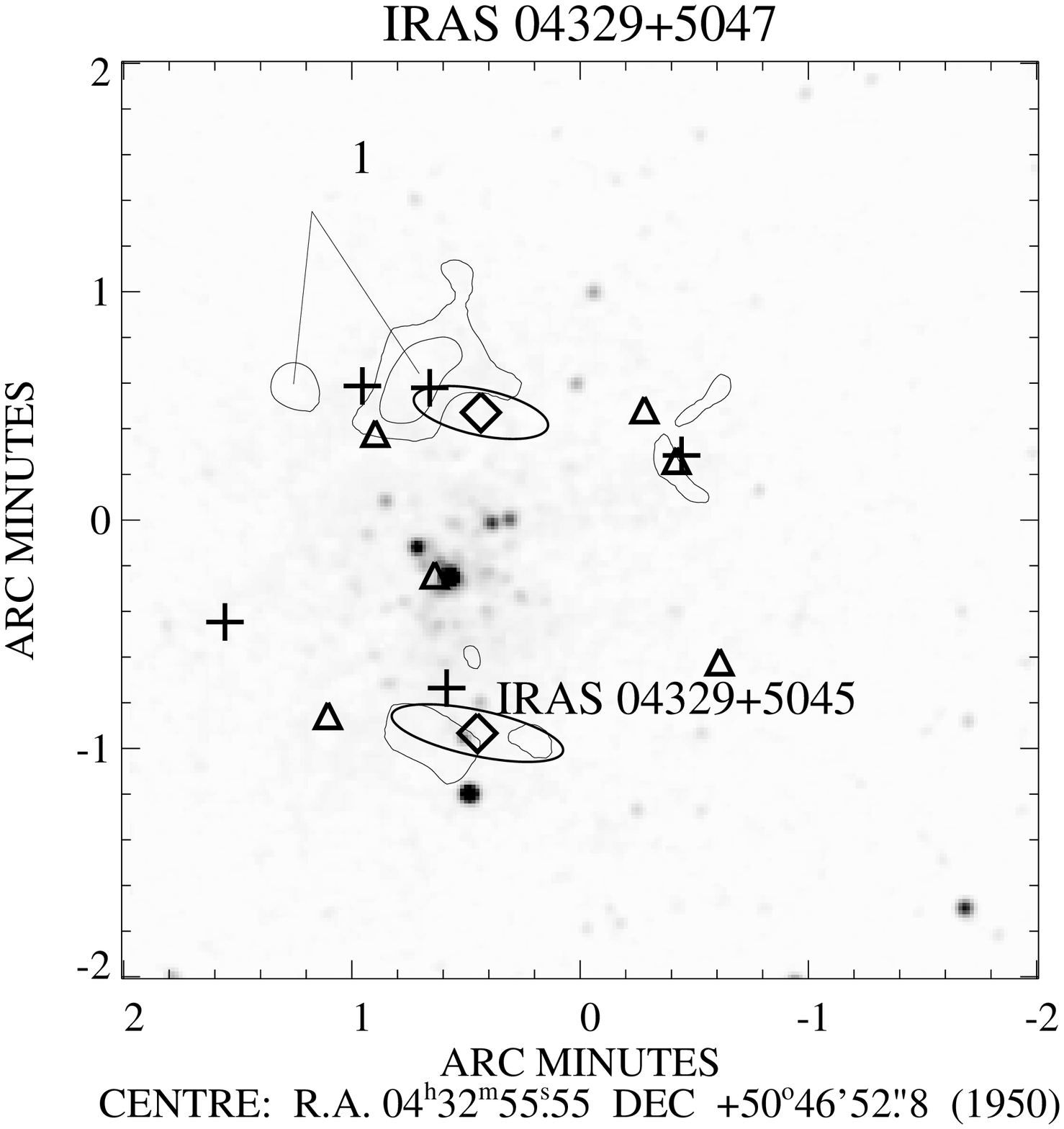}\\
  \hspace*{\fill}contours: (3$\sigma$, $\Delta=1\sigma$, {\boldmath$9\sigma$}, $\Delta=3\sigma$)\hfill\hfill
  contours: (3$\sigma$, 4$\sigma$)\hspace*{\fill}\hspace*{\fill}
\end{figure}
\begin{figure}[htbp]
  \figurenum{1}
  \caption{Continued}
  \includegraphics[bb=35 5 665 520, width= 80mm]{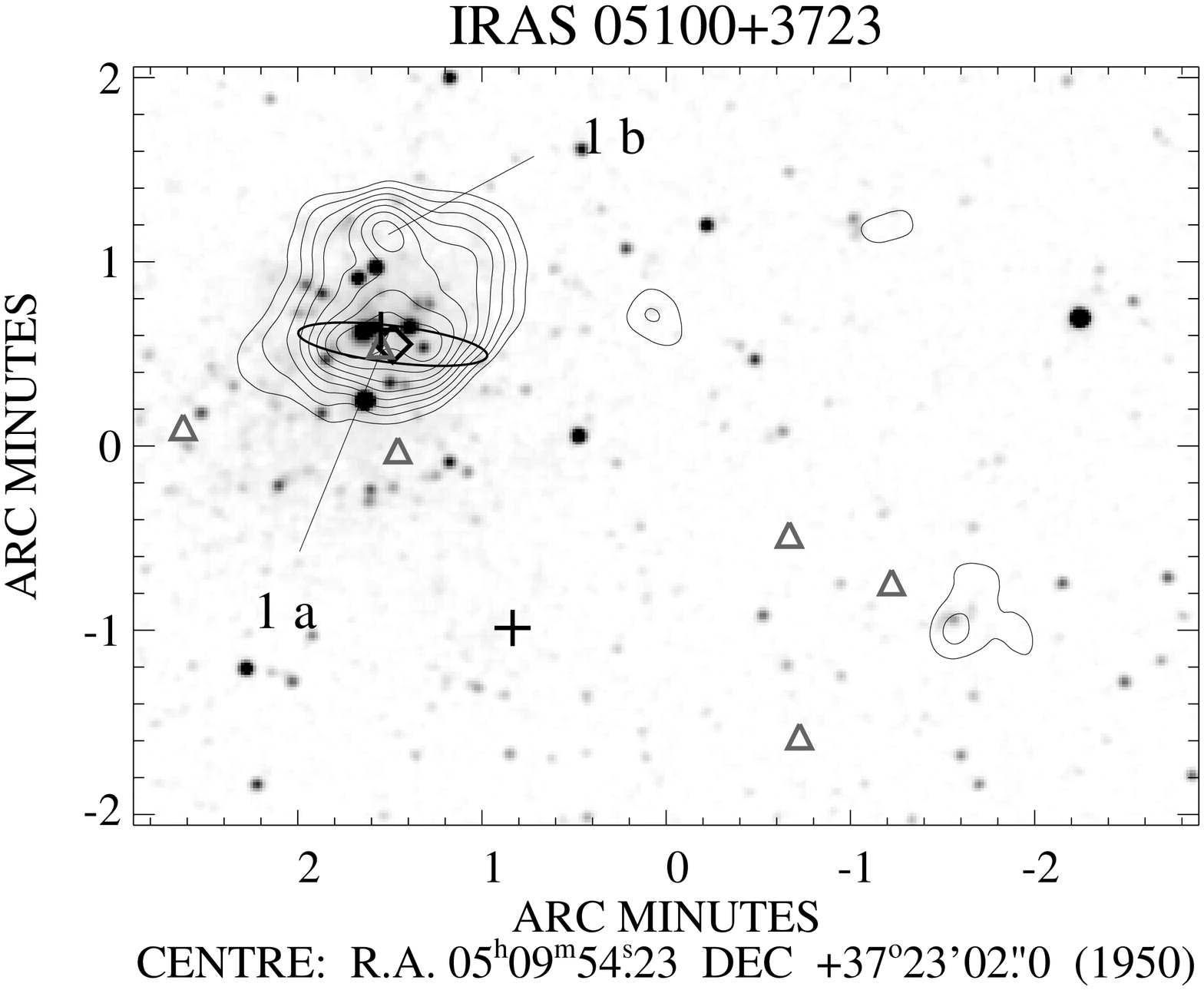}\hfill
  \includegraphics[bb=45 15 580 580, width= 80mm]{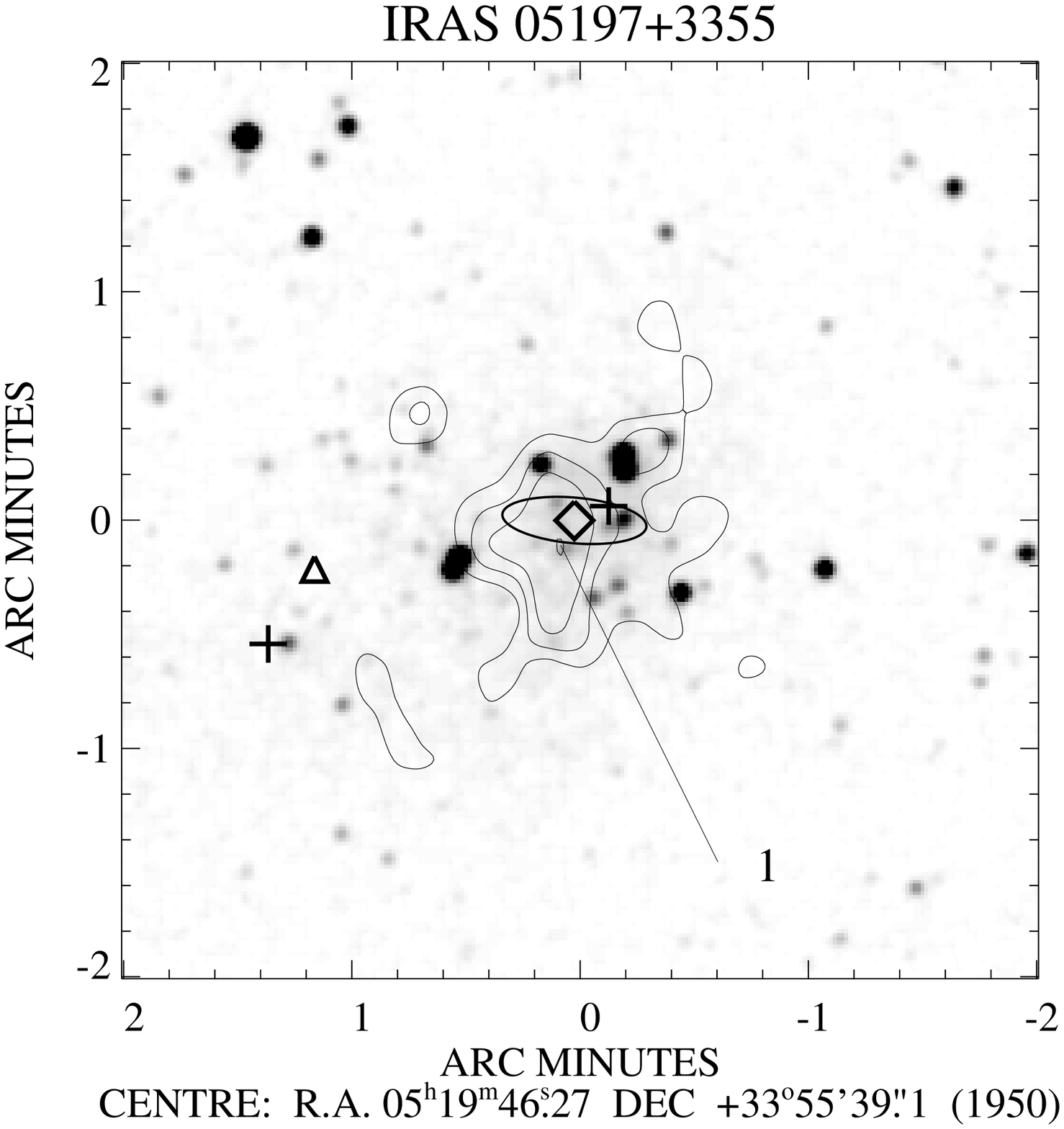}\\
  \hspace*{\fill}
  contours: (3$\sigma$, $\Delta=1\sigma$)\hfill\hfill 
  contours: (3$\sigma$, $\Delta=1\sigma$)\hspace*{\fill}\\[3ex]
  \includegraphics[bb=45 15 555 580, width= 80mm]{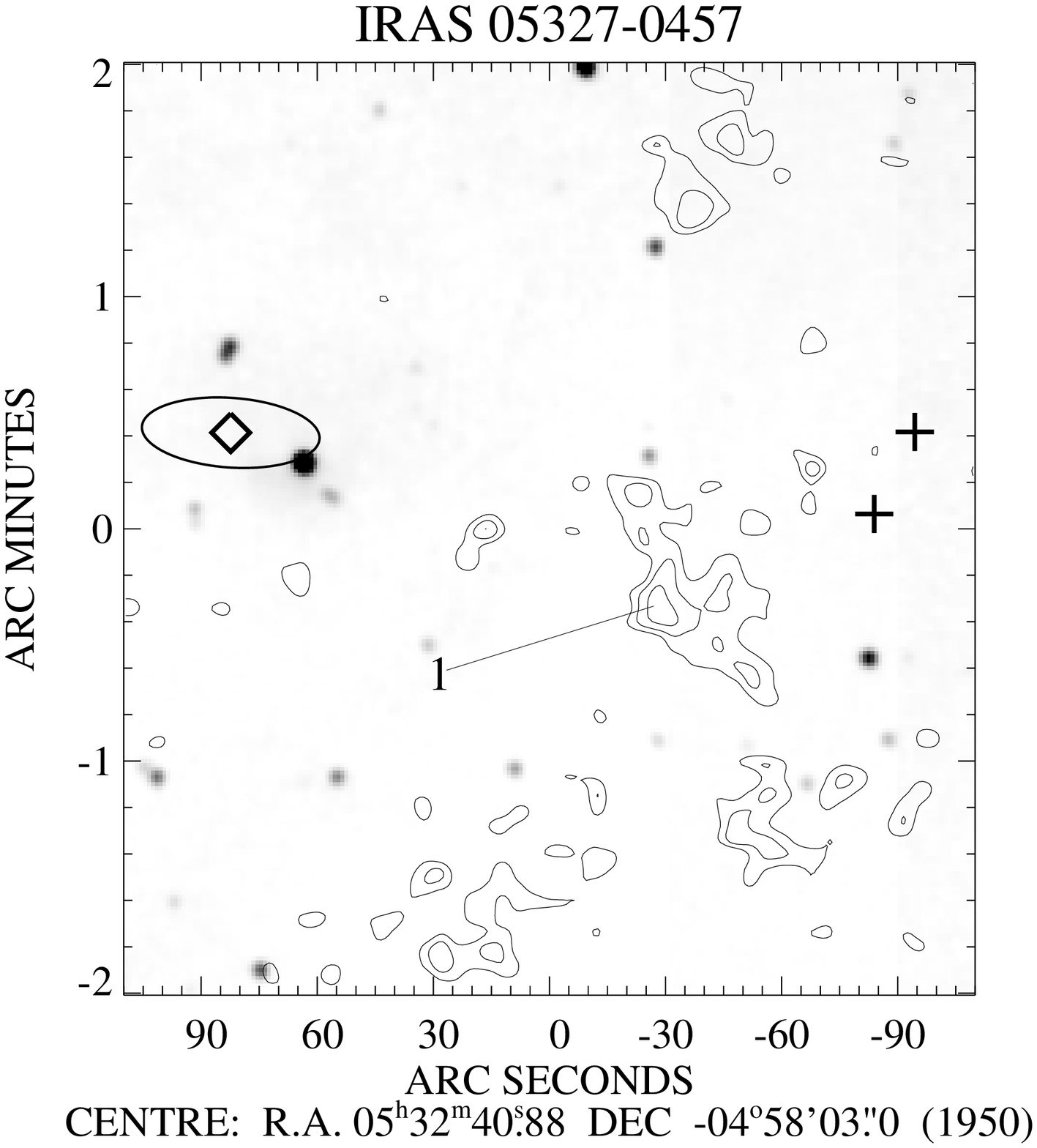}\hfill
  \includegraphics[bb=45 15 580 580, width= 80mm]{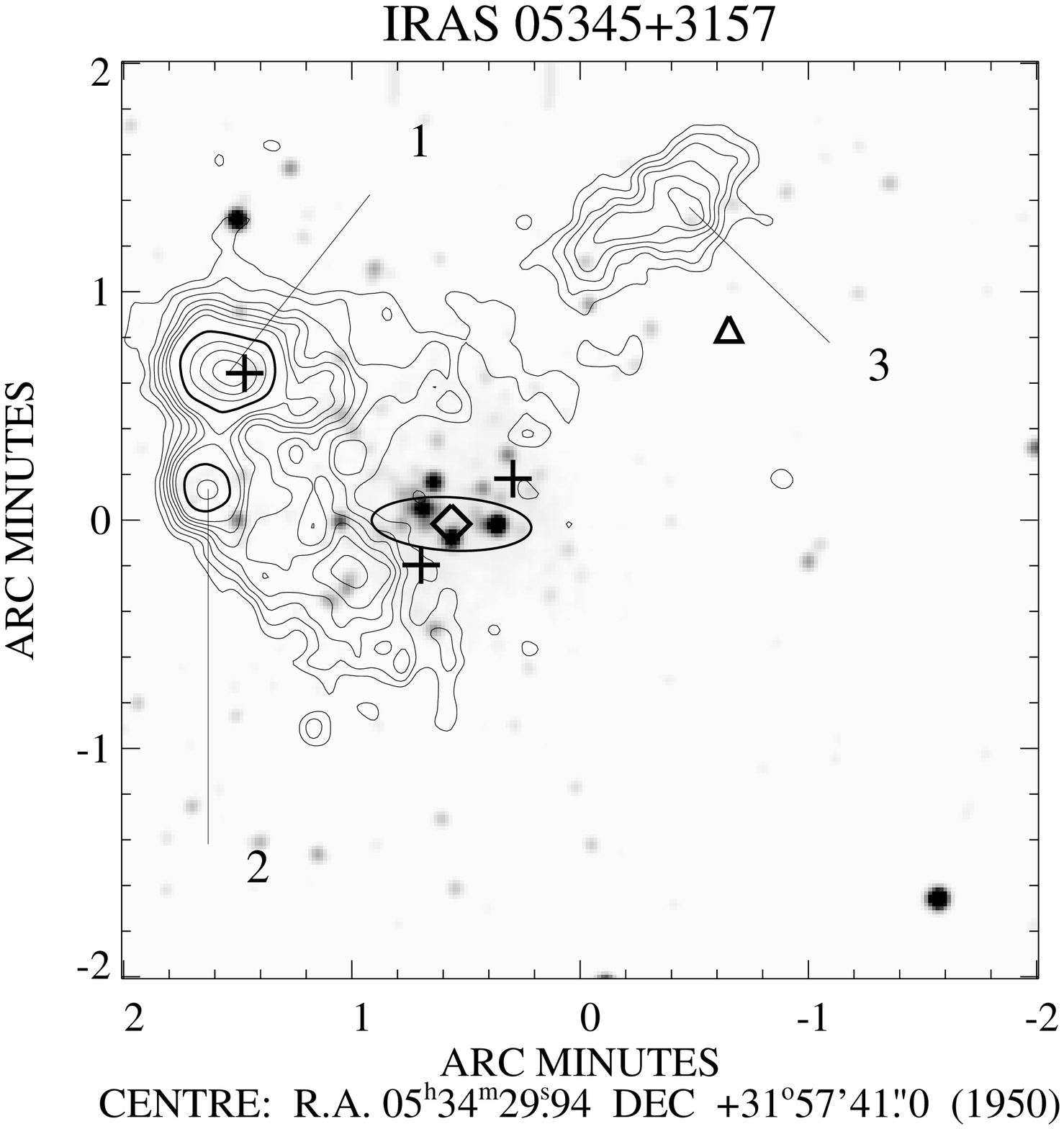}\\
  \hspace*{\fill}\hfill\hfill\hfill\hfill
  contours: (3$\sigma$, $\Delta=1\sigma$)
  \hfill\hfill\hfill\hfill\hfill\hfill\hfill\hfill\hfill
  contours: (3$\sigma$, $\Delta=1\sigma$, {\boldmath$12\sigma$}, $\Delta=3\sigma$)
  \hspace*{\fill}
\end{figure}
\begin{figure}[htbp]
  \figurenum{1}
  \caption{Continued}
  \includegraphics[bb=45 15 580 580, width= 80mm]{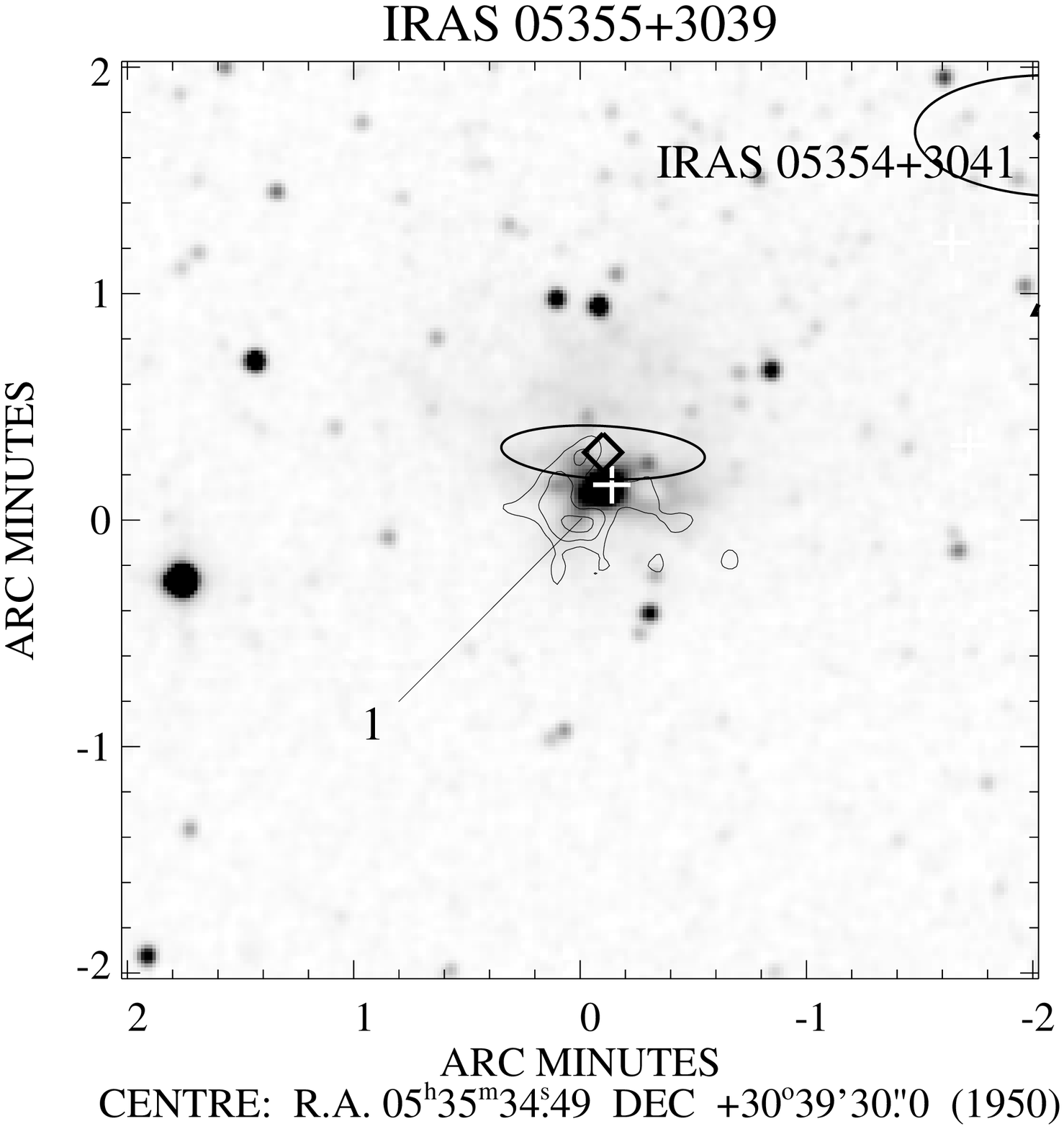}\\
  \hspace*{20mm}
  contours: (3$\sigma$, $\Delta=1\sigma$)\\[3ex]
  \includegraphics[bb=50 25 665 430, width= 150mm]{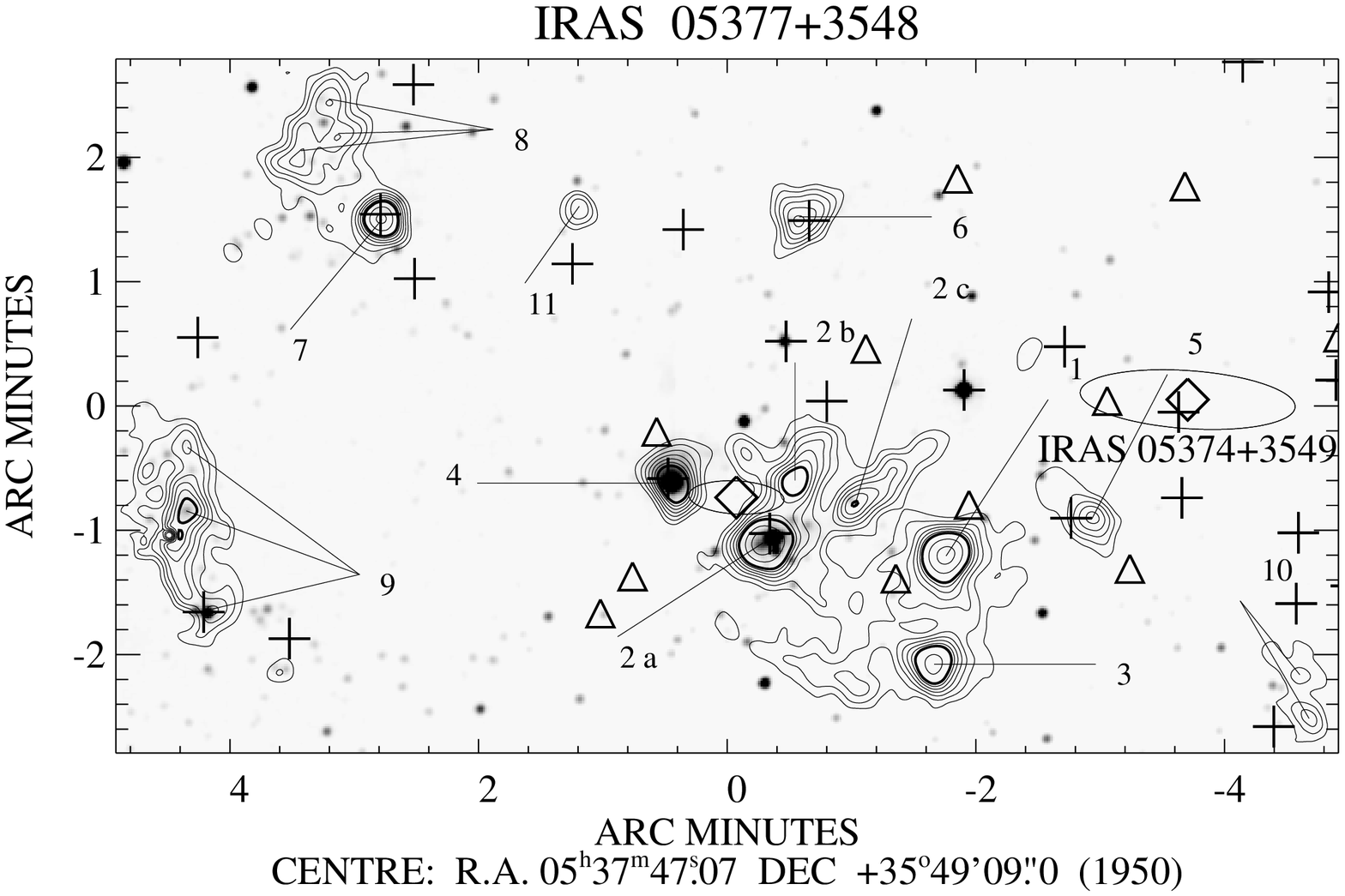}\\
  \hspace*{\fill}contours: (3$\sigma$, $\Delta=1\sigma$, {\boldmath$9\sigma$}, $\Delta=3\sigma$)\hspace*{\fill}\\
\end{figure}
\begin{figure}[htbp]
  \figurenum{1}
  \caption{Continued}
  \hspace*{.5\textwidth}
  \includegraphics[bb=45 15 580 580, width= 80mm]{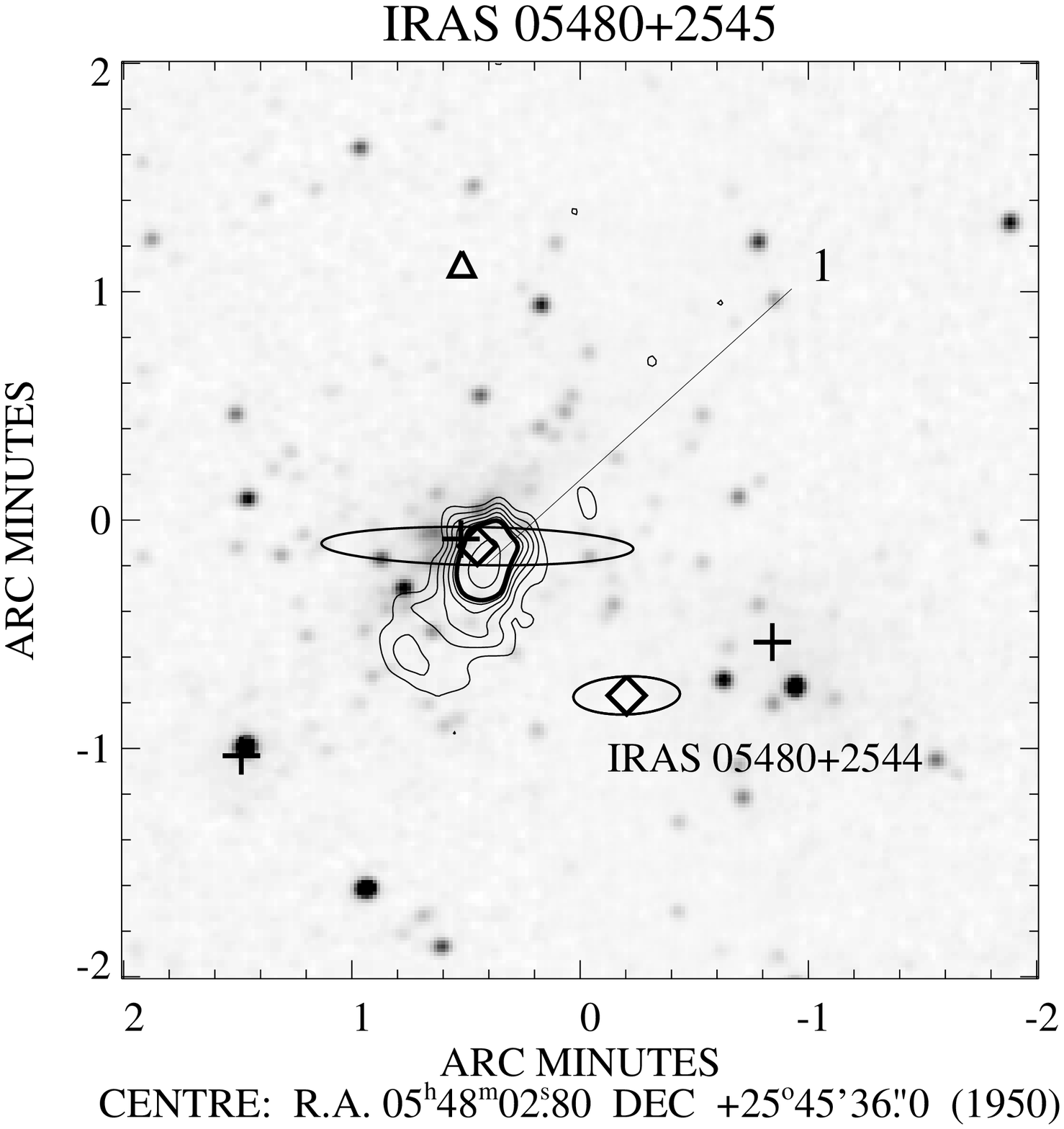}
  \hspace*{.5\textwidth}
  contours: (3$\sigma$, $\Delta=1\sigma$, {\boldmath$8\sigma$}, $\Delta=3\sigma$)\\
  \includegraphics[bb=55 45 340 350, width= 80mm]{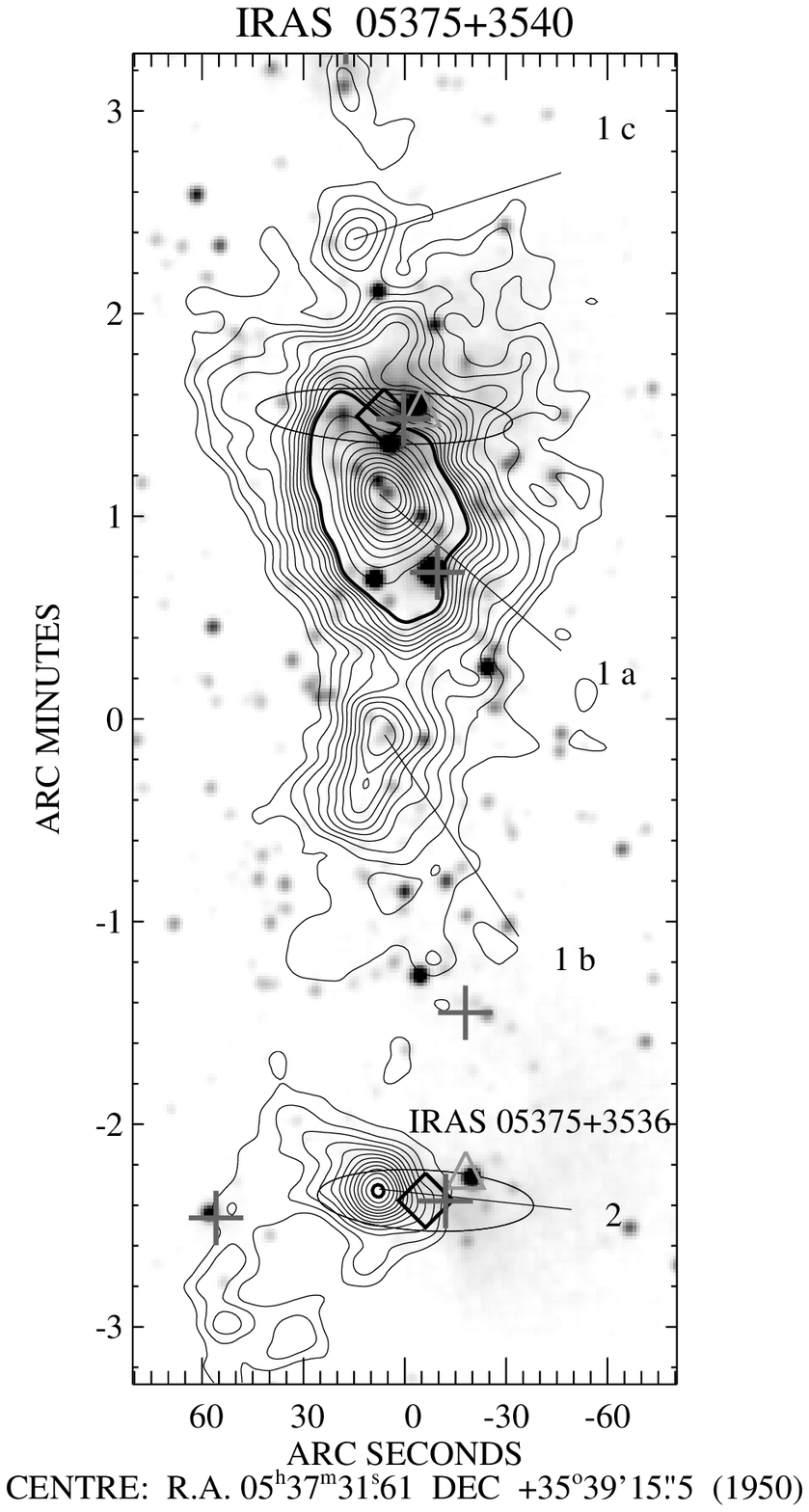}\hfill
  \includegraphics[bb=45 15 580 580, width= 80mm]{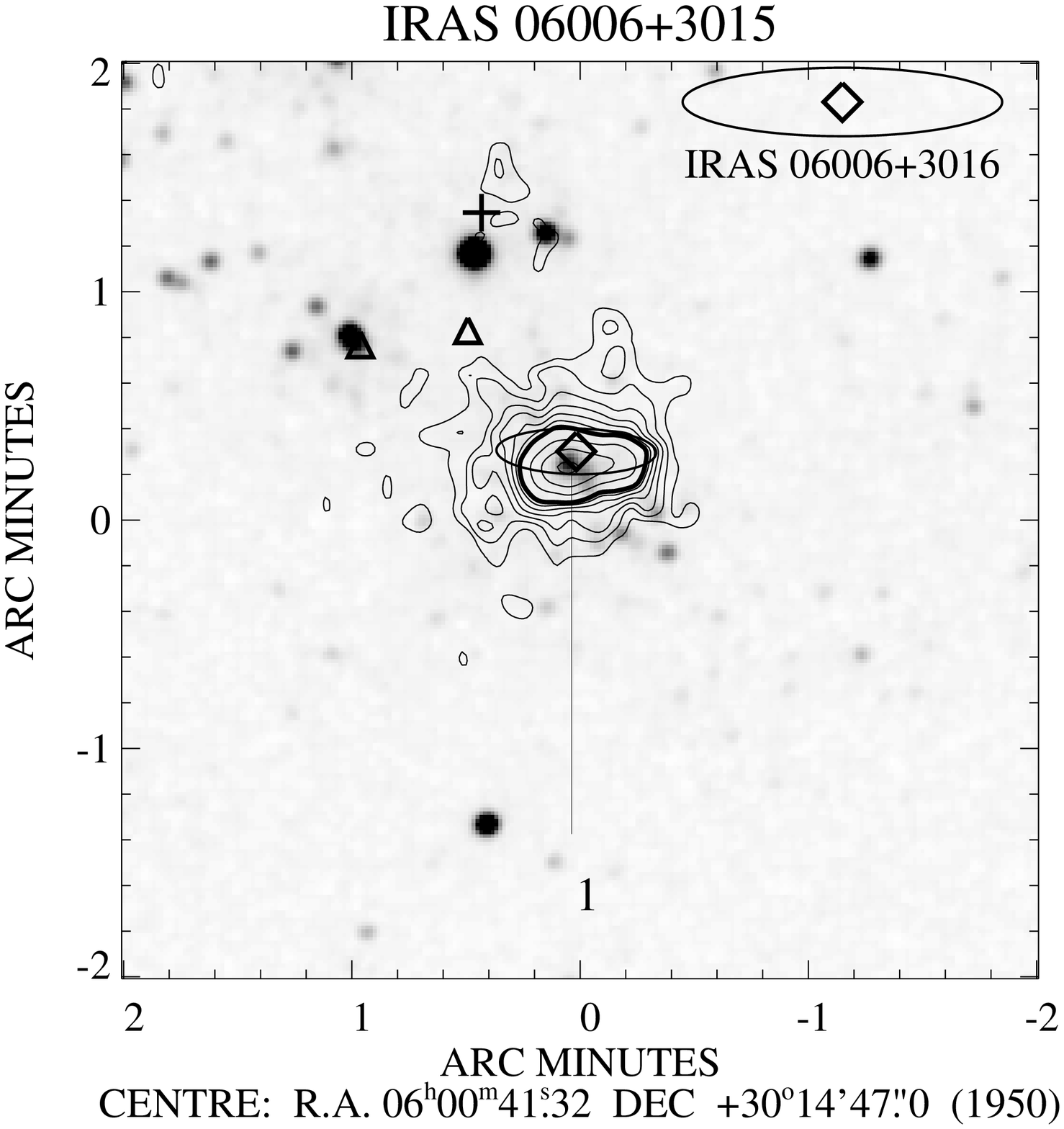}\\
\hspace*{\fill}
  contours: (3$\sigma$, $\Delta=1\sigma$, {\boldmath$15\sigma$}, $\Delta=5\sigma$)
  \hfill\hfill
  contours: (3$\sigma$, $\Delta=1\sigma$, {\boldmath$9\sigma$}, $\Delta=3\sigma$)\hspace*{\fill}
\end{figure}
\begin{figure}[htbp]
  \figurenum{1}
  \caption{Continued}
  \includegraphics[bb=35 10 555 585, width= 70mm]{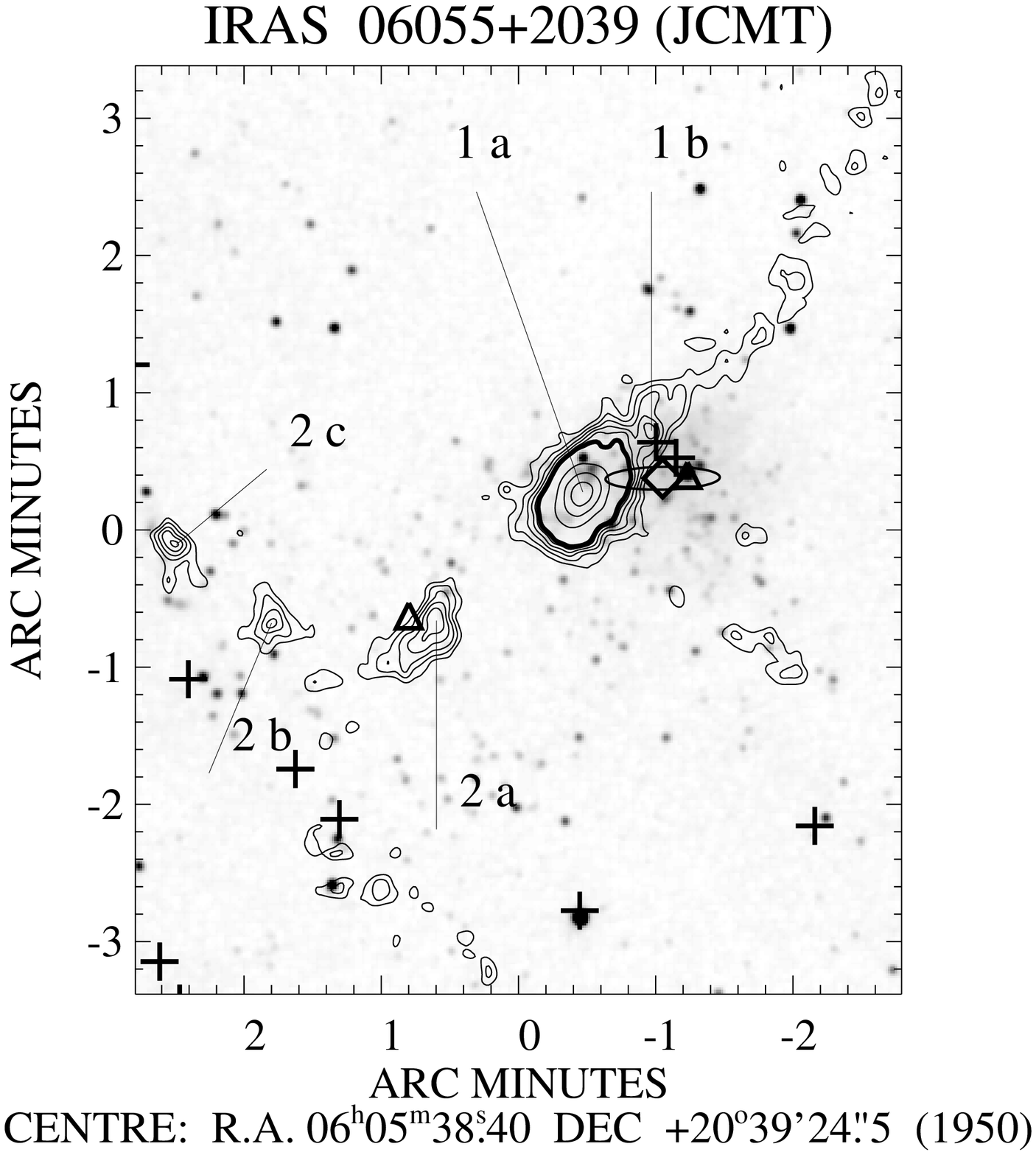}\hfill
  \includegraphics[bb=45 15 580 580, width= 70mm]{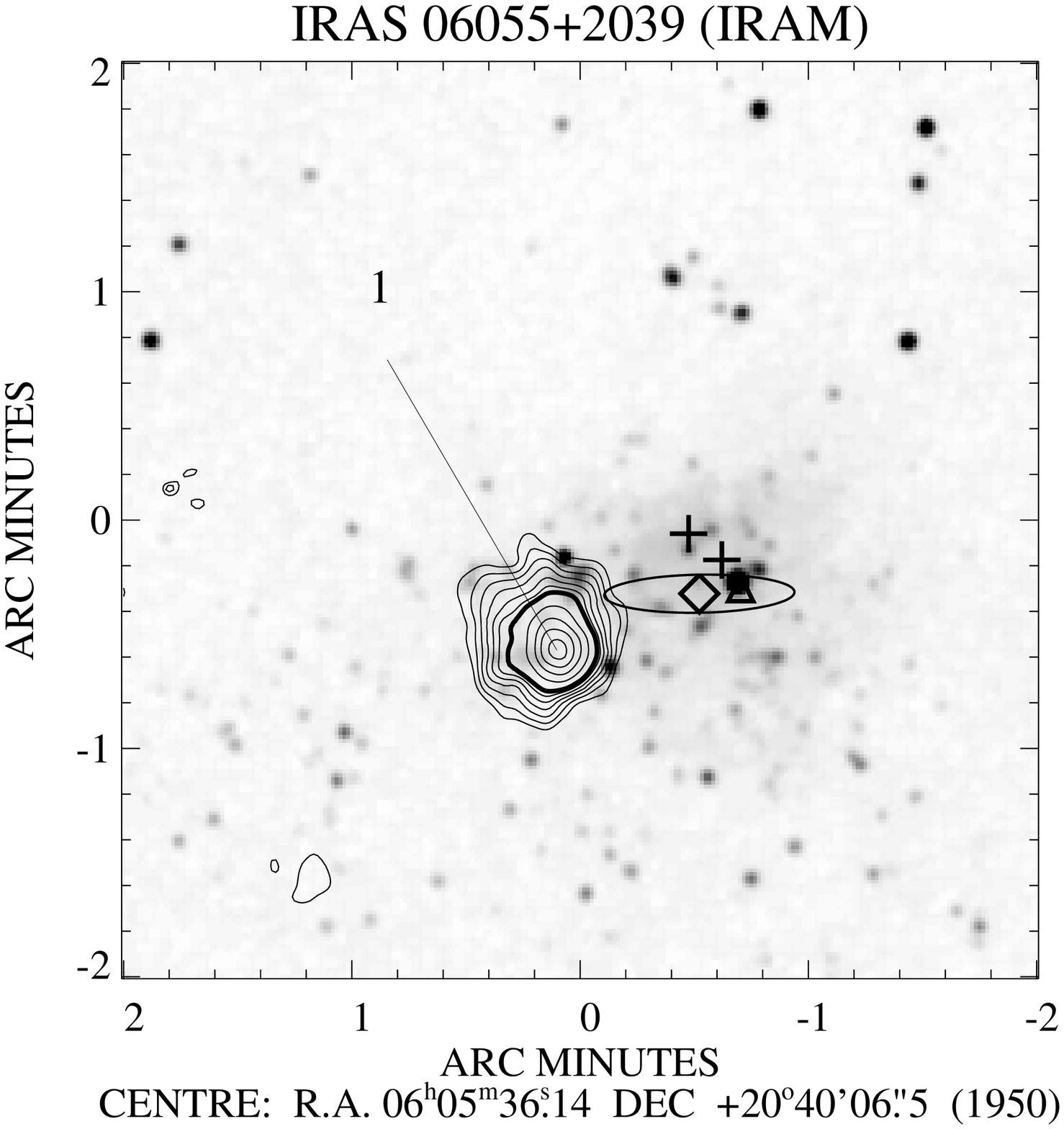}\\
   contours: (3$\sigma$, $\Delta=1\sigma$, {\boldmath$8\sigma$}, $\Delta=10\sigma$)
  \hfill\hfill
  contours: (3$\sigma$, $\Delta=1\sigma$, {\boldmath$9\sigma$}, $\Delta=3\sigma$)\hspace*{\fill}\hspace*{\fill}\\[3ex]
  \includegraphics[bb=50 25 250 320, width= 70mm]{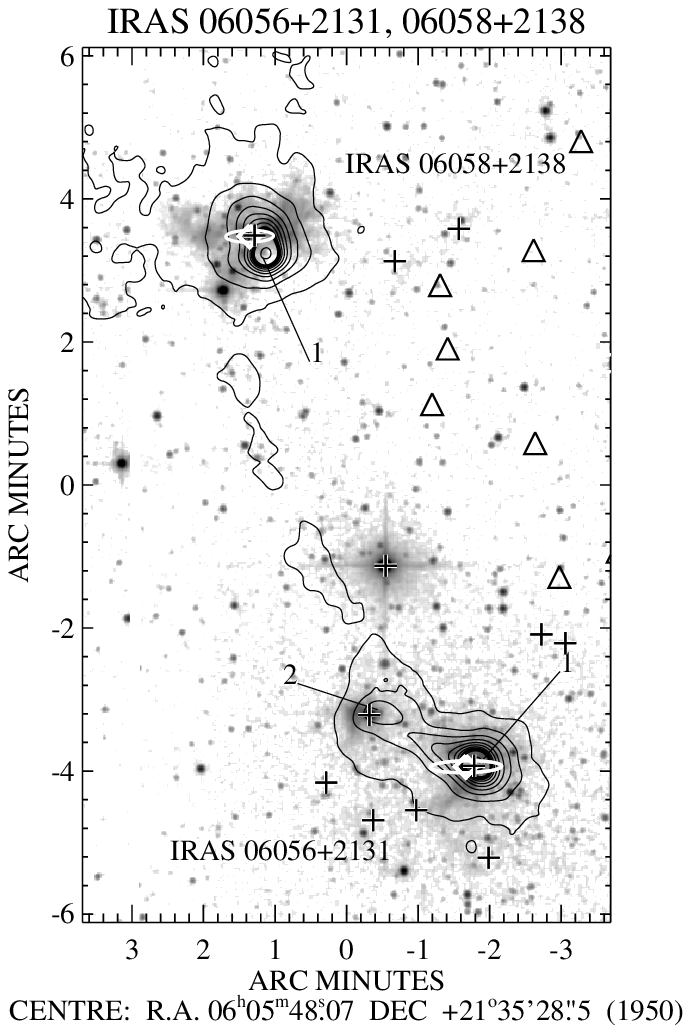}\hfill
  \includegraphics[bb=50 25 250 320, width= 70mm]{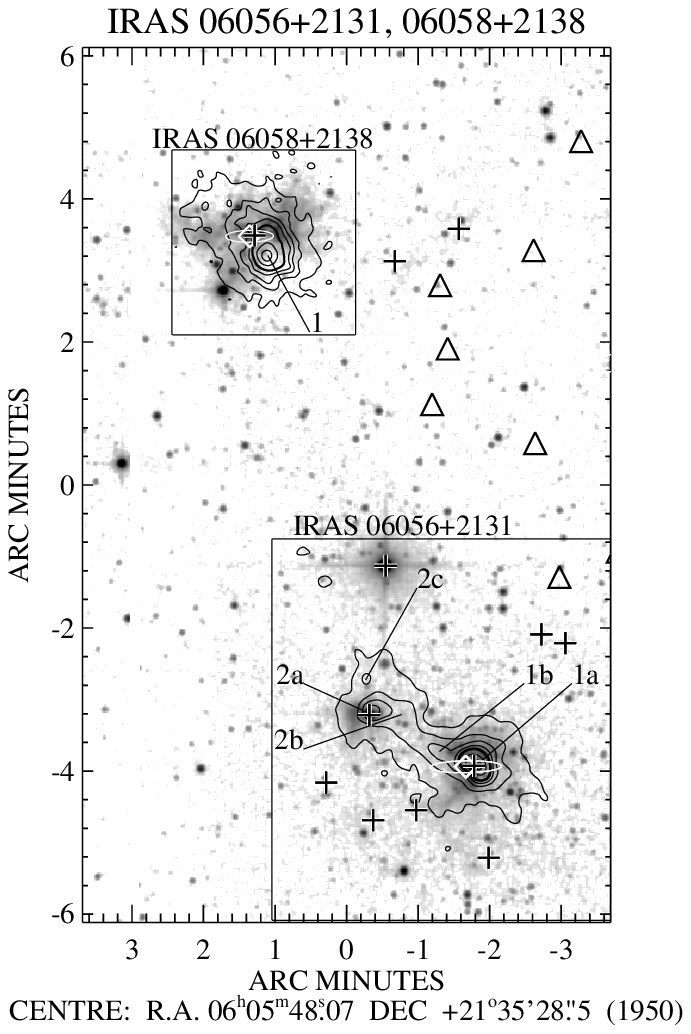}\\[1ex]
  \begin{minipage}[t]{.45\textwidth}
    Combined JCMT maps\\
    $\sigma$ respective to the  IRAS\,06056+2131 map\\
    contours: (3$\sigma$, $\Delta=5\sigma$, {\boldmath$40\sigma$}, $\Delta=20\sigma$)
  \end{minipage}\hfill
  \begin{minipage}[t]{.5\textwidth}
  The boxes give the extent of the IRAM maps.\\
  contours for IRAS\,06056+2131:\\
  (3$\sigma$, $\Delta=6\sigma$, {\boldmath$30\sigma$}, $\Delta=20\sigma$)\\
  contours for IRAS\,06058+2138:\\
  (3$\sigma$, $\Delta=3\sigma$, {\boldmath$15\sigma$}, $\Delta=10\sigma$)
  with respective $\sigma$'s.
  \end{minipage}
\end{figure}
\begin{figure}[htbp]
  \figurenum{1}
  \caption{Continued}
  \includegraphics[bb=45 15 580 580, width= 73mm]{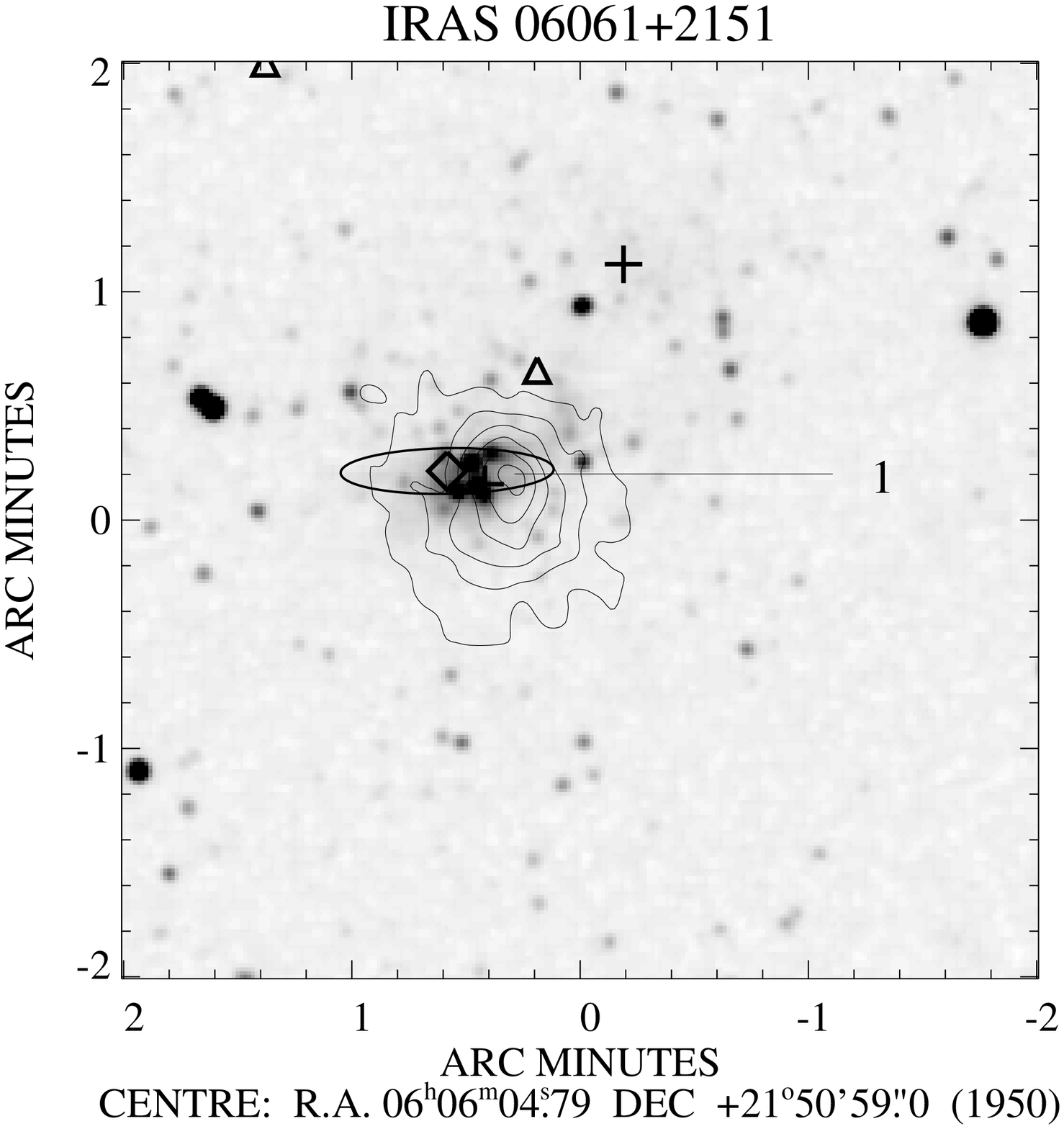}\hfill
  \includegraphics[bb=45 15 665 545, width= 87mm]{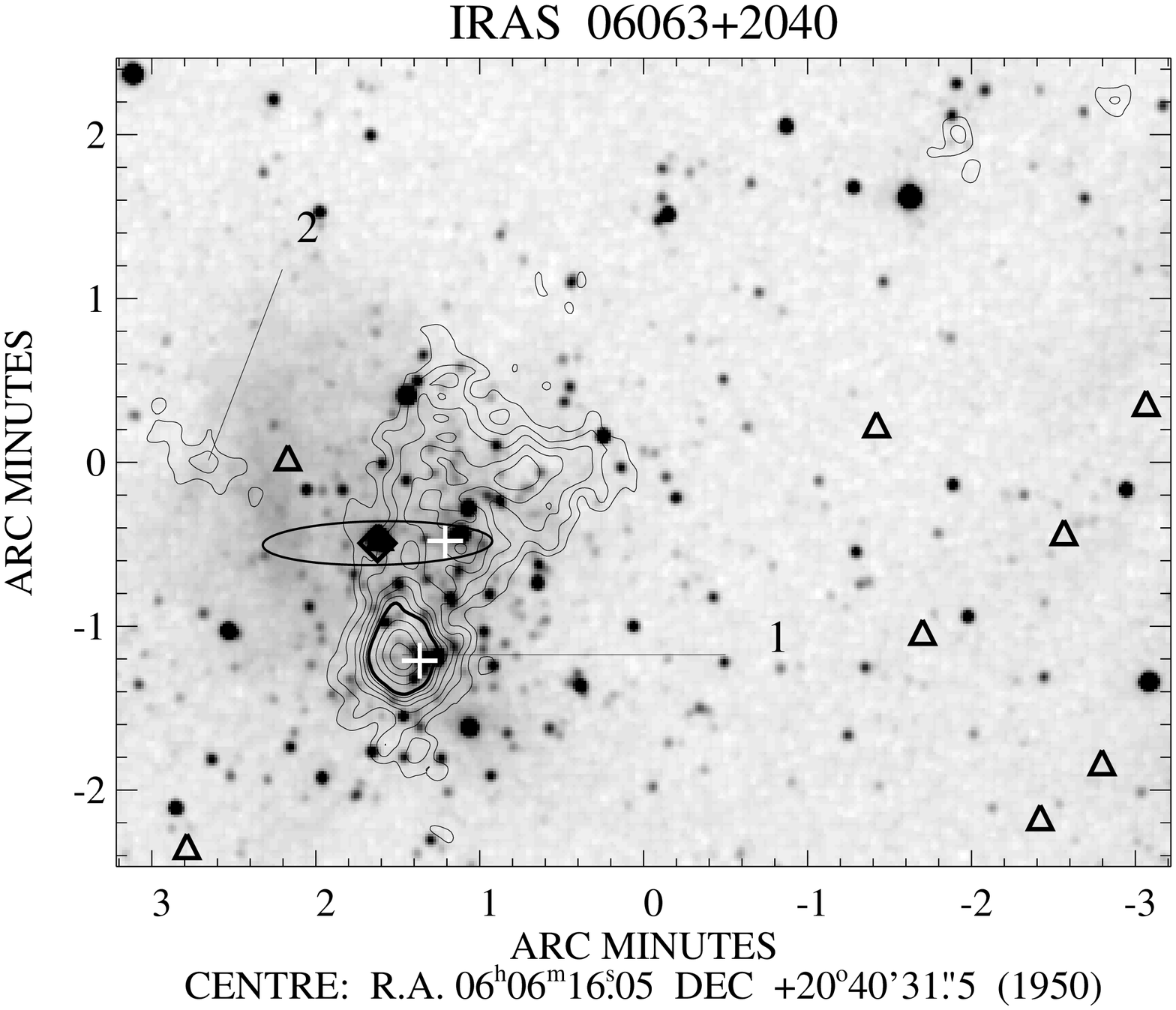}\\
  contours: (3$\sigma$, $\Delta=3\sigma$)\hfill\hfill\hfill\hfill
  contours: (3$\sigma$, $\Delta=1\sigma$, {\boldmath$9\sigma$}, $\Delta=3\sigma$)
  \hspace*{\fill}\hspace*{\fill}
\\[3ex]
  \epsscale{1}
  \includegraphics[bb=45 15 580 580, width= 80mm]{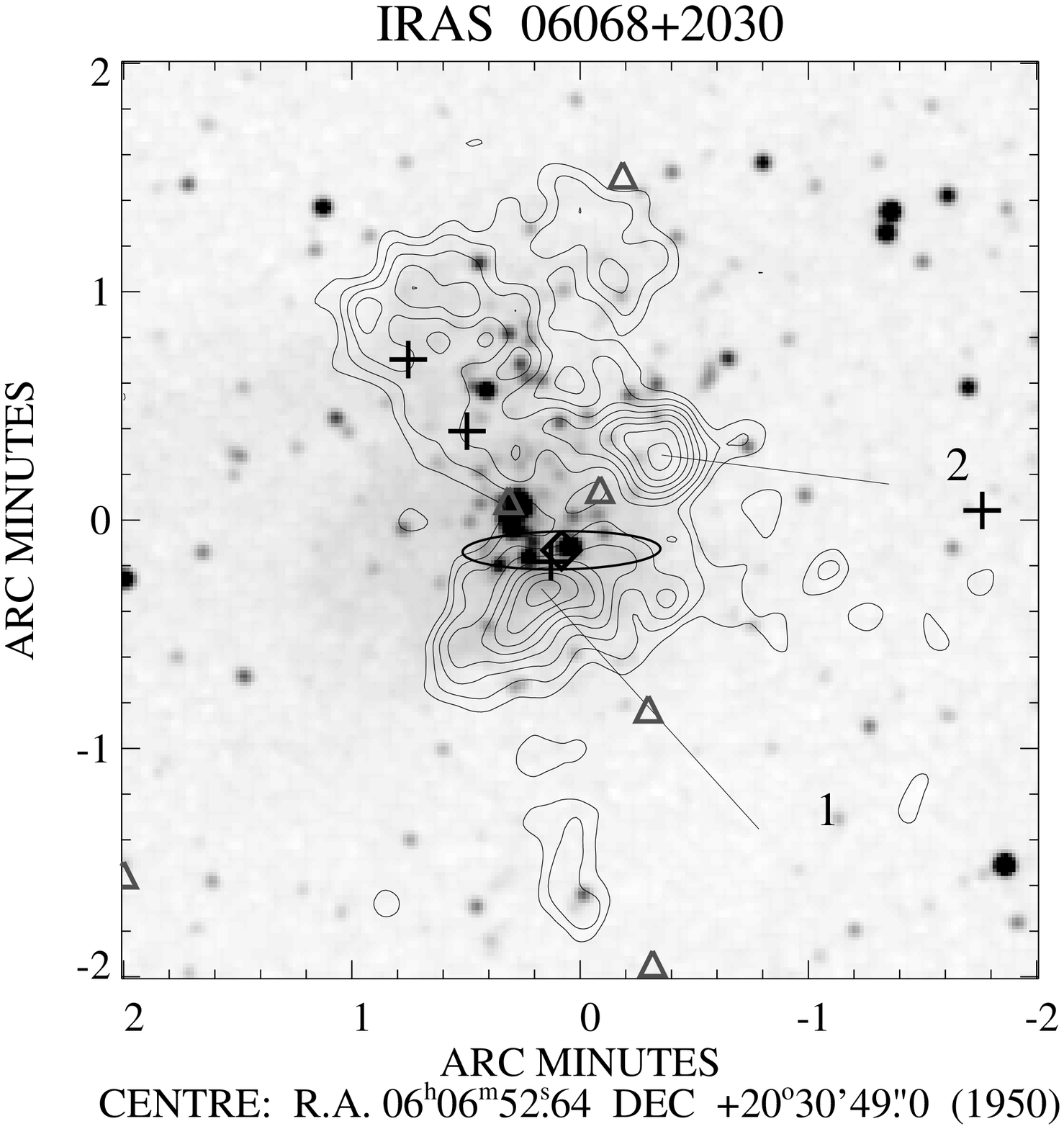}\hfill
  \includegraphics[bb=45 15 580 580, width= 80mm]{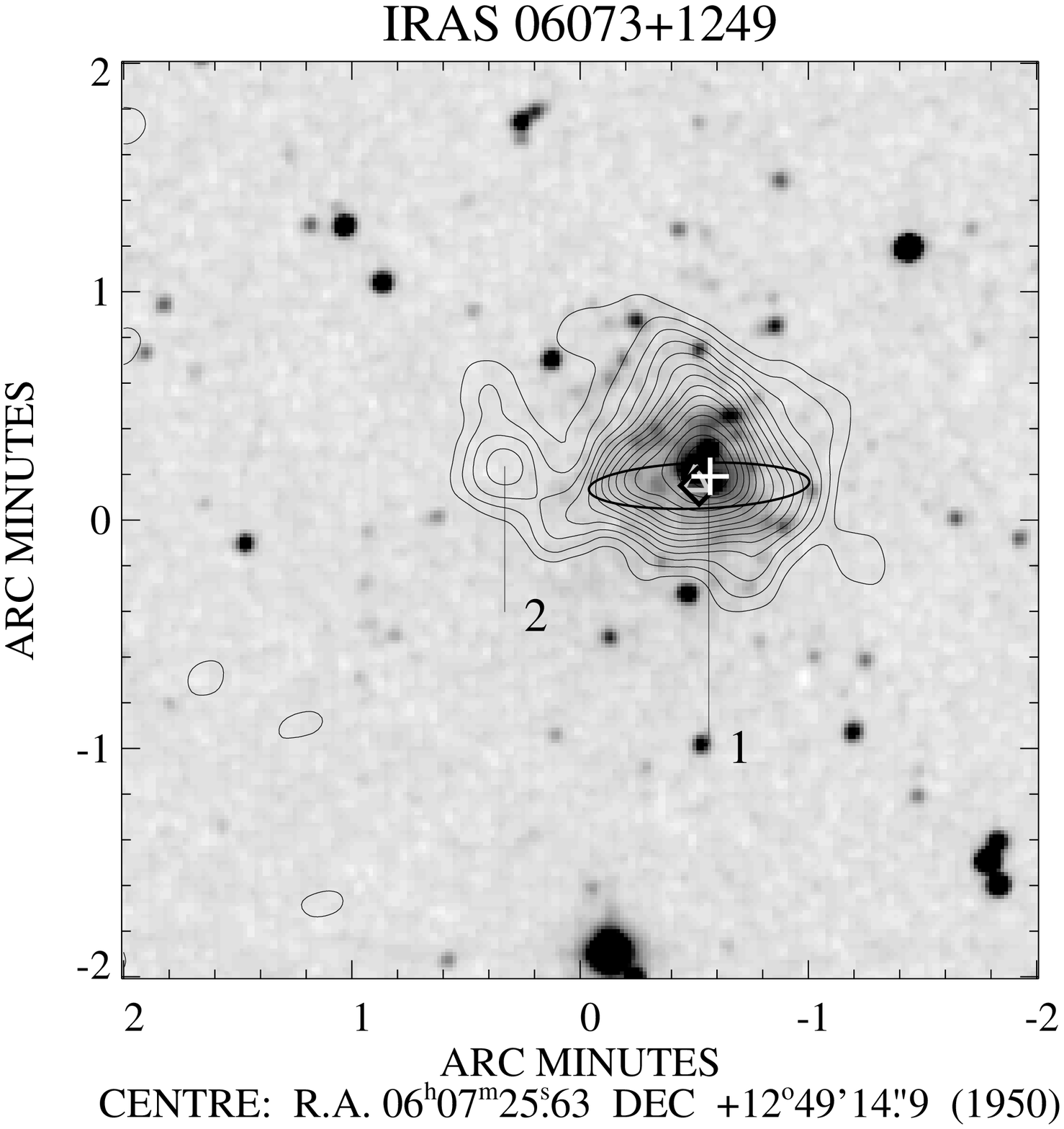}\\
  \hspace*{\fill}
  contours:  (3$\sigma$, $\Delta=1\sigma$)\hfill \hfill\hfill  
  contours: (3$\sigma$, $\Delta=1\sigma$)
  \hspace*{\fill}\hspace*{\fill}
\end{figure}
\begin{figure}[htbp]
  \figurenum{1}
  \caption{Continued}
  \includegraphics[bb=45 15 550 580, width= 80mm]{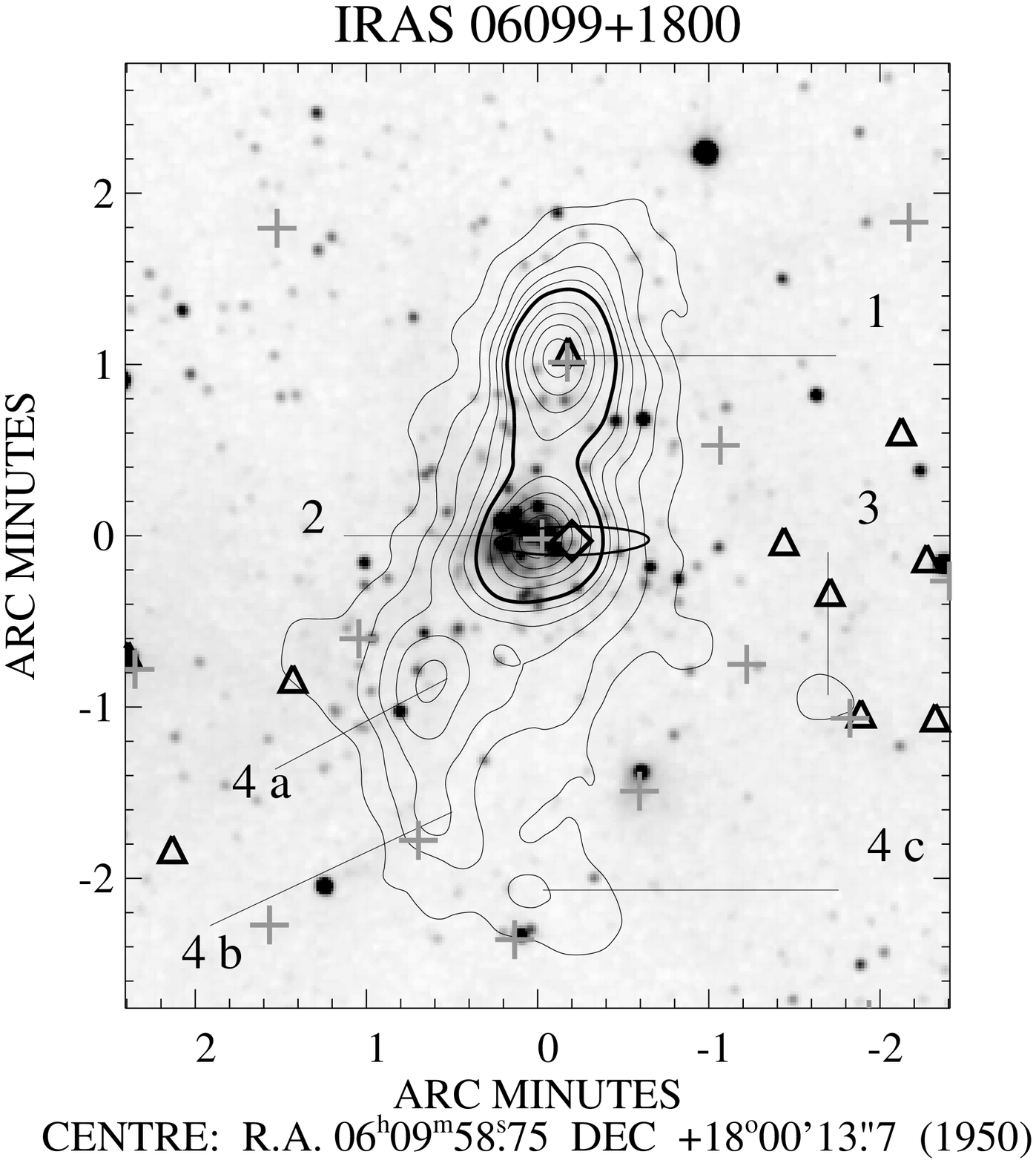}\hfill
  \includegraphics[bb=35 15 580 580, width= 80mm]{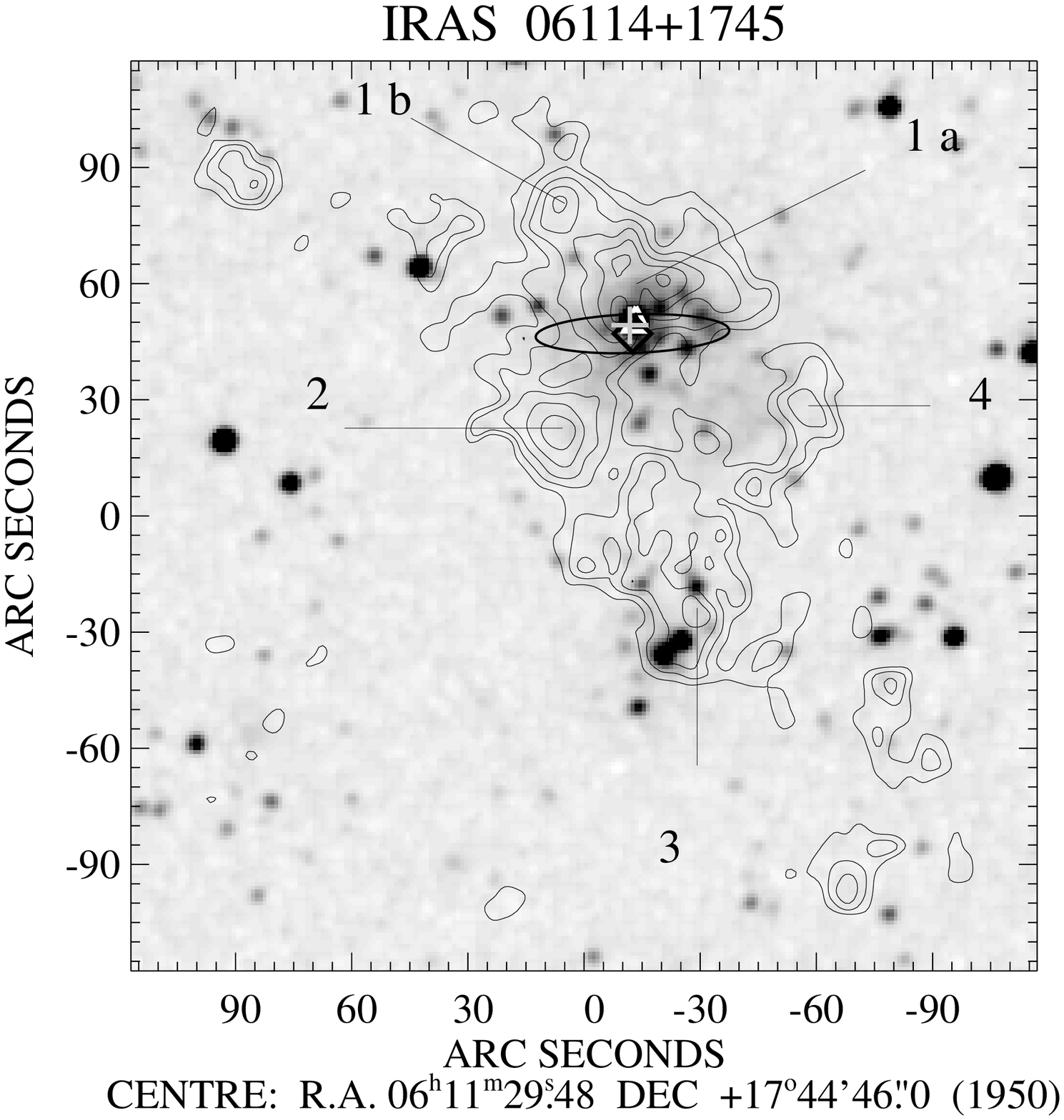}\\[1ex]
  \hspace*{\fill}\hspace*{\fill}
  contours: (3$\sigma$,  $\Delta=2\sigma$, {\boldmath$13\sigma$},  $\Delta=4\sigma$)
  \hfill\hfill\hfill
  contours: (3$\sigma$, $\Delta=1\sigma$)
  \hspace*{\fill}\hspace*{\fill}\\[3ex]
  \includegraphics[bb=45 15 665 425, width= 140mm]{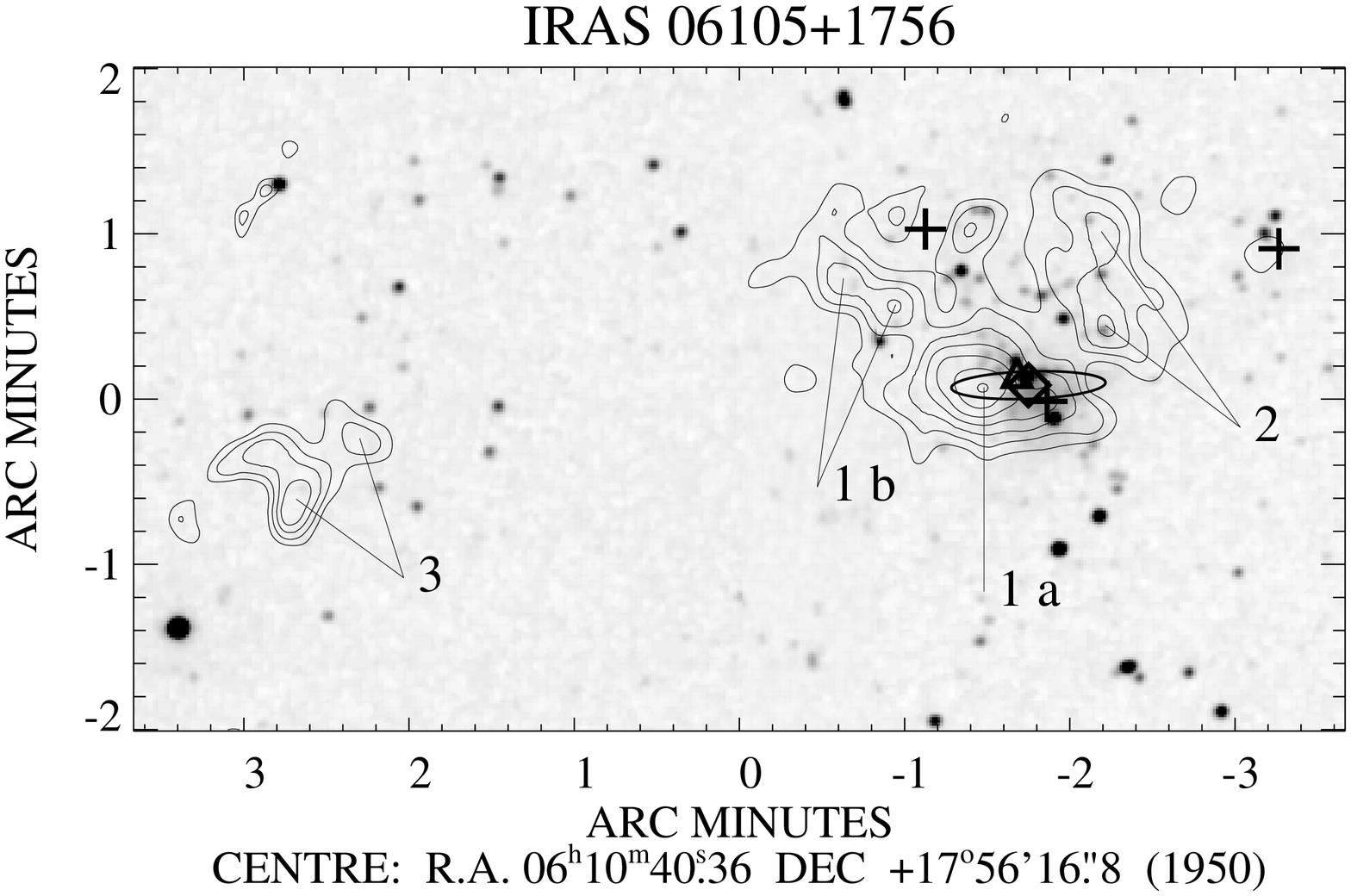}\\
  \hspace*{\fill}
  contours: (3$\sigma$, $\Delta=1\sigma$)
  \hspace*{\fill}
\end{figure}
\begin{figure}[htbp]
  \figurenum{1}
  \caption{Continued}
  \includegraphics[bb=35 15 585 580, width= 80mm]{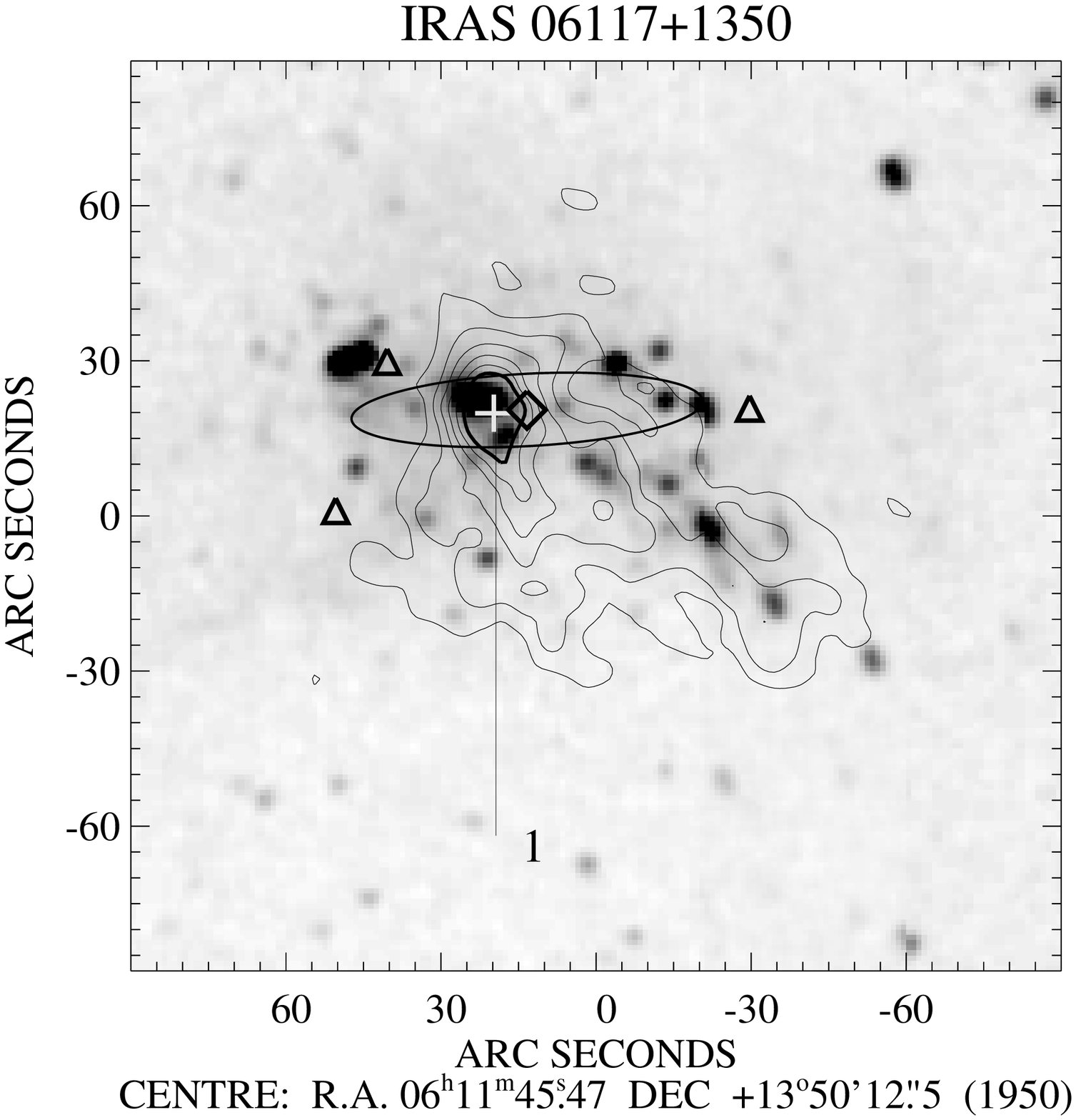}\hfill
  \includegraphics[bb=45 15 580 580, width= 80mm]{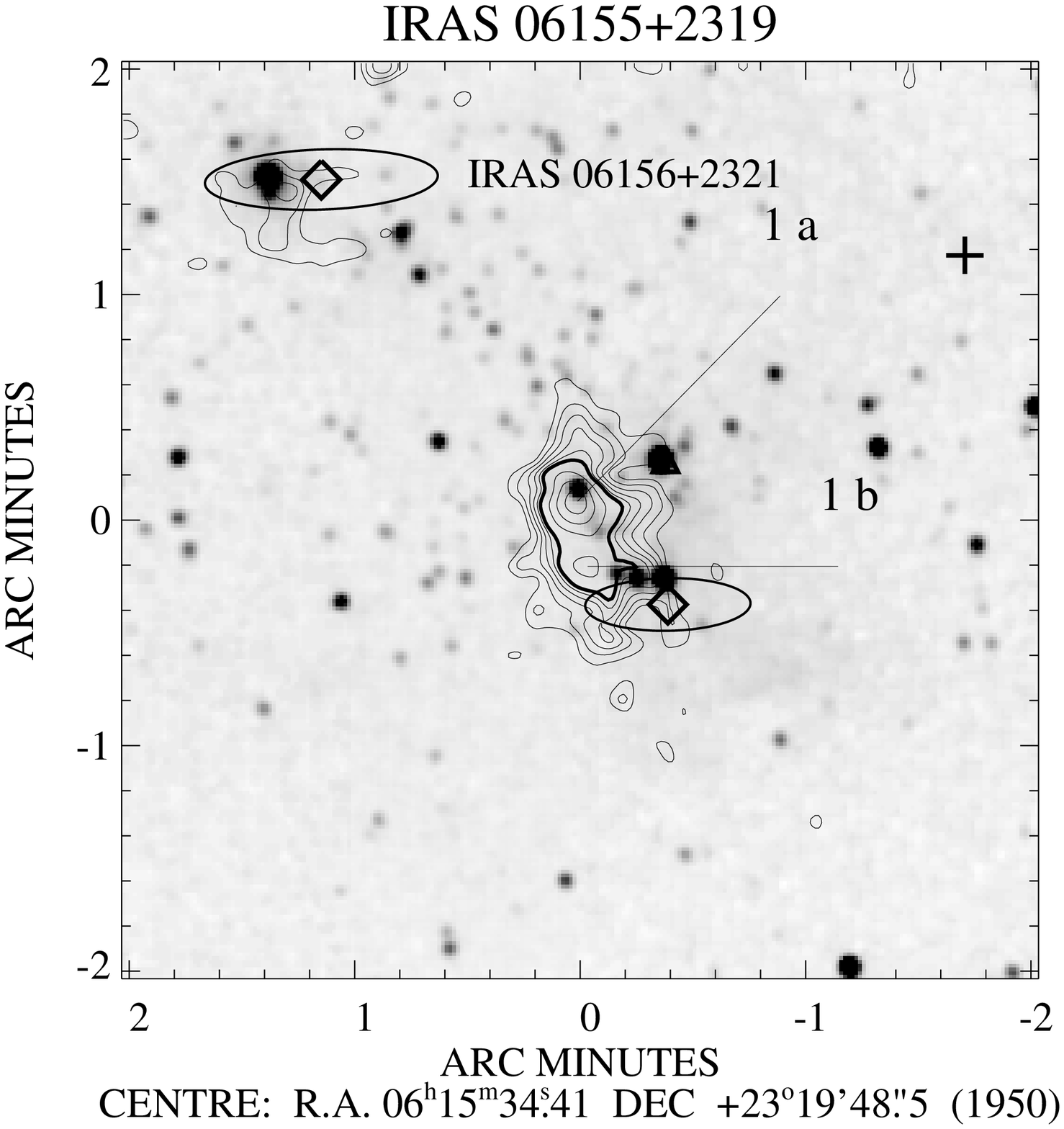}\\[1ex]
  \hspace*{\fill}\hspace*{\fill}
  contours: (3$\sigma$,  $\Delta=1\sigma$, {\boldmath$8\sigma$},  $\Delta=3\sigma$)
  \hfill \hfill \hfill
  contours: (3$\sigma$, $\Delta=1\sigma$, {\boldmath$8\sigma$},  $\Delta=3\sigma$)
  \hspace*{\fill}\\[3ex]
  \includegraphics[bb=45 15 660 580, width= 85mm]{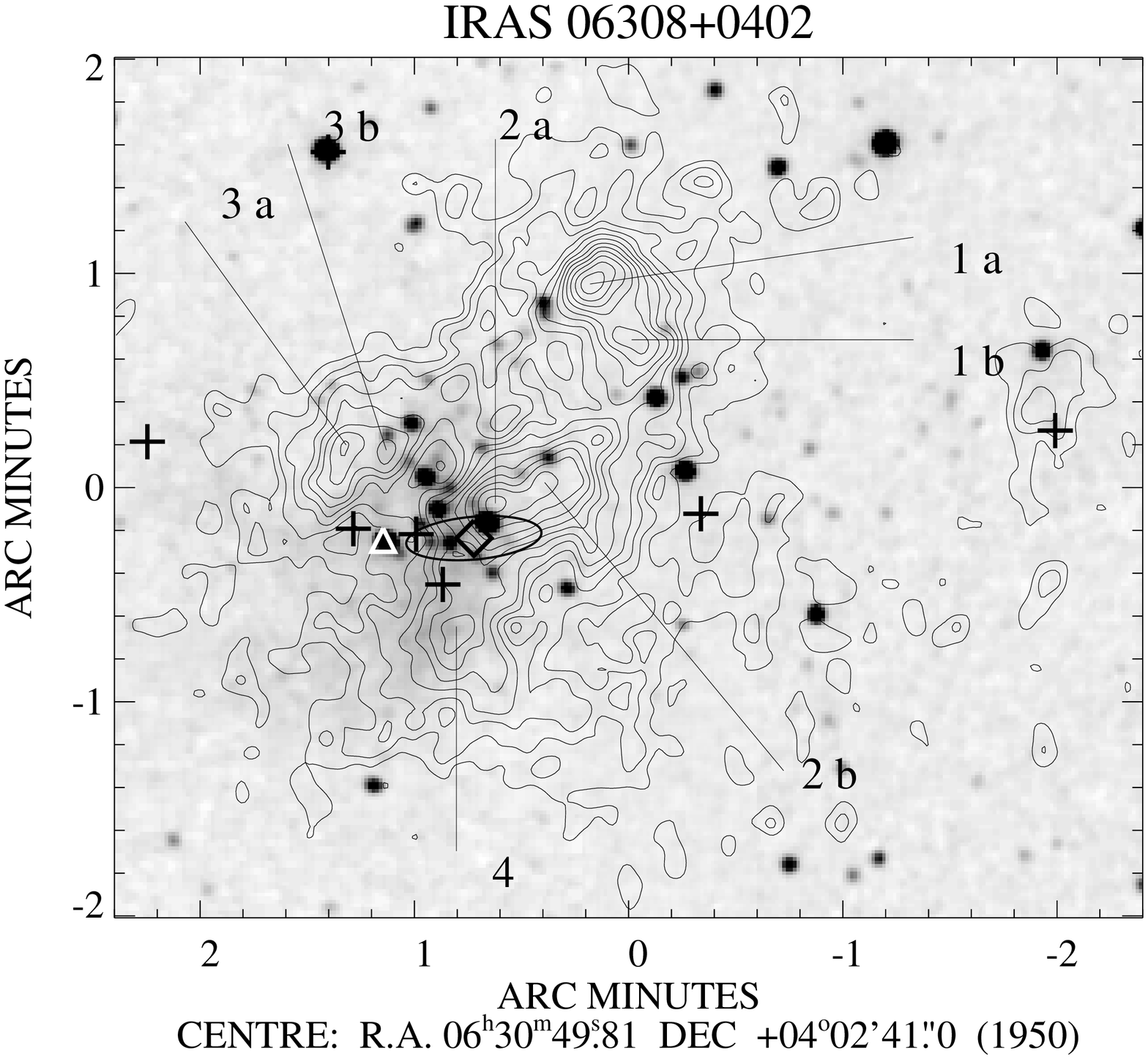}\hfill
  \includegraphics[bb=45 15 580 580, width= 75mm]{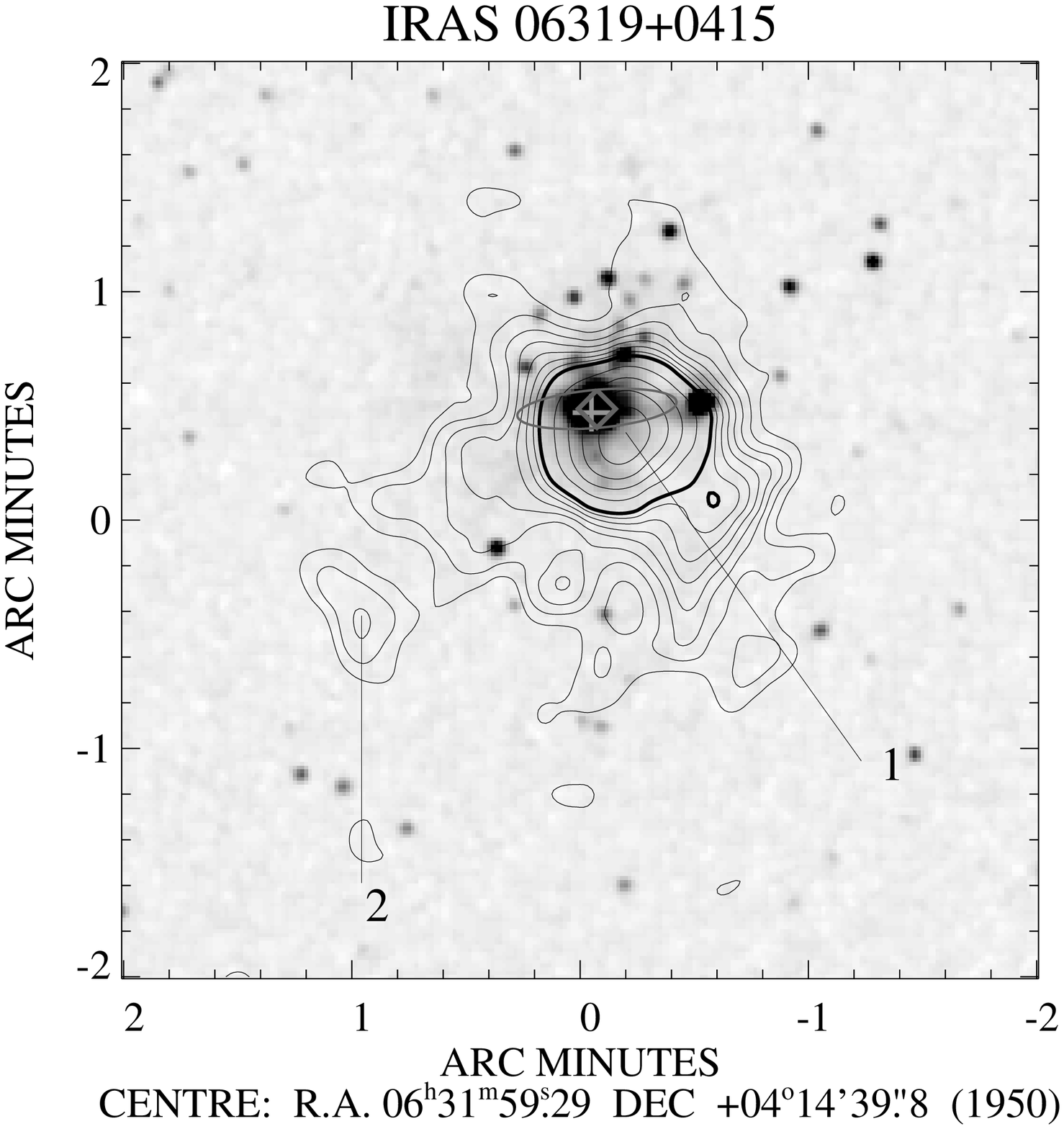}\\[1ex]
  \hspace*{\fill}\hspace*{\fill}\hspace*{\fill}
  contours: (3$\sigma$, $\Delta=1\sigma$)
  \hfill\hfill\hfill\hfill
  contours: (3$\sigma$, $\Delta=1\sigma$, {\boldmath$10\sigma$},  $\Delta=3\sigma$)
  \hspace*{\fill} 
\end{figure}
\begin{figure}[htbp]
  \figurenum{1}
  \caption{Continued}
  \includegraphics[bb=60 30 440 570, width= 75mm]{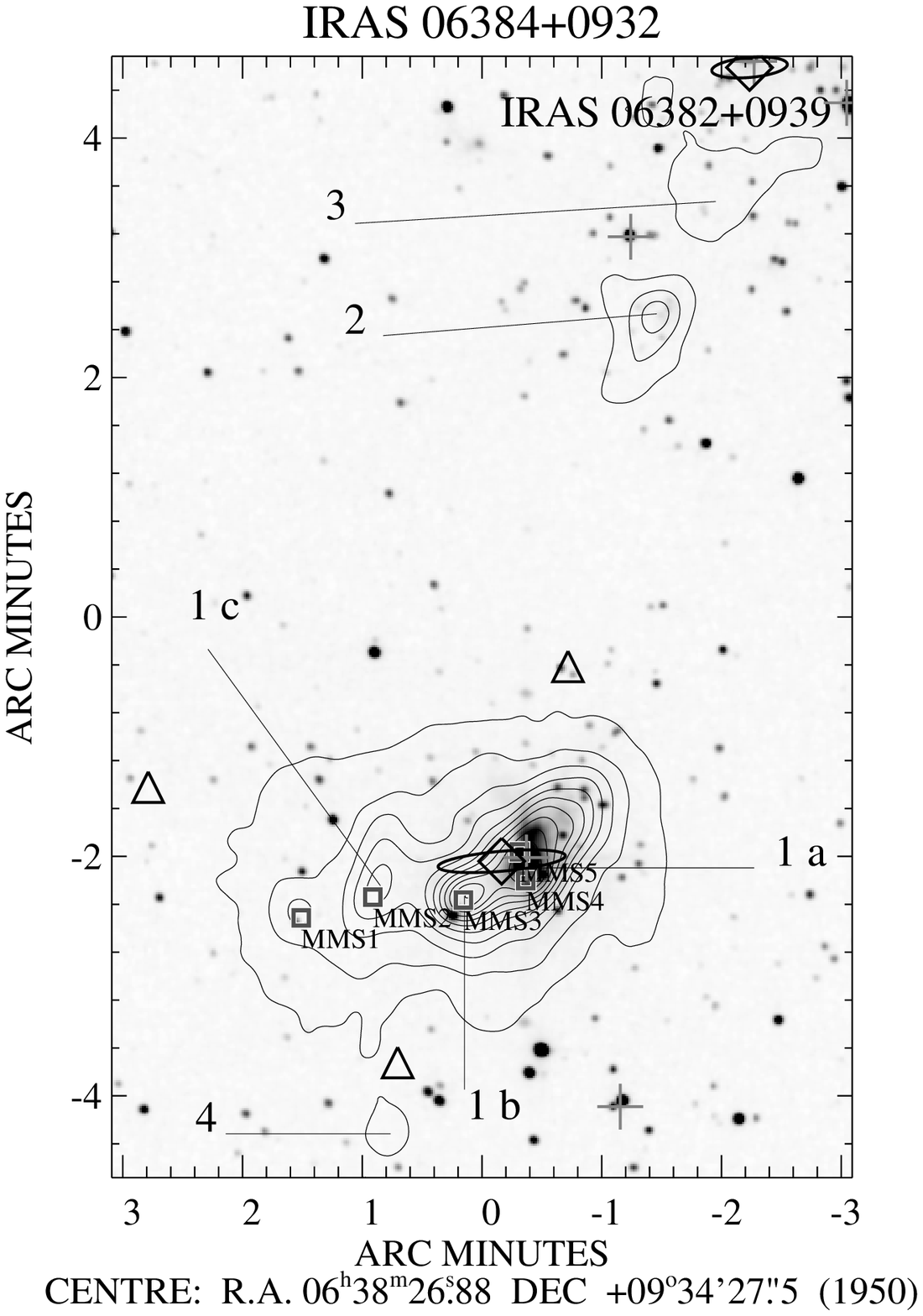}\\[1ex]
  \hfill contours: (3$\sigma$, $\Delta=3\sigma$)\hfill\hfill\hfill\hfill\hfill\hspace*{\fill}\\[1ex]
  \includegraphics[bb=45 15 580 580, width= 75mm]{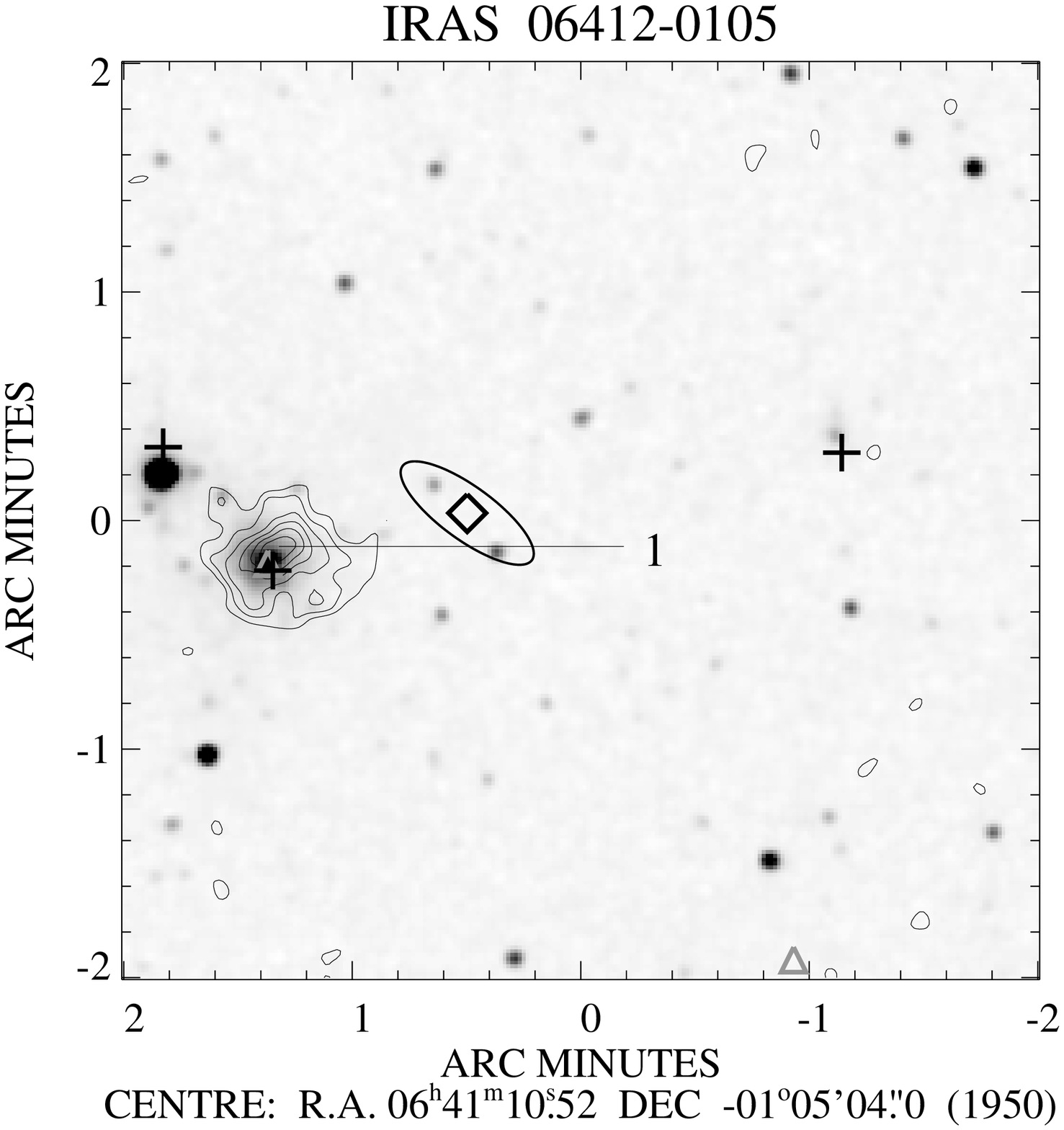}\hfill
  \includegraphics[bb=65 45 350 340, width= 75mm]{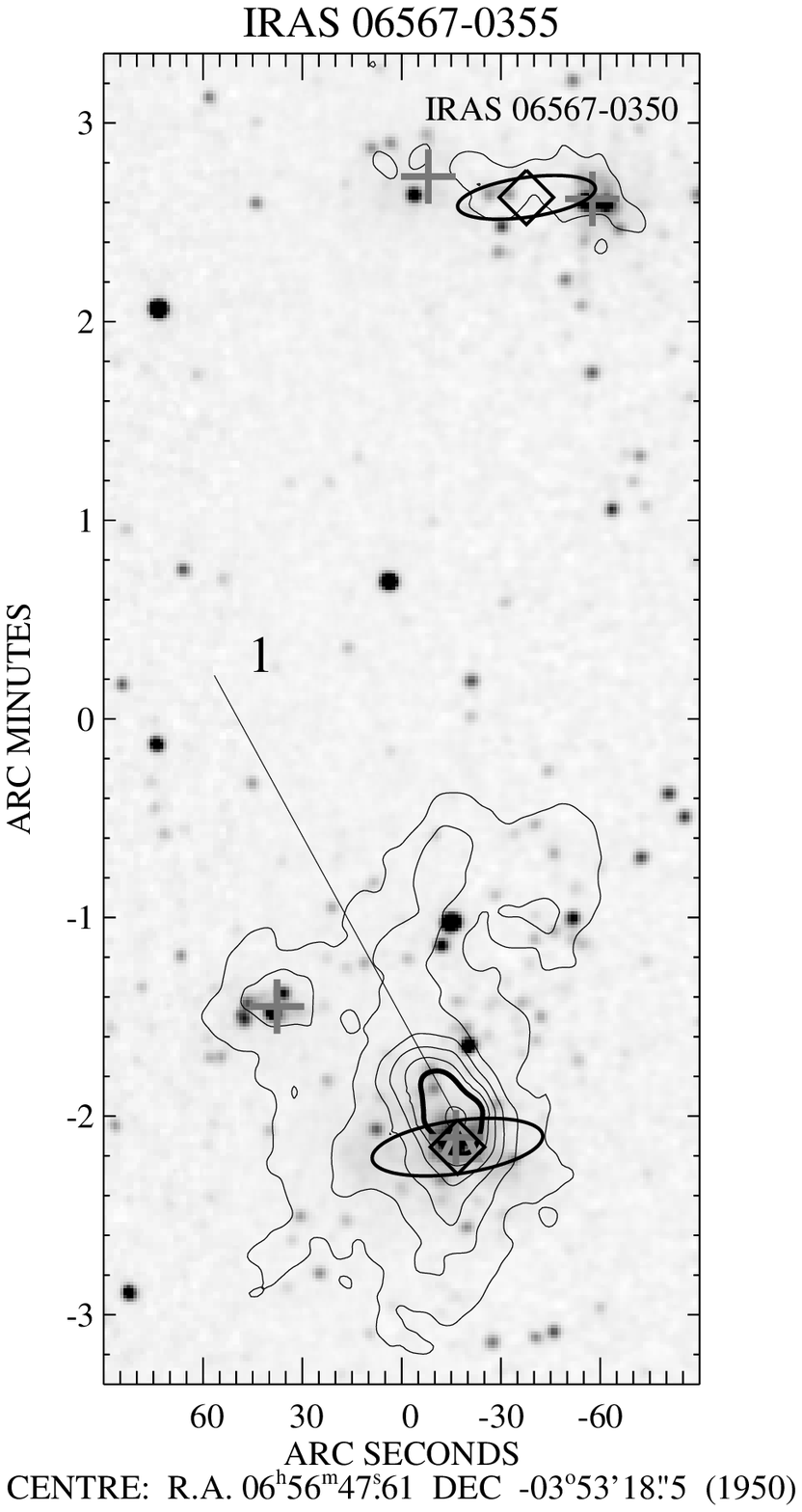}\\[1ex]
  \hspace*{\fill}
  \hfill
  contours: (3$\sigma$, $\Delta=1\sigma$)
  \hfill\hfill\hfill\hfill\hfill\hfill
  contours: (3$\sigma$, $\Delta=3\sigma$, {\boldmath$18\sigma$},  $\Delta=6\sigma$)
  \hspace*{\fill} 
\end{figure}
\begin{figure}[htbp]
  \figurenum{1}
  \caption{Continued}
  \includegraphics[bb=45 15 580 580, width= 70mm]{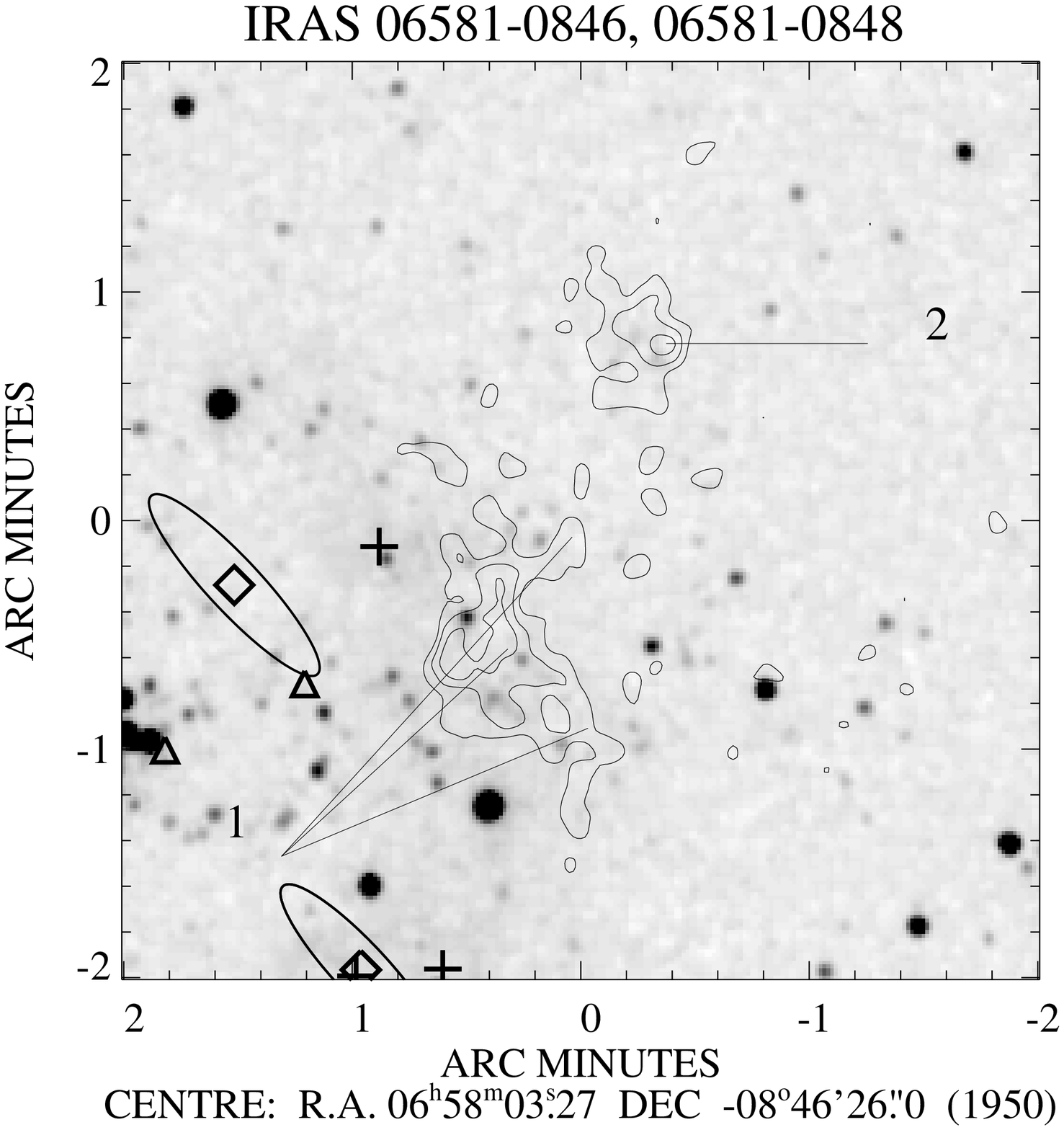}\hfill
  \includegraphics[bb=45 15 580 580, width= 70mm]{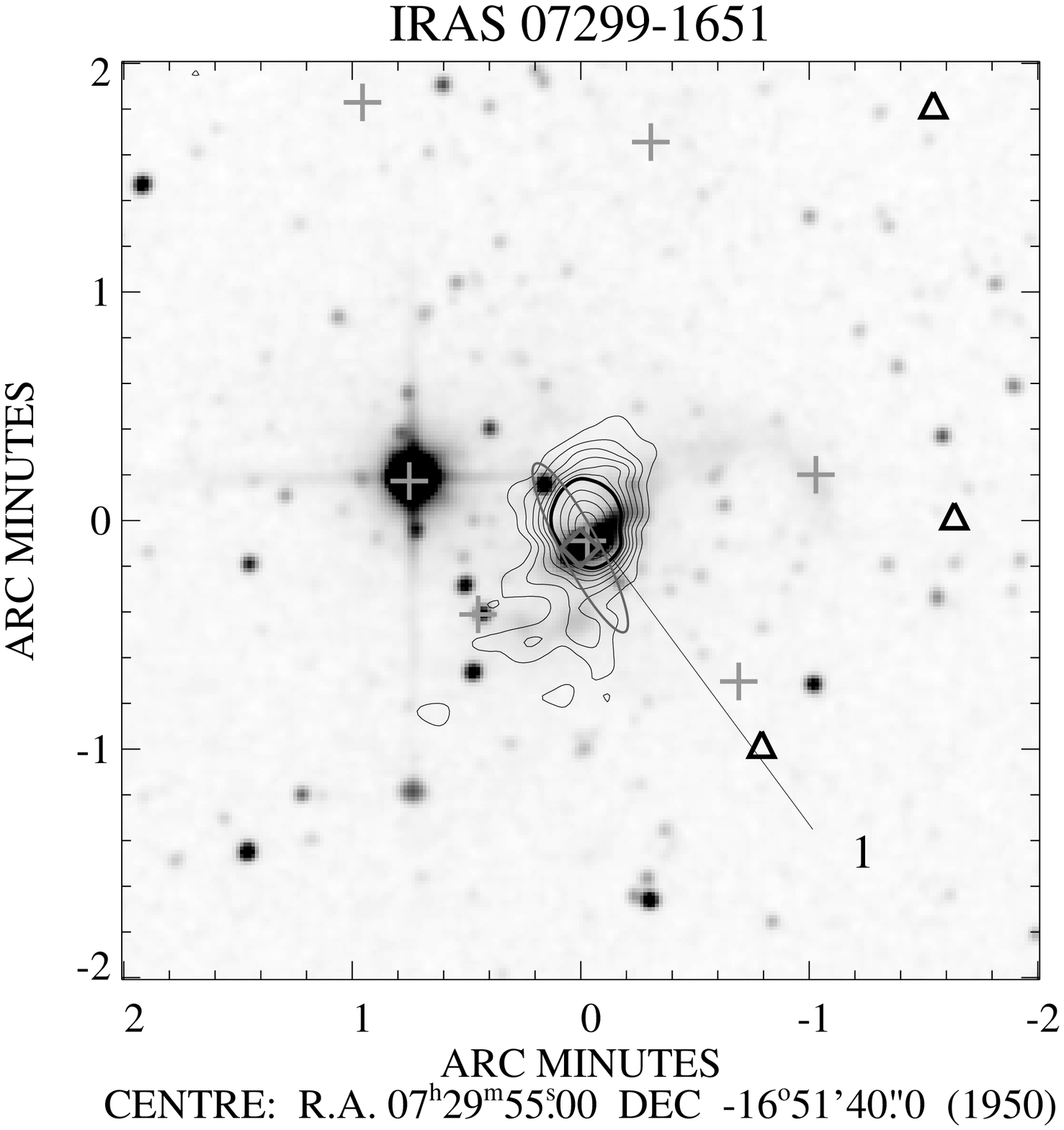}\\
  \hspace*{\fill}\hfill\hfill
  contours: (3$\sigma$, $\Delta=1\sigma$)
  \hfill\hfill\hfill\hfill\hfill\hfill
  contours: (3$\sigma$, $\Delta=1\sigma$, {\boldmath$9\sigma$},  $\Delta=3\sigma$)
  \hspace*{\fill}\hspace*{\fill}\\[3ex]
   \epsscale{1}
  \includegraphics[bb=45 15 655 500, width= 140mm]{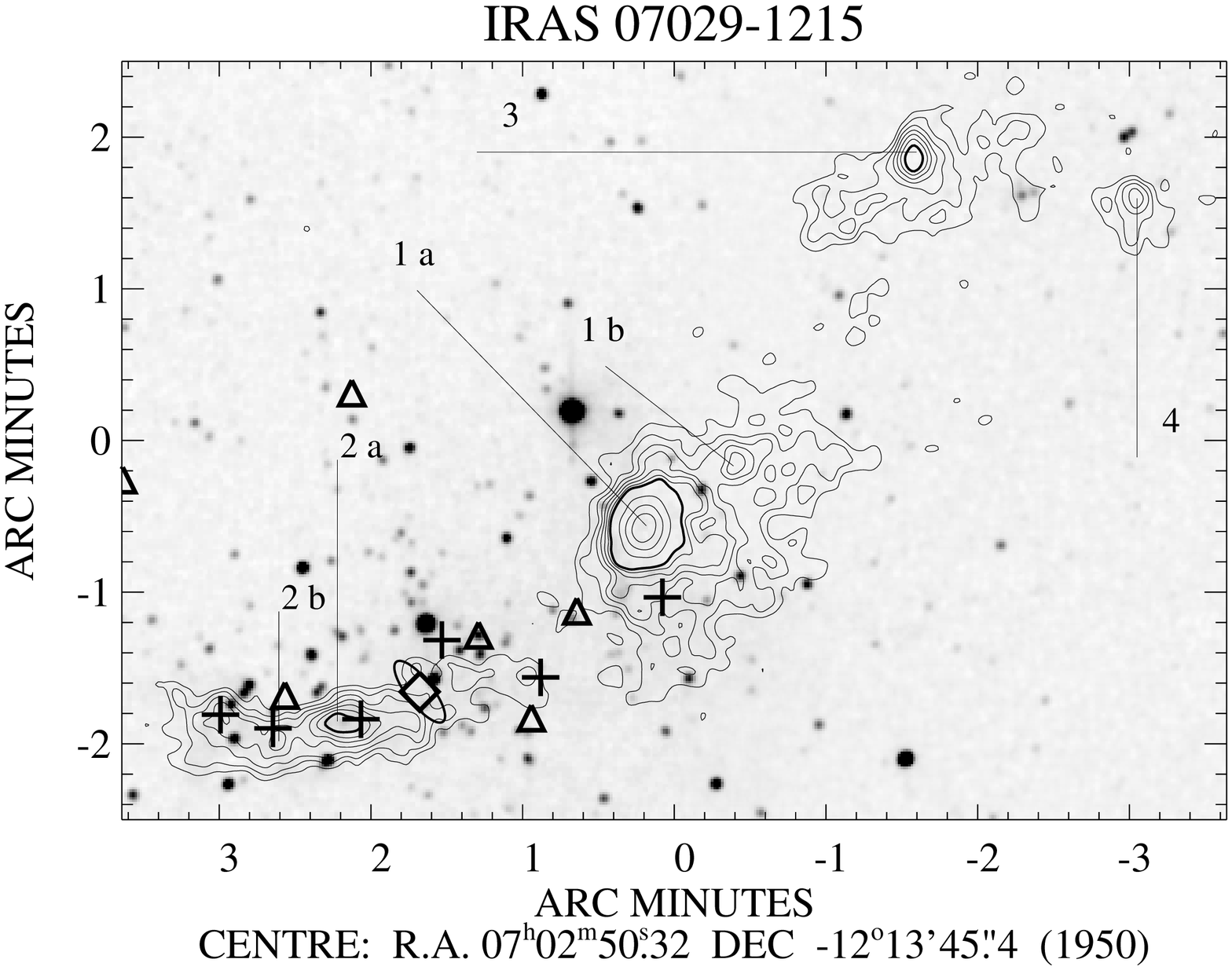}\\
  contours: (3$\sigma$, $\Delta=1\sigma$, {\boldmath$9\sigma$},  $\Delta=5\sigma$)
\end{figure}
\addtocounter{figure}{1}

\begin{figure}[htbp]
  \caption{Histogram showing the numbers of sources in a given mass {\em
      and} distance range.}
  \label{fig:MassvsDist}
  \plotone{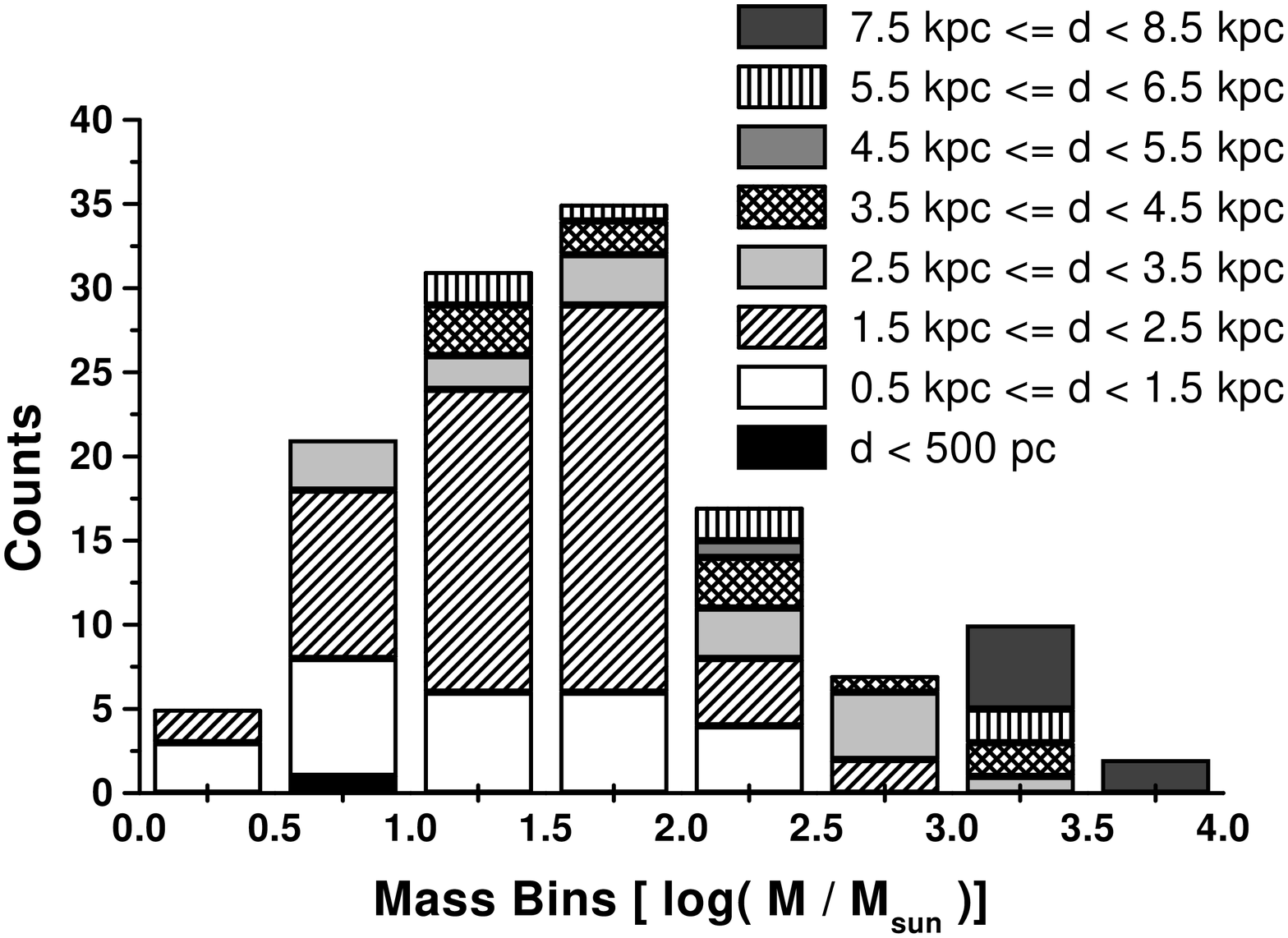}
\end{figure}

\begin{figure}[htbp]
  \caption{Histogram of cloud components and their MSX associations versus mass bins.}
  \label{fig:MSXvsMass}
  \plotone{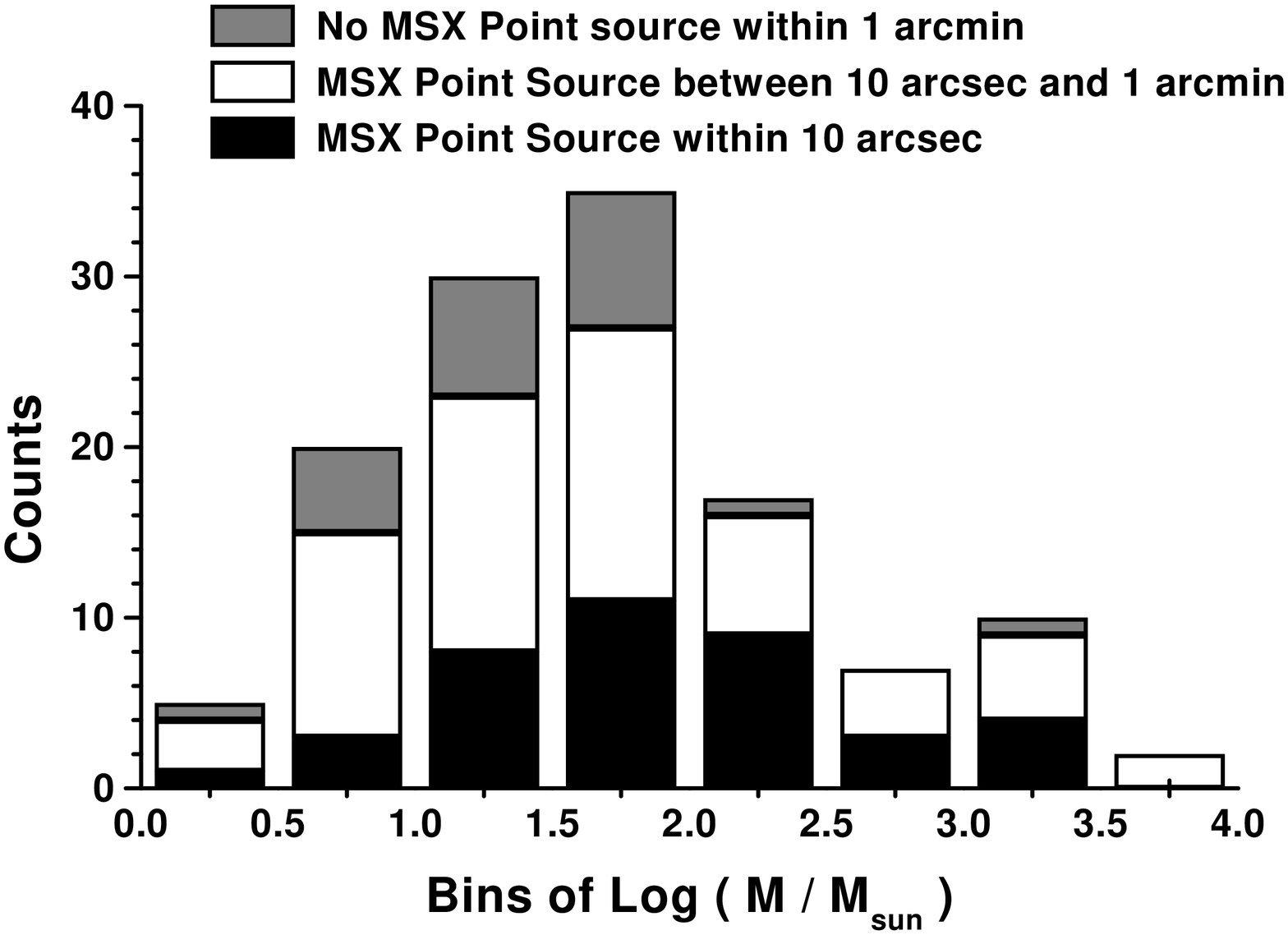}
\end{figure}

\begin{figure}[htbp]
  \caption{The same historgram as in Figure \ref{fig:MSXvsMass}, but
    only for sources closer than 2.5\,kpc.}
  \label{fig:MSXvsMassNear}
  \plotone{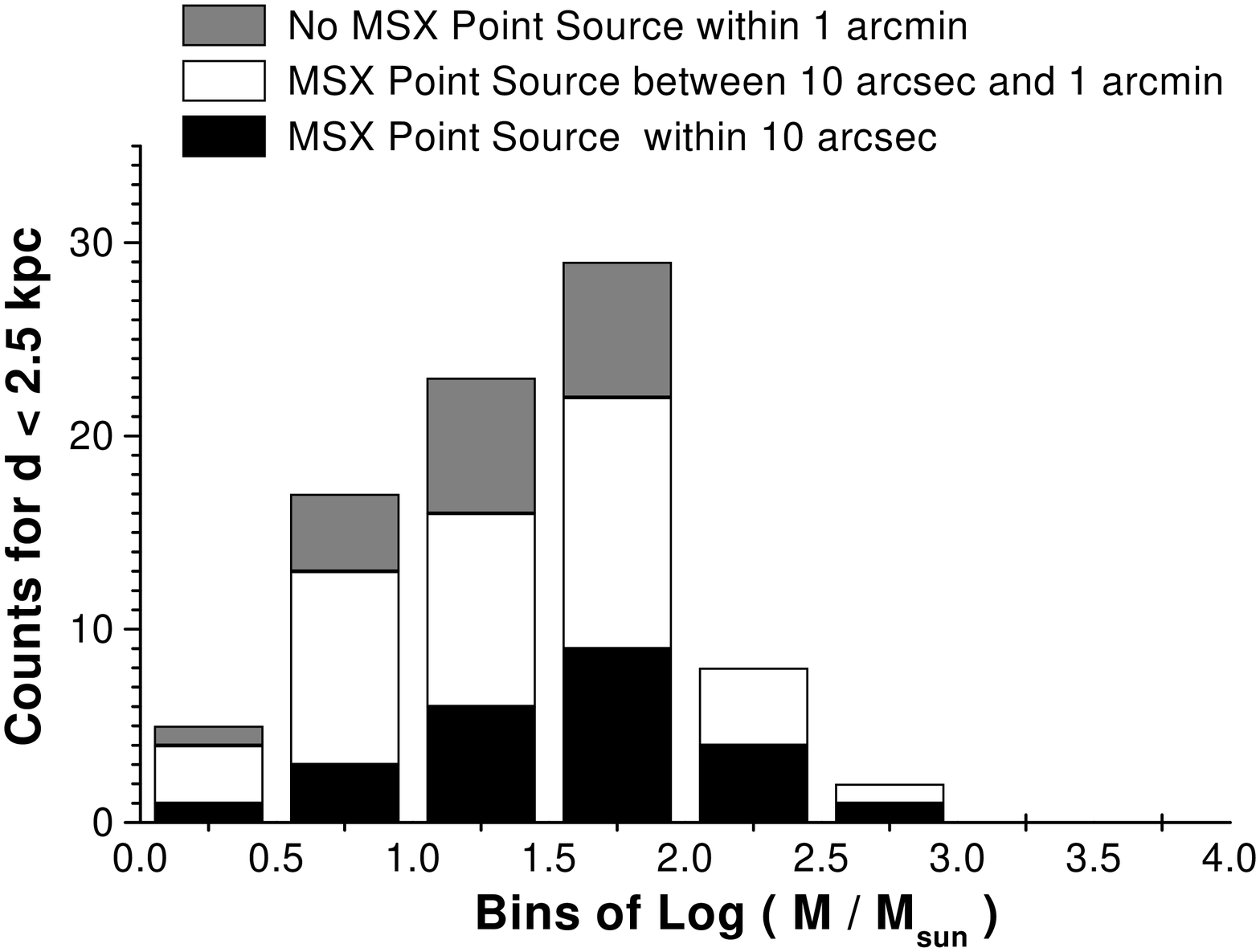}
\end{figure}

\begin{figure}[tbp]
\plotone{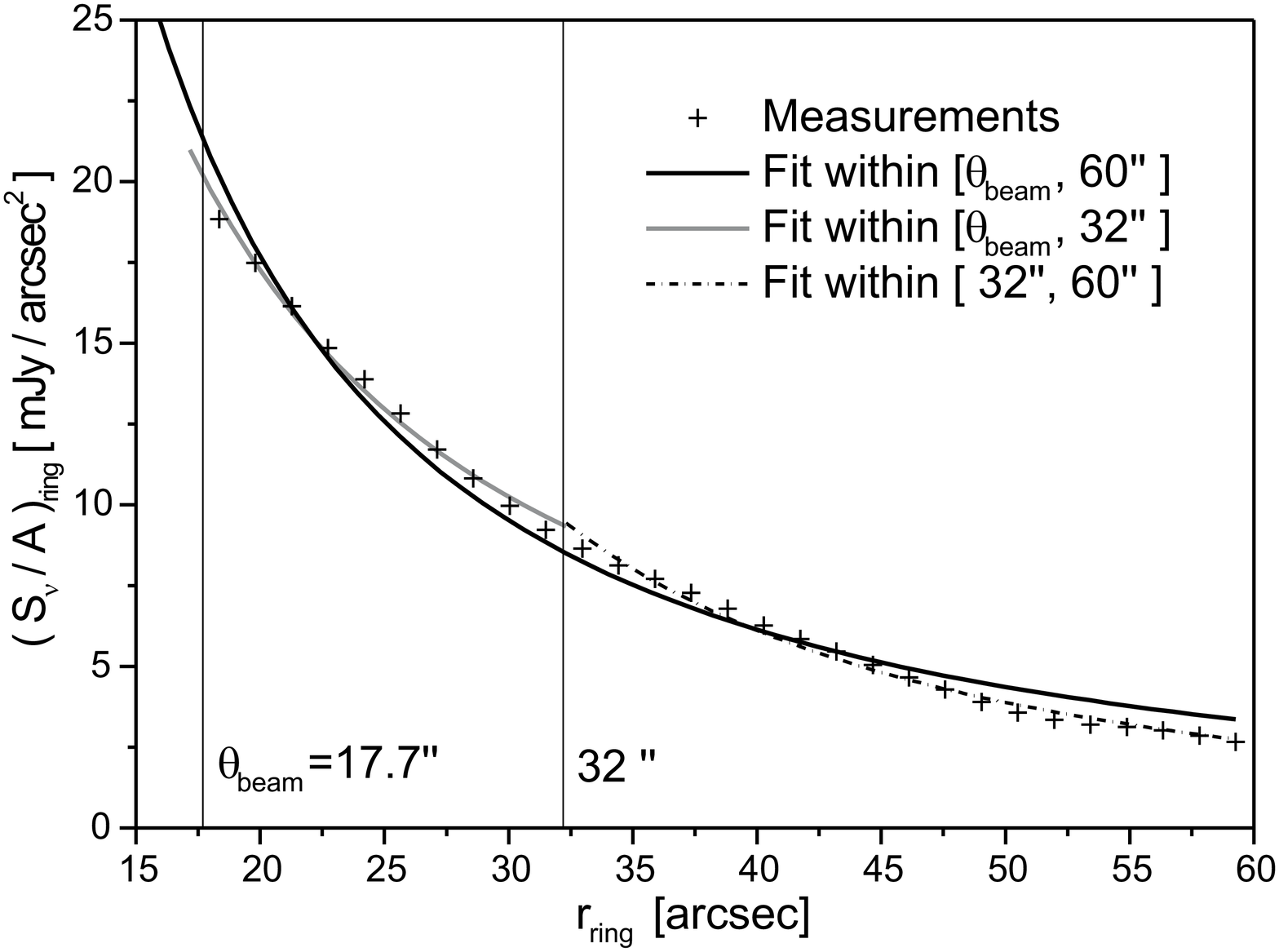}
  \caption{Description of the radial intensity profile by one or two potentials for IRAS~06058+2131 \protect\\
    The shown intensity distribution corresponds to the IRAM map of
    IRAS 06058+2131 \#1 (Figure~\ref{fig:maps}).  While the thick line
    represents the one power-law fit within the whole interval, the
    grey and dotted line correspond to the two power-law fit with a
    break at $32 \arcsec$. For the used method to get the radial
    density distribution see \S~\ref{sec:radial}, the derived
    exponents of the fits can be found in Table~\ref{alphaundp}.
 \label{fig:Iprofil} }
\end{figure}

\clearpage
\newcommand{\detectlable}{Detectable at}
\newlength{\detectwidth}
\settowidth{\detectwidth}{\detectlable}
\newlength{\irasname}
\settowidth{\irasname}{IRAS\,03064+5638\,\#1a}

\begin{deluxetable}{lp{43mm}p{\detectwidth}p{\irasname}}
\tablewidth{\textwidth}
\tablecaption{Tentiative stages of massive star formation\label{tab:stages}}
\tablehead{Stage&Morphology&\detectlable&Example}
\startdata
0: {\PPC}&\raggedright 
massive cloud core without collapse& mm&IRAS\,06073+1249\,\#2  \\
\hline
1: early protocluster&\raggedright 
massive stars have begun to form deeply embedded in the cluster& 
mm&IRAS\,03211+5446\,\#1\\
\hline
2: protocluster&\raggedright 
the forming massive stars begin to clear a cavity, an {\hii} region
begins to evolve&\raggedright 
mm, FIR, radio&IRAS 05197+3355\,\#1\\
\hline
3: evolved protocluster&\raggedright
the cluster starts to emerge, but is still embedded&\raggedright 
mm, FIR, MIR, radio&IRAS\,04329+5047\,\#1\\
\hline
4: young cluster&\raggedright
the cluster has emerged from its parental cloud&\raggedright 
mm, FIR, MIR, NIR&IRAS\,03211+5446\,\#2\\
\hline
5: cluster&\raggedright
the cluster has dispersed its parental cloud&
MIR, NIR&IRAS\,05345+3157 IRAS\,05281+3412
\enddata
\tablecomments{The examples for stage 0 and stage 1 are of course
  pre-protocluster candidates, but the assignment to stage 0 and stage
  1 is arbitrary.}
\end{deluxetable}


%
\clearpage
\noindent

\begin{deluxetable}{lccccl}
\tabletypesize{\small}
\tablewidth{0pt}
\tablecaption{Source List\label{tab:srclst}}
\tablehead{%
IRAS name & D & ref. & L$_{IRAS}$ & $T_{\rm dust}$ & other designations\\
  & [kpc] &  & [L$_{\odot}$] & [K] &}
\startdata
\object[IRAS 01195+6136]{01195+6136} & 2.0 & (1) &$8.5\cdot10^3$ & 26.7&  S187 \\
\object[IRAS 02244+6117]{02244+6117} & 2.2 & (1) &$7.8\cdot10^3$ & 29.5&  AFGL 333, W 4   \\
\object[IRAS 02575+6017]{02575+6017} & 3.9 & (2) &$3.5\cdot10^4$ & 33.0&  AFGL 4029  \\
\object[IRAS 02593+6016]{02593+6016} & 4.0 & (2) &$4.6\cdot10^4$ & 29.7&  AFGL 416, S201, G138.5+1.6\\
\object[IRAS 03064+5638]{03064+5638} & 4.1 & (1) &$1.6\cdot10^4$ & 28.2&  AFGL 5090   \\
\object[IRAS 03211+5446]{03211+5446} & 3.1 & (1) &$1.3\cdot10^4$ & 30.8&  AFGL 5094  \\
\object[IRAS 03236+5836]{03236+5836} & 1.0 & (3) &$2.7\cdot10^3$ & 36.1&   AFGL 490  \\
\object[IRAS 03595+5110]{03595+5110} & 3.3 & (1) &$1.3\cdot10^4$ & 30.8&   AFGL 5111, NGC 1491, S206\\
\object[IRAS 04073+5102]{04073+5102} & 8.2 & (4) &$1.7\cdot10^5$ & 8.4 &  AFGL 550, S209 \\
\object[IRAS 04269+3510]{04269+3510} & 0.8 & (1) &$6.8\cdot10^3$ & 31.5&  AFGL 585, G164.4-9.0, LkH$\alpha$\,101  \\
\object[IRAS 04324+5106]{04324+5106} & 6.0 & (1) &$5.9\cdot10^4$ & 37.5&  AFGL 5124   \\
\object[IRAS 04329+5047]{04329+5047} & 6.0 & (1) &$3.1\cdot10^4$ & 29.0&  AFGL 5125, S211   \\
\object[IRAS 05100+3723]{05100+3723} & 2.6 & (1) &$7.6\cdot10^3$ & 32.3&  LBN 784, AFGL 5137, S228   \\
\object[IRAS 05197+3355]{05197+3355} & 1.8 & (2) &$2.5\cdot10^3$ & 28.8&  S230   \\
\object[IRAS 05281+3412]{05281+3412} & 1.8 & (1) &$5.9\cdot10^3$ & 34.9&  AFGL 5144, G173.9+0.3, NGC 1931, S237\\
\object[IRAS 05327-0457]{05327$-$0457} & 0.45 & (5) &$1.6\cdot10^4$ & 28.7&   S279  \\
\object[IRAS 05341-0530]{05341$-$0530} & 1.8 & (1) &$8.3\cdot10^2$ & 22.6   &  \\
\object[IRAS 05345+3157]{05345+3157} & 1.8 & (1) &$2.8\cdot10^3$ & 31.2&   AFGL 5157, NGC 1985 \\
\object[IRAS 05355+3039]{05355+3039} & 1.8 & (1) &$2.4\cdot10^3$ & 29.2&   AFGL 5158 \\
\object[IRAS 05375+3540]{05375+3540} & 1.8 & (1) &$1.3\cdot10^4$ & 38.0&   S235B \\
\object[IRAS 05377+3548]{05377+3548} & 1.8 & (1) &$6.3\cdot10^3$ & 27.7&  S235, G173.71+2.70\\
\object[IRAS 05480+2544]{05480+2544} & 2.1 & (1) &$3.5\cdot10^3$ & 27.4&   BFS 48\\
\object[IRAS 05480+2545]{05480+2545} & 2.1 & (1) &$4.2\cdot10^3$ & 31.0&   BFS 48\\
\object[IRAS 06006+3015]{06006+3015} & 4.7 & (1) &$1.3\cdot10^4$ & 26.4&   AFGL 5176, S241\\
\object[IRAS 06013+3030]{06013+3030} & 4.7 & (1) &$2.1\cdot10^4$ & 30.0&   CED 061 \\
\object[IRAS 06055+2039]{06055+2039} & 2.0 & (6) &$1.1\cdot10^4$ & 31.2&   S252A \\
\object[IRAS 06056+2131]{06056+2131} & 2.0 & (6) &$2.0\cdot10^4$ & 32.3&   AFGL 6366S \\
\object[IRAS 06058+2138]{06058+2138} & 2.0 & (6) &$1.0\cdot10^4$ & 30.7&   AFGL 5180, S247 \\
\object[IRAS 06061+2151]{06061+2151} & 2.0 & (6) &$9.6\cdot10^3$ & 34.3&   AFGL 5182, S247 \\
\object[IRAS 06063+2040]{06063+2040} & 2.0 & (6) &$1.2\cdot10^4$ & 30.6&   AFGL 5183, S252\,C \\
\object[IRAS 06068+2030]{06068+2030} & 2.0 & (6) &$7.5\cdot10^3$ & 32.0&   AFGL 5184, NGC 2175, S252\,E\\
\object[IRAS 06073+1249]{06073+1249} & 6.0 & (2) &$4.0\cdot10^3$ & 38.3&   AFGL 5185, S270 \\
\object[IRAS 06099+1800]{06099+1800} & 0.8 & (1) &$6.0\cdot10^3$ & 31.1&   AFGL 896, S255 \\
\object[IRAS 06105+1756]{06105+1756} & 2.5 & (1) &$7.0\cdot10^3$ & 32.1&   S258 \\
\object[IRAS 06114+1745]{06114+1745} & 2.5 & (1) &$8.6\cdot10^3$ & 34.0&   AFGL 5188 \\
\object[IRAS 06117+1350]{06117+1350} & 3.8 & (1) &$5.1\cdot10^4$ & 33.5&   AFGL 902, S269  \\
\object[IRAS 06155+2319]{06155+2319} & 1.6 & (1) &$3.7\cdot10^3$ & 29.5&  BFS 51  \\
\object[IRAS 06308+0402]{06308+0402} & 1.6 & (1) &$3.9\cdot10^3$ & 31.8&   RNO 73 \\
\object[IRAS 06319+0415]{06319+0415} & 1.6 & (1) &$8.9\cdot10^3$ & 36.9&   AFGL 961\\
\object[IRAS 06380+0949]{06380+0949} & 0.8 & (1) &$4.0\cdot10^2$ & 25.4&   AFGL 4519S \\
\object[IRAS 06384+0932]{06384+0932} & 0.76 \tablenotemark{p} & (1) &$2.3\cdot10^3$ & 30.9&  AFGL 989, NGC 2264--1, W 217 \\
\object[IRAS 06412-0105]{06412-0105} & 1.1 & (1) &$2.0\cdot10^3$ & 39.6&   \\
\object[IRAS 06567-0355]{06567-0355} & 2.6 & (1) &$1.8\cdot10^4$ & 38.9&  NS 14, BFS 57 \\
\object[IRAS 06581-0846]{06581-0846} & 4.0 & (1) &$6.2\cdot10^3$ & 18.1&  BFS 64 \\
\object[IRAS 06581-0848]{06581-0848} & 4.0 & (1) &$8.9\cdot10^3$ & 23.4&  BFS 64 \\
\object[IRAS 07029-1215]{07029-1215} & 1.0 & (1) &$1.7\cdot10^3$ & 29.1&  AFGL 5222, S297, vdB 94 \\
\object[IRAS 07299-1651]{07299-1651} & 1.4 & (1) &$6.2\cdot10^3$ & 36.8&  AFGL 5234, S302, DG 121

\enddata
\tablenotetext{p}{photometric distance}
\tablerefs{%
  (1) \citet{Henning92b}, (2) \citet{Bronf96}, (3)
  \citet{KatharinaDOK}, (4) \citet{Mueller2002} (5)
  \citet{Szymczak2000}, (6) \citet{Valdettaro2001}}
\tablecomments{\ Dust temperature from \citet{Schreyer96}, luminosities are estimated according to
  \citet{Schreyer96}.}
\end{deluxetable}


\begin{deluxetable}{lccccc|l|lll}
\rotate 
\tablewidth{0pt} \tablecaption{Source Associations --- The
  association of the 47 sources in the sample with molecular line
  emission, outflow detection, and maser emission (last three columns)
  according to \citet{Schreyer96} (NH$_3$ (1,1),NH$_3$ (2,2),
  HCO$^{+}$ ($J=1\rightarrow 0$)), \citet{Pirogov99} (HCN $J=1\rightarrow 0$), \citet{Bronf96,KatharinaDOK} (CS
  ($J=2\rightarrow 1$)) and references given in the table. ++ --
  detection; N -- no detection; \ldots -- no information available
    \label{tab:Literatur}}
\tablehead{%
\textsc{IRAS Name} & \textsc{NH$_3$(1,1)} & \textsc{NH$_3$(2,2)} & \textsc{HCO$^{+}$} & \textsc{HCN} & \textsc{CS} & Outflows & \textsc{H$_2$O} & \textsc{OH} & \textsc{CH$_3$OH}  \\
 & [++/N] & [++/N] & [++/N] & [++/N] & [++/N] &  [++/N] &  [++/N] &  [++/N] &  [++/N]}
\startdata
01195+6136   & N & N & N & N & \nodata & ++  (7) & N (1) & \nodata & N  (5) \\
02244+6117   & ++ & N & ++ & ++ & \nodata & ++  (7) & N  (1) & \nodata & N  (5) \\
02575+6017   & ++ & ++ & ++ & ++ \tablenotemark{S} & ++  (2) & ++  (7,8,9) & ++  (1,11) & N  (11,13) & N  (5) \\
02593+6016   & ++ & N & ++ & ++ \tablenotemark{S} & ++  (2) & N  (8) & ++  (1,11) & ++  (11,10) & N  (5) \\
03064+5638   & N & N & N & ++ \tablenotemark{S} & \nodata & N  (8) & N  (1) & N  (11) & N  (5) \\
03211+5446   & N & N & N & N & \nodata & N  (8) & ++  (11) & N  (11) & N  (5) \\
03236+5836   & ++ & N & ++ & ++ \tablenotemark{S} & ++  (3) & ++  (7) & ++  (1,11) & ++  (11) & N  (5) \\
03595+5110   & N & N & N & N & \nodata & \nodata & N  (1) & \nodata & N  (5) \\
04073+5102   & N & N & N & N & \nodata & N  (9)& N  (1) & \nodata & \nodata \\
04269+3510   & N & N & ++ & ++ & \nodata & ++  (7) & N  (1) & N  (13) & N  (5) \\
04324+5106   & ++ & N & ++ & ++ \tablenotemark{S} & \nodata & N  (8) & ++  (11) & N  (11) & N  (5) \\
04329+5047   & N & N & N & ++ \tablenotemark{S} & \nodata & N  (9)& N  (1) & \nodata & N  (5) \\
05100+3723   & N & N & ++ & ++ & ++  (2) & N  (8) & N  (1) & \nodata & N  (5) \\
05197+3355   & N & N & N & ++ & \nodata & N  (9)& N  (1) & \nodata & N  (5) \\
05281+3412   & N & N & N & N & N  (2) & N  (8) & N  (1) & \nodata & N  (5) \\
05327-0457   & ++ & N & ++ & \nodata & \nodata & \nodata & ++  (1,6) & N  (13) & N  (5) \\
05341-0530   & N & N & ++ & \nodata & \nodata & ++  (7) & ++  (1) & \nodata & N  (5) \\
05345+3157   & N & N & ++ & ++ & \nodata & ++  (7,8,9) & ++  (1,6,11) & \nodata & ++  (11) \\
05355+3039   & N & N & ++ & ++ & \nodata & N  (8) & N  (1) & \nodata & N  (5) \\
05375+3540   & ++ & ++ & ++ & ++ \tablenotemark{S} & ++  (2) & ++  (7) & ++  (1,6,11,12) & N  (11,13) & N  (5) \\
05377+3548   & ++ & ++ & ++ & ++ \tablenotemark{S} & ++  (4) & ++  (7) & ++  (1) & \nodata & N  (5) \\
05480+2544   & ++ & N & ++ & ++ \tablenotemark{S} & \nodata & N  (9)& N  (1) & \nodata & N  (5) \\
05480+2545   & ++ & ++ & ++ & ++ \tablenotemark{S} & ++  (2) & N  (9)& N  (1) & \nodata & N  (5) \\
06006+3015   & ++ & ++ & ++ & ++ \tablenotemark{S} & ++  (4) & N  (9)& ++  (1) & N  (11) & N  (5) \\
06013+3030   & N & N & N & N & \nodata & N  (9)& N  (1) & \nodata & N  (5) \\
06055+2039   & ++ & ++ & ++ & ++ \tablenotemark{S} & ++  (2) & N  (9)& ++  (1,6,12) & ++  (10) & ++  (5) \\
06056+2131   & ++ & ++ & ++ & ++ \tablenotemark{S} & ++  (2) & ++  (7,8,9) & ++  (1,11,10) & N  (10) & ++  (5) \\
06058+2138   & ++ & ++ & ++ & ++ \tablenotemark{S} & ++  (2) & ++  (7,8,9) & ++  (1,6,11,12,10) & N  (10) & ++  (5) \\
06061+2151   & ++ & ++ & ++ & ++ \tablenotemark{S} & ++  (2) & N  (8) & ++  (1,6,11) & N  (10) & ++  (5) \\
06063+2040   & N & N & ++ & ++ \tablenotemark{S} & ++  (2) & N  (8) & N  (1) & N  (13) & N  (5) \\
06068+2030   & N & N & ++ & N & \nodata & N  (8) & N  (1) & \nodata & N  (5) \\
06073+1249   & N & N & N & ++ \tablenotemark{S} & ++  (2) & N  (8) & N  (1) & \nodata & N  (5) \\
06099+1800   & ++ & ++ & ++ & ++ \tablenotemark{S} & \nodata & ++  (7) & ++  (1,6,11,12) & ++  (10) & ++  (15) \\
06105+1756   & N & N & ++ & ++ & ++  (2) & ++  (7) & N  (1) & \nodata & N  (5) \\
06114+1745   & ++ & N & ++ & ++ \tablenotemark{S} & ++  (2) & N  (8) & N  (1) & \nodata & N  (11,5) \\
06117+1350   & N & N & ++ & ++ \tablenotemark{S} & ++  (2) & N  (9)& ++  (1,6,11,12,13) & ++  (10) & ++  (5) \\
06155+2319   & N & N & ++ & ++ \tablenotemark{S} & ++  (2) & N  (9)& N  (1) & \nodata & N  (5) \\
06308+0402   & N & N & ++ & ++ \tablenotemark{S} & \nodata & ++  (7,9) & ++  (1) & \nodata & N  (5) \\
06319+0415   & ++ & ++ & ++ & ++ \tablenotemark{S} & \nodata & ++  (7) & N  (1) & \nodata & N  (5) \\
06380+0949   & N & N & N & N & \nodata & N  (9)& N  (1) & \nodata & N  (5) \\
06384+0932   & ++ & ++ & ++ & ++ \tablenotemark{S} & ++  (3) & ++  (7) & ++  (1,11) & ++  (11) & N  (5) \\
06412-0105   & N & N & \nodata & \nodata & N  (2) & \nodata & N  (1) & \nodata & N  (5) \\
06567-0355   & ++ & ++ & \nodata & \nodata & ++  (2) & ++  (7) & N  (1) & \nodata & N  (5) \\
06581-0848   & N & N & \nodata & \nodata & \nodata & \nodata & N  (1) & \nodata & N  (5) \\
06581-0846   & N & N & \nodata & \nodata & \nodata & \nodata & N  (1) & \nodata & N  (5) \\
07029-1215   & N & N & \nodata & \nodata & \nodata & \nodata & N  (1) & \nodata & N  (5) \\
07299-1651   & N & N & \nodata & \nodata & ++  (2) & \nodata & ++  (1) & N  (13) & ++  (5)
\enddata
\tablenotetext{S}{\ strong sources ($\rm > 3\, K\,km\,s^{-1}$,\citealt{Yun99})}
\tablerefs{%
(1) \citet{Henning92b}, (2) \citet{Bronf96}, (3) \citet{KatharinaDOK}
(4) \citet{Mueller2002} (5) \citet{Szymczak2000}, (6)
\citet{Valdettaro2001}, (7) \citet{Wu1996,Wu96}, (8) \citet{Snell88},
(9) \citet{Snell90}, (10) \citet{Chan1996}, (11) \citet{Wout93}, (12)
\citet{Palagi93}, (13) \citet{Baudry1997}}
\end{deluxetable}


\begin{deluxetable}{cllrrrrr}
\tablewidth{0pt}
\tablecaption{The Observing Runs\label{tab:beam}}
\tablehead{%
\#&telescope&instrument&\colhead{$\nu$}&\colhead{$\Delta\nu$}&\colhead{date}&\colhead{FWHM
}&\colhead{rms}\\
 & & &  $[{\rm GHz}]$ & $[{\rm GHz}]$ & & [{\arcsec}] & [mJy/beam]}
\startdata
1 & IRAM & MAMBO 37ch-bol & 235 & 80 & December 1998 & 14.1 & 3.3\\
2 & IRAM & MAMBO 37ch-bol & 235 & 80 & April 1999 & 17.7 & 20.0\\ 
3 & JCMT & SCUBA 37ch-bol & 350 & 30 & Service 1999 & 14.8 & 47.0\\
4 & JCMT & SCUBA 37ch-bol & 350 & 30 & October 1999 & 16.0 & 23.7\\
5 & JCMT & SCUBA 37ch-bol & 350 & 30 & September 2000 & 15.7 & 17.3\\
6 & JCMT & SCUBA 37ch-bol & 350 & 30 & Service 2000 & 15.2 & 22.0\\
7 & SMT  & 19ch MPIfR bol & 345 & 50 & March 2001 & 26.0 & 430.3
\enddata
\tablecomments{Wavelenghts and Bandwidths in \um:\\
235\,GHz, 80\,GHz: 1270\um, 430\um\\
345\,GHz, 30\,GHz: 870\um, 70\um\\
350\,GHz, 50\,GHz: 850\um, 125\um}
\end{deluxetable}


\clearpage

\begin{deluxetable}{lccccccc}
\tabletypesize{\small}
\tablewidth{420.0pt}

\tablecaption{For each source component the following quantities are given: the mean integrated flux density $F_{\nu }^{int}$ with its systematic and statistic error  $\vartriangle F_{\nu }^{int}$; the mean integration area $A^{int}$, resulting from three polygons; the peak flux density $F_{\nu }^{Peak}$; the $\sigma$ level of the map; and the component extension $\theta_{Source}$ derived from $A^{int}$ \label{tab:Messergeb} }

\tablehead{
\colhead{IRAS Name}      & \colhead{Component} &
\colhead{ $F_{\nu }^{int}$}   & \colhead{$\vartriangle F_{\nu }^{int}$}  & \colhead{$A^{int}$} &
\colhead{$F_{\nu }^{Peak}$}  & \colhead{$\sigma$} & \colhead{$\theta_{Source}$} \\
\colhead{ }      & \colhead{ } &
\colhead{[Jy]}   & \colhead{[Jy]}  &  \colhead{[$\arcsec$]$^2$}  & \colhead{$\left[ \frac{\rm Jy}{\theta_{beam}} \right]$} &
\colhead{$\left[ \frac{\rm mJy}{\theta_{beam}} \right]$}  & \colhead{[pc]}  }

\startdata
01195+6136 \tablenotemark{( 5 )} & 1   & 0.18 & 0.26 & 3050 & 0.09 & 13 & 0.58  \\
02244+6117 \tablenotemark{( 2 )} & 1 a & 0.56 & 0.22 & 1320 & 0.19 & 18 & 0.39  \\
02244+6117 \tablenotemark{( 2 )} & 1 b & 0.45 & 0.19 & 1310 & 0.15 & 18 & 0.39  \\
02244+6117 \tablenotemark{( 2 )} & 1 c & 0.33 & 0.15 & 800 & 0.15 & 18 & 0.28  \\
02244+6117 \tablenotemark{( 2 )} & 1 d & 0.17 & 0.12 & 670 & 0.15 & 18 & 0.25  \\
02244+6117 \tablenotemark{( 2 )} & 2 a & 0.29 & 0.15 & 680 & 0.19 & 18 & 0.25  \\
02244+6117 \tablenotemark{( 2 )} & 2 b & 0.08 & 0.09 & 190 & 0.17 & 18 & 0.19 \tablenotemark{B} \\
02244+6117 \tablenotemark{( 2 )} & 2 c & 0.13 & 0.11 & 390 & 0.15 & 18 & 0.14  \\
02244+6117 \tablenotemark{( 2 )} & 3   & 0.15 & 0.14 & 670 & 0.14 & 18 & 0.25  \\
02575+6017 \tablenotemark{( 1 )} & 1 a & 0.71 & 0.19 & 1480 & 0.18 & 3 & 0.78  \\
02575+6017 \tablenotemark{( 1 )} & 1 b & 0.06 & 0.04 & 480 & 0.04 & 3 & 0.38  \\
02575+6017 \tablenotemark{( 1 )} & 2   & 0.04 & 0.03 & 790 & 0.02 & 3 & 0.54  \\
02593+6016 \tablenotemark{( 1 )} & 1   & 0.50 & 0.09 & 2170 & 0.11 & 3 & 0.98 \\
02593+6016 \tablenotemark{( 1 )} & 2   & 0.05 & 0.03 & 640 & 0.04 & 3 & 0.48 \\
02593+6016 \tablenotemark{( 1 )} & 3   & 0.14 & 0.06 & 1640 & 0.03 & 3 & 0.84 \\
03064+5638  \tablenotemark{( 7 )} & 1 a&10.82 & 7.28 & 5300 & 3.12 & 594 & 1.55 \\
03064+5638 \tablenotemark{( 7 )} & 1 b & 7.63 & 3.89 & 2280 & 3.31 & 594 & 0.94 \\
03211+5446 \tablenotemark{( 7 )} & 1   &13.46 & 4.65 & 7780 & 2.48 & 371 & 1.44 \\
03211+5446 \tablenotemark{( 7 )} & 2   &11.39 & 5.38 & 8240 & 2.55 & 371 & 1.49 \\
03236+5836 \tablenotemark{( 3 )} & 1 a & 6.68 & 0.74 & 1220 & 2.21 & 35 & 0.18 \\
03236+5836 \tablenotemark{( 3 )} & 1 b & 1.22 & 0.50 & 870 & 0.43 & 35 & 0.14 \\
03236+5836 \tablenotemark{( 3 )} & 2 a & 2.13 & 0.79 & 1610 & 0.58 & 35 & 0.21 \\
03236+5836 \tablenotemark{( 3 )} & 2 b & 2.06 & 0.70 & 1840 & 0.31 & 35 & 0.22 \\
03595+5110 \tablenotemark{( 7 )} & 1 a & 1.12 & 0.87 & 790 & 1.43 & 239 & 0.29 \\
03595+5110 \tablenotemark{( 7 )} & 1 b & 1.06 & 0.74 & 600 & 1.43 & 239 & 0.15 \\
04073+5102 \tablenotemark{( 7 )} & 1   & 7.73 & 3.29 & 2700 & 3.60 & 448 & 2.09 \\
04073+5102 \tablenotemark{( 7 )} & 2   & 5.46 & 3.37 & 3210 & 2.40 & 448 & 2.32 \\
04073+5102 \tablenotemark{( 7 )} & 3   & 9.66 & 4.15 & 4040 & 2.79 & 448 & 2.66  \\
04073+5102 \tablenotemark{( 7 )} & 4   & 4.35 & 2.72 & 1850 & 2.69 & 448 & 1.63  \\
04073+5102 \tablenotemark{( 7 )} & 5   & 2.32 & 2.84 & 2130 & 2.40 & 448 & 1.80  \\
04073+5102 \tablenotemark{( 7 )} & 6   & 2.72 & 2.93 & 2000 & 2.05 & 448 & 1.72  \\
04073+5102 \tablenotemark{( 7 )} & 7   & 3.58 & 4.11 & 3400 & 2.01 & 448 & 2.40  \\
04269+3510 \tablenotemark{( 7 )} & 1   & 1.89 & 1.15 & 570 & 2.81 & 330 & 0.03  \\
04269+3510 \tablenotemark{( 7 )} & 2   & 2.25 & 1.05 & 650 & 2.86 & 330 & 0.05  \\
04324+5106 \tablenotemark{( 2 )} & 1   & 0.23 & 0.14 & 370 & 0.28 & 22 & 0.37  \\
04324+5106 \tablenotemark{( 2 )} & 2   & 0.04 & 0.09 & 140 & 0.19 & 22 & 0.51 \tablenotemark{B} \\
04324+5106 \tablenotemark{( 2 )} & 3   & 0.01 & 0.05 & 80 & 0.10 & 22 & 0.51 \tablenotemark{B} \\
04324+5106 \tablenotemark{( 2 )} & 4   & 0.01 & 0.07 & 120 & 0.28 & 22 & 0.51 \tablenotemark{B} \\
04329+5047 \tablenotemark{( 7 )} & 1   &10.39 & 4.03 & 7050 & 1.86 & 354 & 2.65\\
05100+3723 \tablenotemark{( 3 )} & 1 a & 3.38 & 0.86 & 3850 & 0.22 & 22 & 0.86 \\
05100+3723 \tablenotemark{( 3 )} & 1 b & 1.75 & 0.74 & 2660 & 0.19 & 22 & 0.71 \\
05197+3355 \tablenotemark{( 7 )} & 1   &11.42 & 3.73 & 4280 & 3.01 & 421 & 0.60 \\
05281+3412 \tablenotemark{( 5 )} & $\cdots$ & $\cdots$ & $\cdots$ & $\cdots$ & $\cdots$ & 15 & $\cdots$ \\
05327$-$0457 \tablenotemark{( 2 )} & 1   & 0.09 & 0.12 & 740 & 0.08 & 13 & 0.22 \\
05341$-$0530 \tablenotemark{( 7 )} & $\cdots$ & $\cdots$ & $\cdots$ & $\cdots$ & $\cdots$ & 258 & $\cdots$ \\
05345+3157 \tablenotemark{( 5 )} & 1   & 1.99 & 0.44 & 5730 & 0.31 & 15 & 0.73 \\
05345+3157 \tablenotemark{( 5 )} & 2   & 1.59 & 0.50 & 7810 & 0.21 & 15 & 0.86 \\
05345+3157 \tablenotemark{( 5 )} & 3   & 1.16 & 0.62 & 8480 & 0.11 & 15 & 0.90 \\
05355+3039 \tablenotemark{( 5 )} & 1   & 0.53 & 0.34 & 2720 & 0.12 & 18 & 0.49 \\
05375+3540 \tablenotemark{( 6 )} & 1 a & 9.72 & 0.73 & 4560 & 1.57 & 22 & 0.65 \\
05375+3540 \tablenotemark{( 6 )} & 1 b & 1.43 & 0.40 & 2050 & 0.28 & 22 & 0.43 \\
05375+3540 \tablenotemark{( 6 )} & 1 c & 0.20 & 0.24 & 560 & 0.16 & 22 & 0.19 \\
05375+3540 \tablenotemark{( 6 )} & 2   & 0.97 & 0.39 & 1360 & 0.34 & 22 & 0.34 \\
05377+3548 \tablenotemark{( 6 )} & 1   & 2.40 & 0.68 & 3280 & 0.40 & 25 & 0.55 \\
05377+3548 \tablenotemark{( 6 )} & 2 a & 1.97 & 0.72 & 2480 & 0.43 & 25 & 0.47 \\
05377+3548 \tablenotemark{( 6 )} & 2 b & 2.13 & 0.71 & 3480 & 0.25 & 25 & 0.57 \\
05377+3548 \tablenotemark{( 6 )} & 2 c & 2.06 & 0.97 & 4020 & 0.21 & 25 & 0.61 \\
05377+3548 \tablenotemark{( 6 )} & 3   & 2.47 & 0.88 & 5070 & 0.32 & 25 & 0.69 \\
05377+3548 \tablenotemark{( 6 )} & 4   & 1.51 & 0.65 & 3130 & 0.30 & 25 & 0.53 \\
05377+3548 \tablenotemark{( 6 )} & 5   & 1.29 & 0.65 & 3610 & 0.17 & 25 & 0.58 \\
05377+3548 \tablenotemark{( 6 )} & 6   & 1.15 & 0.51 & 2520 & 0.19 & 25 & 0.48 \\
05377+3548 \tablenotemark{( 6 )} & 7   & 1.38 & 0.52 & 3060 & 0.37 & 25 & 0.53 \\
05377+3548 \tablenotemark{( 6 )} & 8   & 3.59 & 0.79 & 6810 & 0.17 & 25 & 0.80 \\
05377+3548 \tablenotemark{( 6 )} & 9   & 4.91 & 1.41 & 9250 & 0.24 & 25 & 0.94 \\
05377+3548 \tablenotemark{( 6 )} & 10  & 1.85 & 0.91 & 6040 & 0.12 & 25 & 0.75 \\
05377+3548 \tablenotemark{( 6 )} & 11  & 0.43 & 0.60 & 2850 & 0.13 & 25 & 0.51 \\
05480+2545/4 \tablenotemark{( 2 )} & 1 & 0.32 & 0.20 & 460 & 0.35 & 26 & 0.17 \\
06006+3015 \tablenotemark{( 2 )} & 1   & 0.40 & 0.15 & 1020 & 0.25 & 15 & 0.71 \\
06013+3030 \tablenotemark{( 7 )} & $\cdots$ & $\cdots$ & $\cdots$ & $\cdots$ & $\cdots$ & 437 &\\
06055+2039 \tablenotemark{( 2 )} & 1   & 0.74 & 0.22 & 1160 & 0.41 & 15 & 0.33 \\
06055+2039 \tablenotemark{( 5 )} & 1 a & 2.86 & 0.34 & 2260 & 0.86 & 19 & 0.50 \\
06055+2039 \tablenotemark{( 5 )} & 1 b & 0.31 & 0.17 & 620 & 0.18 & 19 & 0.23 \\
06055+2039 \tablenotemark{( 5 )} & 2 a & 1.08 & 0.50 & 4000 & 0.16 & 19 & 0.67 \\
06055+2039 \tablenotemark{( 5 )} & 2 b & 0.58 & 0.43 & 2550 & 0.12 & 19 & 0.53 \\
06055+2039 \tablenotemark{( 5 )} & 2 c & 0.30 & 0.41 & 2450 & 0.15 & 19 & 0.52 \\
06056+2131 \tablenotemark{( 1 )} & 1 a & 1.00 & 0.11 & 1360 & 0.33 & 4 & 0.38 \\
06056+2131 \tablenotemark{( 1 )} & 1 b & 0.21 & 0.07 & 910 & 0.08 & 4 & 0.30 \\
06056+2131 \tablenotemark{( 1 )} & 2 a & 0.22 & 0.08 & 750 & 0.11 & 4 & 0.27 \\
06056+2131 \tablenotemark{( 1 )} & 2 b & 0.10 & 0.07 & 450 & 0.06 & 4 & 0.19 \\
06056+2131 \tablenotemark{( 1 )} & 2 c & 0.05 & 0.04 & 500 & 0.05 & 4 & 0.20 \\
06056+2131 \tablenotemark{( 3 )} & 1   &13.72 & 1.70 & 3000 & 2.19 & 37 & 0.58 \\
06056+2131 \tablenotemark{( 3 )} & 2   & 5.28 & 1.00 & 2040 & 0.76 & 37 & 0.47 \\
06058+2138 \tablenotemark{( 2 )} & 1   & 5.26 & 0.91 & 12360 & 0.69 & 21 & 1.20 \\
06058+2138 \tablenotemark{( 3 )} & 1   &17.28 & 2.17 & 4220 & 2.23 & 73 & 0.70 \\
06061+2151 \tablenotemark{( 2 )} & 1   & 0.61 & 0.56 & 2260 & 0.31 & 19 & 0.49 \\
06063+2040 \tablenotemark{( 6 )} & 1   & 1.43 & 0.32 & 1170 & 0.49 & 24 & 0.34 \\
06063+2040 \tablenotemark{( 6 )} & 2   & 0.22 & 0.24 & 700 & 0.12 & 24 & 0.25 \\
06068+2030 \tablenotemark{( 6 )} & 1   & 2.44 & 1.17 & 10540 & 0.16 & 17 & 1.11 \\
06068+2030 \tablenotemark{( 6 )} & 2   & 1.42 & 0.44 & 4360 & 0.16 & 17 & 0.71 \\
06073+1249 \tablenotemark{( 7 )} & 1   & 9.52 & 1.74 & 2780 & 3.68 & 240 & 1.25 \\
06073+1249 \tablenotemark{( 7 )} & 2   & 1.10 & 0.91 & 810 & 1.40 & 240 & 0.44 \\
06099+1800 \tablenotemark{( 7 )} & 1   &29.51 & 7.39 & 1650 & 15.89 & 501 & 0.15 \\
06099+1800 \tablenotemark{( 7 )} & 2   &29.77 & 9.69 & 1680 & 16.92 & 501 & 0.15 \\
06099+1800 \tablenotemark{( 7 )} & 3   & 1.77 & 1.81 & 820 & 2.11 & 501 & 0.07  \\
06099+1800 \tablenotemark{( 7 )} & 4 a &14.25 & 3.13 & 1920 & 6.02 & 501 & 0.16  \\
06099+1800 \tablenotemark{( 7 )} & 4 b & 1.22 & 1.38 & 280 & 4.17 & 501 & 0.10 \tablenotemark{B} \\
06099+1800 \tablenotemark{( 7 )} & 4 c & 3.54 & 2.60 & 1070 & 2.89 & 501 & 0.10  \\
06105+1756 \tablenotemark{( 7 )} & 1 a & 9.56 & 2.22 & 5800 & 2.06 & 230 & 0.99  \\
06105+1756 \tablenotemark{( 7 )} & 1 b & 7.51 & 2.87 & 7300 & 1.38 & 230 & 1.13  \\
06105+1756 \tablenotemark{( 7 )} & 2   &12.56 & 3.13 & 9630 & 1.39 & 230 & 1.30  \\
06105+1756 \tablenotemark{( 7 )} & 3   & 5.67 & 2.69 & 5940 & 1.56 & 230 & 1.01  \\
06114+1745 \tablenotemark{( 5 )} & 1 a & 0.37 & 0.21 & 850 & 0.13 & 15 & 0.35  \\
06114+1745 \tablenotemark{( 5 )} & 1 b & 0.16 & 0.14 & 490 & 0.11 & 15 & 0.24  \\
06114+1745 \tablenotemark{( 5 )} & 2   & 0.27 & 0.17 & 1000 & 0.12 & 15 & 0.39  \\
06114+1745 \tablenotemark{( 5 )} & 3   & 0.13 & 0.13 & 530 & 0.10 & 15 & 0.25  \\
06114+1745 \tablenotemark{( 5 )} & 4   & 0.33 & 0.23 & 1680 & 0.09 & 15 & 0.26  \\
06117+1350 \tablenotemark{( 2 )} & 1   & 0.45 & 0.28 & 640 & 0.43 & 35 & 0.41  \\
06155+2319 \tablenotemark{( 2 )} & 1 a & 0.29 & 0.11 & 480 & 0.28 & 15 & 0.13  \\
06155+2319 \tablenotemark{( 2 )} & 1 b & 0.07 & 0.06 & 200 & 0.16 & 15 & 0.14 \tablenotemark{B} \\
06308+0402 \tablenotemark{( 4 )} & 1 a & 0.62 & 0.26 & 800 & 0.30 & 21 & 0.21  \\
06308+0402 \tablenotemark{( 4 )} & 1 b & 0.36 & 0.17 & 530 & 0.24 & 21 & 0.16  \\
06308+0402 \tablenotemark{( 4 )} & 2 a & 0.45 & 0.21 & 590 & 0.29 & 21 & 0.17  \\
06308+0402 \tablenotemark{( 4 )} & 2 b & 0.55 & 0.23 & 670 & 0.29 & 21 & 0.19  \\
06308+0402 \tablenotemark{( 4 )} & 3 a & 0.15 & 0.10 & 160 & 0.21 & 21 & 0.12 \tablenotemark{B} \\
06308+0402 \tablenotemark{( 4 )} & 3 b & 0.12 & 0.12 & 270 & 0.20 & 21 & 0.07  \\
06308+0402 \tablenotemark{( 4 )} & 4   & 0.25 & 0.22 & 580 & 0.20 & 21 & 0.17  \\
06319+0415 \tablenotemark{( 7 )} & 1   &27.07 & 5.08 & 2060 & 13.06 & 667 & 0.34  \\
06319+0415 \tablenotemark{( 7 )} & 2   & 1.14 & 2.16 & 570 & 3.63 & 667 & 0.06  \\
06380+0949 \tablenotemark{( 7 )} & $\cdots$ & $\cdots$ & $\cdots$ & $\cdots$ & $\cdots$ & 724 & $\cdots$ \\
06384+0932 \tablenotemark{( 3 )} & 1 a &41.74 & 12.59 & 12700 & 2.06 & 68 & 0.47  \\
06384+0932 \tablenotemark{( 3 )} & 1 b &57.28 & 13.12 & 43690 & 2.05 & 68 & 0.87  \\
06384+0932 \tablenotemark{( 3 )} & 1 c & 7.34 & 1.85 & 1960 & 1.03 & 68 & 0.18  \\
06384+0932 \tablenotemark{( 3 )} & 2   &15.01 & 4.17 & 15600 & 0.70 & 68 & 0.52  \\
06384+0932 \tablenotemark{( 3 )} & 3   &17.75 & 4.40 & 17320 & 0.37 & 68 & 0.54  \\
06384+0932 \tablenotemark{( 3 )} & 4   &17.78 & 12.91 & 17320 & 0.36 & 68 & 0.54 \\
06412$-$0105 \tablenotemark{( 5 )} & 1   & 0.47 & 0.29 & 2820 & 0.11 & 14 & 0.31 \\
06567$-$0355 \tablenotemark{( 4 )} & 1   & 4.00 & 0.84 & 2460 & 0.74 & 32 & 0.67 \\
06581$-$0846/8 \tablenotemark{( 5 )} & 1 & 1.68 & 0.93 & 11210 & 0.11 & 18 & 2.31 \\
06581$-$0846/8 \tablenotemark{( 5 )} & 2 & 0.68 & 0.77 & 5060 & 0.11 & 18 & 1.54 \\
07029$-$1215 \tablenotemark{( 4 )} & 1 a & 1.99 & 0.50 & 1660 & 0.58 & 18 & 0.21 \\
07029$-$1215 \tablenotemark{( 4 )} & 1 b & 0.36 & 0.16 & 640 & 0.15 & 18 & 0.12 \\
07029$-$1215 \tablenotemark{( 4 )} & 2 a & 0.51 & 0.20 & 930 & 0.21 & 18 & 0.15 \\
07029$-$1215 \tablenotemark{( 4 )} & 2 b & 0.16 & 0.12 & 270 & 0.16 & 18 & 0.05 \\
07029$-$1215 \tablenotemark{( 4 )} & 3   & 0.46 & 0.18 & 880 & 0.20 & 18 & 0.15 \\
07029$-$1215 \tablenotemark{( 4 )} & 4   & 0.18 & 0.13 & 490 & 0.13 & 18 & 0.09 \\
07299$-$1651 \tablenotemark{( 2 )} & 1   & 0.45 & 0.15 & 720 & 0.35 & 16 & 0.17 \\
07299$-$1651 \tablenotemark{( 5 )} & $\cdots$ & $\cdots$ & $\cdots$ & $\cdots$ & $\cdots$ & 14 & $\cdots$ \\
\enddata

\tablenotetext{( 1 ) -- ( 7 ) }{1. -  7. Run}
\tablenotetext{B}{\ indicates that beam area is larger than the source extent, and it is used for further calculations (see text)}
 
\end{deluxetable}


\begin{deluxetable}{lcccccr} \tabletypesize{\small} \tablewidth{0pt}
\tablecaption{Gas masses $M^{gas}$, column densities $N(H)$ and number
  densities $n(H)$ calculated for each source component at $T_{dust}
  =20$\,K \label{tab:massenetc} } \tablehead{
  \colhead{IRAS name} & \colhead{Component} & \colhead{$M^{gas}$}  & \colhead{$N(H)$}  & \colhead{$n(H)$}&\multicolumn{2}{c}{Peak position (B1950)}\\
  \colhead{ }& \colhead{ } & \colhead{[M$_{\odot}$]} & \colhead{
    $10^{22}$ [1/cm$^2$ ]} & \colhead{$10^{4}$
    [1/cm$^3$]}&\colhead{RA}&\colhead{DEC}} \startdata
01195+6136 \tablenotemark{( 5 )} & 1    &    5 &  0.9 &  0.5 &01:19:56.3&$+61$:34:23\\
02244+6117 \tablenotemark{( 2 )} & 1 a  &   89 &  6.7 &  5.5 &02:24:20.8&$+61$:16:27\\
02244+6117 \tablenotemark{( 2 )}  & 1 b &   71 &  5.3 &  4.4 &02:24:24.3&$+61$:16:11\\
02244+6117 \tablenotemark{( 2 )}  & 1 c &   53 &  5.5 &  6.3 &02:24:18.3&$+61$:15:54\\
02244+6117 \tablenotemark{( 2 )}  & 1 d &   27 &  5.3 &  6.9 &02:24:17.8&$+61$:15:42\\
02244+6117 \tablenotemark{( 2 )}  & 2 a &   45 &  6.8 &  8.8 &02:24:17.9&$+61$:14:29\\
02244+6117 \tablenotemark{( 2 )}  & 2 b &   12 &  6.1 & 10.4 \tablenotemark{B}  &02:24:17.2&$+61$:13:43\\
02244+6117 \tablenotemark{( 2 )}  & 2 c &   21 &  5.4 & 12.1 &02:24:16.1&$+61$:13:00\\
02244+6117 \tablenotemark{( 2 )}  & 3   &   23 &  5.1 &  6.8 &02:24:22.1&$+61$:13:46\\
02575+6017 \tablenotemark{( 1 )}  & 1 a &  350 & 10.1 &  4.2 &02:57:34.7&$+60$:17:30\\
02575+6017 \tablenotemark{( 1 )}  & 1 b &   30 &  2.1 &  1.8 &02:57:39.0&$+60$:16:49\\
02575+6017 \tablenotemark{( 1 )}  & 2   &   18 &  1.0 &  0.6 &02:57:35.6&$+60$:14:08\\
02593+6016 \tablenotemark{( 1 )}  & 1   &  260 &  6.5 &  2.2 &02:59:23.3&$+60$:16:19\\
02593+6016 \tablenotemark{( 1 )}  & 2   &   27 &  2.0 &  1.4 &02:59:28.9&$+60$:16:54\\
02593+6016 \tablenotemark{( 1 )}  & 3   &   75 &  1.6 &  0.6 &02:59:03.9&$+60$:16:30\\
03064+5638 \tablenotemark{( 7 )}  & 1 a & 1400 & 12.7 &  2.7 &03:06:26.9&$+56$:39:19\\
03064+5638 \tablenotemark{( 7 )}  & 1 b & 1000 & 13.5 &  4.6 &03:06:28.8&$+56$:39:08\\
03211+5446 \tablenotemark{( 7 )}  & 1   & 1000 & 10.1 &  2.3 &03:21:04.7&$+54$:47:07\\
03211+5446 \tablenotemark{( 7 )}  & 2   &  870 & 10.3 &  2.3 &03:21:03.6&$+54$:46:54\\
03236+5836 \tablenotemark{( 3 )}  & 1 a &   49 & 25.7 & 47.0 &03:23:39.0&$+58$:36:38\\
03236+5836 \tablenotemark{( 3 )}  & 1 b &    9 &  5.0 & 11.2 &03:23:40.0&$+58$:35:22\\
03236+5836 \tablenotemark{( 3 )}  & 2 a &   16 &  6.7 & 10.5 &03:23:24.7&$+58$:33:14\\
03236+5836 \tablenotemark{( 3 )}  & 2 b &   15 &  3.7 &  5.3 &03:23:31.5&$+58$:33:53\\
03595+5110 \tablenotemark{( 7 )}  & 1 a &   97 &  5.8 &  6.4 &03:59:25.9&$+51$:11:14\\
03595+5110 \tablenotemark{( 7 )}  & 1 b &   92 &  5.8 & 12.8 &03:59:27.3&$+51$:11:28\\
04073+5102 \tablenotemark{( 7 )}  & 1   & 4100 & 14.6 &  2.3 &04:07:17.7&$+51$:02:44\\
04073+5102 \tablenotemark{( 7 )}  & 2   & 2900 &  9.8 &  1.4 &04:07:16.2&$+51$:01:34\\
04073+5102 \tablenotemark{( 7 )}  & 3   & 5200 & 11.4 &  1.4 &04:07:24.8&$+51$:01:04\\
04073+5102 \tablenotemark{( 7 )}  & 4   & 2300 & 10.9 &  2.2 &04:07:21.0&$+51$:01:04\\
04073+5102 \tablenotemark{( 7 )}  & 5   & 1200 &  9.8 &  1.8 &04:07:16.0&$+51$:00:08\\
04073+5102 \tablenotemark{( 7 )}  & 6   & 1500 &  8.3 &  1.6 &04:07:28.4&$+51$:00:17\\
04073+5102 \tablenotemark{( 7 )}  & 7   & 1900 &  8.2 &  1.1 &04:07:25.3&$+51$:02:14\\
04269+3510 \tablenotemark{( 7 )}  & 1   &   10 & 11.4 &144.2 &04:26:57.8&$+35$:10:03\\
04269+3510 \tablenotemark{( 7 )}  & 2   &   11 & 11.6 & 78.6 &04:26:58.6&$+35$:10:38\\
04324+5106 \tablenotemark{( 2 )}  & 1   &  270 &  9.9 &  8.7 &04:32:30.1&$+51$:06:54\\
04324+5106 \tablenotemark{( 2 )}  & 2   &   44 &  6.8 &  4.3 \tablenotemark{B}&04:32:29.7&$+51$:06:15\\
04324+5106 \tablenotemark{( 2 )}  & 3   &   13 &  3.6 &  2.2 \tablenotemark{B}&04:32:31.7&$+51$:06:13\\
04324+5106 \tablenotemark{( 2 )}  & 4   &   16 &  9.9 &  6.3 \tablenotemark{B}&04:32:31.4&$+51$:06:30\\
04329+5047 \tablenotemark{( 7 )}  & 1   & 3000 &  7.5 &  0.9 &04:33:00.2&$+50$:47:28\\
05100+3723 \tablenotemark{( 3 )}  & 1 a &  170 &  2.5 &  1.0 &05:10:02.7&$+37$:23:36\\
05100+3723 \tablenotemark{( 3 )}  & 1 b &   87 &  2.2 &  1.0 &05:10:01.9&$+37$:24:11\\
05197+3355 \tablenotemark{( 7 )}  & 1   &  300 & 12.2 &  6.6 &05:19:49.7&$+33$:55:32\\
05327$-$0457 \tablenotemark{( 2 )}  & 1   &    9 &  2.8& 4.1 &05:32:38.9&$-04$:58:22\\
05345+3157 \tablenotemark{( 5 )}  & 1   &   48 &  3.2 &  1.4 &05:34:37.1&$+31$:58:20\\
05345+3157 \tablenotemark{( 5 )}  & 2   &   38 &  2.2 &  0.8 &05:34:37.6&$+31$:57:49\\
05345+3157 \tablenotemark{( 5 )}  & 3   &   28 &  1.2 &  0.4 &05:34:27.6&$+31$:59:01\\
05355+3039 \tablenotemark{( 5 )}  & 1   &   13 &  1.2 &  0.8 &05:35:34.5&$+30$:39:28\\
05375+3540 \tablenotemark{( 6 )}  & 1 a &  230 & 17.3 &  8.6 &05:37:32.1&$+35$:40:22\\
05375+3540 \tablenotemark{( 6 )}  & 1 b &   34 &  3.1 &  2.4 &05:37:32.2&$+35$:39:10\\
05375+3540 \tablenotemark{( 6 )}  & 1 c &    5 &  1.8 &  3.0 &05:37:32.8&$+35$:41:38\\
05375+3540 \tablenotemark{( 6 )}  & 2   &   23 &  3.7 &  3.6 &05:37:32.3&$+35$:36:56\\
05377+3548 \tablenotemark{( 6 )}  & 1   &   57 &  4.4 &  2.6 &05:37:38.4&$+35$:47:57\\
05377+3548 \tablenotemark{( 6 )}  & 2 a &   47 &  4.7 &  3.3 &05:37:45.7&$+35$:48:03\\
05377+3548 \tablenotemark{( 6 )}  & 2 b &   51 &  2.8 &  1.6 &05:37:44.4&$+35$:48:33\\
05377+3548 \tablenotemark{( 6 )}  & 2 c &   49 &  2.4 &  1.3 &05:37:42.0&$+35$:48:22\\
05377+3548 \tablenotemark{( 6 )}  & 3   &   59 &  3.6 &  1.7 &05:37:38.9&$+35$:47:04\\
05377+3548 \tablenotemark{( 6 )}  & 4   &   36 &  3.3 &  2.0 &05:37:49.2&$+35$:48:31\\
05377+3548 \tablenotemark{( 6 )}  & 5   &   31 &  1.9 &  1.0 &05:37:32.6&$+35$:48:15\\
05377+3548 \tablenotemark{( 6 )}  & 6   &   28 &  2.2 &  1.5 &05:37:44.2&$+35$:50:38\\
05377+3548 \tablenotemark{( 6 )}  & 7   &   33 &  4.1 &  2.5 &05:37:00.8&$+35$:50:39\\
05377+3548 \tablenotemark{( 6 )}  & 8   &   86 &  1.9 &  0.7 &05:37:04.1&$+35$:51:07\\
05377+3548 \tablenotemark{( 6 )}  & 9   &  120 &  2.6 &  0.9 &05:37:08.5&$+35$:48:19\\
05377+3548 \tablenotemark{( 6 )}  & 10  &   44 &  1.3 &  0.6 &05:37:23.9&$+35$:46:38\\
05377+3548 \tablenotemark{( 6 )}  & 11  &   10 &  1.4 &  0.9 &05:37:53.0&$+35$:50:44\\
05480+2545/4 \tablenotemark{( 2 )}  & 1 &   46 & 12.4 & 24.0 &05:51:10.6& $+25$:46:08 \\
06006+3015 \tablenotemark{( 2 )}  & 1   &  290 &  8.8 &  4.0 &06:00:41.6& $+30$:15:01 \\
06055+2039 \tablenotemark{( 2 )}  & 1   &   97 & 14.6 & 14.3 &06:05:36.6& $+20$:39:33 \\
06055+2039 \tablenotemark{( 5 )}  & 1 a &   84 &  8.9 &  5.8 &06:05:36.5& $+20$:39:40 \\
06055+2039 \tablenotemark{( 5 )}  & 1 b &    9 &  1.8 &  2.6 &06:05:32.3& $+20$:40:07 \\
06055+2039 \tablenotemark{( 5 )}  & 2 a &   32 &  1.6 &  0.8 &06:05:41.0& $+20$:38:46 \\
06055+2039 \tablenotemark{( 5 )}  & 2 b &   17 &  1.3 &  0.8 &06:05:46.1& $+20$:38:43 \\
06055+2039 \tablenotemark{( 5 )}  & 2 c &    9 &  1.5 &  0.9 &06:05:49.1& $+20$:39:18 \\
06056+2131 \tablenotemark{( 1 )}  & 1 a &  130 & 18.7 & 15.9 &06:05:40.3& $+21$:31:34 \\
06056+2131 \tablenotemark{( 1 )}  & 1 b &   27 &  4.4 &  4.7 &06:05:42.5& $+21$:31:45 \\
06056+2131 \tablenotemark{( 1 )}  & 2 a &   29 &  6.0 &  7.3 &06:05:46.5& $+21$:32:19 \\
06056+2131 \tablenotemark{( 1 )}  & 2 b &   13 &  3.3 &  5.7 &06:05:44.4& $+21$:32:14 \\
06056+2131 \tablenotemark{( 1 )}  & 2 c &    7 &  2.6 &  4.2 &06:05:46.9& $+21$:32:47 \\
06056+2131 \tablenotemark{( 3 )}  & 1   &  410 & 25.4 & 14.2 &06:05:40.2& $+21$:31:35 \\
06056+2131 \tablenotemark{( 3 )}  & 2   &  160 &  8.8 &  6.0 &06:05:46.1& $+21$:32:20 \\
06058+2138 \tablenotemark{( 2 )}  & 1   &  690 & 24.9 &  6.7 &06:05:52.8& $+21$:38:41 \\
06058+2138 \tablenotemark{( 3 )}  & 1   &  510 & 25.9 & 12.1 &06:05:53.0& $+21$:38:41 \\
06061+2151 \tablenotemark{( 2 )}  & 1   &   80 & 11.3 &  7.4 &06:06:06.0& $+21$:51:10 \\
06063+2040 \tablenotemark{( 6 )}  & 1   &   42 &  5.4 &  5.1 &06:06:22.4& $+20$:39:20 \\
06063+2040 \tablenotemark{( 6 )}  & 2   &    6 &  1.3 &  1.7 &06:06:27.4& $+20$:40:32 \\
06068+2030 \tablenotemark{( 6 )}  & 1   &   72 &  1.8 &  0.5 &06:06:53.3& $+20$:30:30\\
06068+2030 \tablenotemark{( 6 )}  & 2   &   42 &  1.8 &  0.8 &06:06:51.2& $+20$:31:05\\
06073+1249 \tablenotemark{( 7 )}  & 1   & 1700 & 14.9 &  3.9 &06:07:23.4& $+12$:49:28\\
06073+1249 \tablenotemark{( 7 )}  & 2   &  200 &  5.7 &  4.2 &06:07:27.0& $+12$:49:28\\
06099+1800 \tablenotemark{( 7 )}  & 1   &  150 & 64.6 &143.1 &06:09:58.3& $+18$:01:16\\
06099+1800 \tablenotemark{( 7 )}  & 2   &  150 & 68.8 &150.9 &06:09:58.7& $+18$:00:12\\
06099+1800 \tablenotemark{( 7 )}  & 3   &    9 &  8.6 & 37.1 &06:09:51.7& $+17$:59:16\\
06099+1800 \tablenotemark{( 7 )}  & 4 a &   73 & 24.5 & 48.7 &06:10:01.4& $+17$:59:24\\
06099+1800 \tablenotemark{( 7 )}  & 4 b &    6 & 17.0 & 54.4\tablenotemark{B} &06:10:01.4& $+17$:58:37  \\
06099+1800 \tablenotemark{( 7 )}  & 4 c &   18 & 11.8 & 37.6 &06:09:59.0& $+17$:58:09\\
06105+1756 \tablenotemark{( 7 )}  & 1 a &  480 &  8.4 &  2.7 &06:10:34.2& $+17$:56:21\\
06105+1756 \tablenotemark{( 7 )}  & 1 b &  380 &  5.6 &  1.6 &06:10:36.4& $+17$:56:51\\
06105+1756 \tablenotemark{( 7 )}  & 2   &  630 &  5.6 &  1.4 &06:10:31.1& $+17$:56:43\\
06105+1756 \tablenotemark{( 7 )}  & 3   &  280 &  6.3 &  2.0 &06:10:51.7& $+17$:55:37\\
06114+1745 \tablenotemark{( 5 )}  & 1 a &   17 &  1.3 &  1.2 &06:08:33.4& $+17$:46:30\\
06114+1745 \tablenotemark{( 5 )}  & 1 b &    7 &  1.1 &  1.5 &06:08:34.8& $+17$:46:51\\
06114+1745 \tablenotemark{( 5 )}  & 2   &   12 &  1.2 &  1.0 &06:08:34.8& $+17$:45:52\\
06114+1745 \tablenotemark{( 5 )}  & 3   &    6 &  1.1 &  1.4 &06:08:32.4& $+17$:45:05\\
06114+1745 \tablenotemark{( 5 )}  & 4   &    8 &  0.9 &  1.1 &06:08:30.3& $+17$:45:56\\
06117+1350 \tablenotemark{( 2 )}  & 1   &  210 & 15.5 & 12.3 &06:11:46.7& $+13$:50:34\\
06155+2039 \tablenotemark{( 2 )}  & 1 a &   24 & 10.1 & 24.7 &06:15:34.4& $+23$:19:54\\
06155+2039 \tablenotemark{( 2 )}  & 1 b &    6 &  5.8 & 13.8\tablenotemark{B}&06:15:34.3& $+23$:19:36\\
06308+0402 \tablenotemark{( 4 )}  & 1 a &   12 &  3.0 &  4.6 &06:30:50.5& $+04$:03:38\\
06308+0402 \tablenotemark{( 4 )}  & 1 b &    7 &  2.4 &  4.8 &06:30:49.8& $+04$:03:23\\
06308+0402 \tablenotemark{( 4 )}  & 2 a &    8 &  2.9 &  5.5 &06:30:52.3& $+04$:02:36\\
06308+0402 \tablenotemark{( 4 )}  & 2 b &   10 &  2.9 &  4.9 &06:30:51.4& $+04$:02:42\\
06308+0402 \tablenotemark{( 4 )}  & 3 a &    3 &  2.1 &  5.4\tablenotemark{B} &06:30:55.1& $+04$:02:52  \\
06308+0402 \tablenotemark{( 4 )}  & 3 b &    2 &  2.0 &  8.6 &06:30:54.4& $+04$:02:51\\
06308+0402 \tablenotemark{( 4 )}  & 4   &    5 &  2.0 &  3.8 &06:30:53.0& $+04$:02:02\\
06319+0415 \tablenotemark{( 7 )}  & 1   &  550 & 53.1 & 50.3 &06:31:58.5& $+04$:15:02\\
06319+0415 \tablenotemark{( 7 )}  & 2   &   23 & 14.8 & 86.7 &06:32:03.1& $+04$:14:13\\
06384+0932 \tablenotemark{( 3 )}  & 1 a &  180 & 24.0 & 16.7 &06:38:25.3& $+09$:32:21\\
06384+0932 \tablenotemark{( 3 )}  & 1 b &  240 & 23.8 &  8.9 &06:38:27.3& $+09$:32:08\\
06384+0932 \tablenotemark{( 3 )}  & 1 c &   31 & 11.9 & 22.0 &06:38:30.4& $+09$:32:12\\
06384+0932 \tablenotemark{( 3 )}  & 2   &   64 &  8.1 &  5.1 &06:38:21.0& $+09$:36:59\\
06384+0932 \tablenotemark{( 3 )}  & 3   &   76 &  4.3 &  2.6 &06:38:19.0& $+09$:37:54\\
06384+0932 \tablenotemark{( 3 )}  & 4   &   76 &  4.2 &  2.5 &06:38:30.0& $+09$:30:08\\
06412$-$0105 \tablenotemark{( 5 )}  & 1   &    4 &  1.1 &  1.2 &06:38:43.3& $-01$:02:18\\
06567$-$0355 \tablenotemark{( 4 )}  & 1   &  200 &  7.4 &  3.6 &06:56:46.5& $-03$:55:19\\
06581$-$0846/8 \tablenotemark{( 5 )}  & 1 &  200 &  1.1 &  0.2 &06:55:41.8& $-08$:42:57\\
06581$-$0846/8 \tablenotemark{( 5 )}  & 2 &   81 &  1.1 &  0.2 &06:55:38.0& $-08$:41:34\\
07029$-$1215 \tablenotemark{( 4 )}  & 1 a &   15 &  5.8 &  8.8 &07:00:31.3& $-12$:09:54\\
07029$-$1215 \tablenotemark{( 4 )}  & 1 b &    3 &  1.5 &  4.2 &07:00:28.9& $-12$:09:31\\
07029$-$1215 \tablenotemark{( 4 )}  & 2 a &    4 &  2.1 &  4.5 &07:00:39.6& $-12$:11:10\\
07029$-$1215 \tablenotemark{( 4 )}  & 2 b &    1 &  1.6 & 11.0 &07:00:41.3& $-12$:11:18\\
07029$-$1215 \tablenotemark{( 4 )}  & 3   &    4 &  2.0 &  4.5 &07:00:23.9& $-12$:07:30\\
07029$-$1215 \tablenotemark{( 4 )}  & 4   &    1 &  1.3 &  4.5 &07:04:18.0& $-12$:07:46\\
07299$-$1651 \tablenotemark{( 2 )}  & 1   &   28 & 12.7 & 24.6 &07:29:54.9& $-16$:51:41\\

\enddata

\tablenotetext{( 1 ) -- ( 7 )}{1. --  7. Run}
\tablenotetext{B}{beam area used for N(H), n(H) because of too small source extents (see text)}

\end{deluxetable}


\clearpage

\begin{deluxetable}{ccccccc}
  \rotate
 \tablewidth{0pt} \tablecaption{
    Exponents of the radial intensity and density profiles for IRAS 06058+2131 \protect\\
    The fitted radial intensity profile exponent is $\alpha$, the
    resulting radial density profile exponent is $p$. The error of $p$
    is about 0.3. The mean values and the standard deviation of the
    surveys by \citet{Beuther2002II} and \citet{Mueller2002} are given
    for comparison.  The noted interval limits refer to the smallest
    and largest \emph{mean} ring radii.
 \label{alphaundp}}

\tablehead{
& \colhead{IRAM} & \colhead{IRAM} & \colhead{IRAM} & \colhead{JCMT} & \colhead{JCMT} & \colhead{JCMT}  \\ 
r & [$18.3 \arcsec$, $59.3 \arcsec$] & [$18.3 \arcsec$, $31.5 \arcsec$] & [$33 \arcsec$, $59.3 \arcsec$] & [$15.7 \arcsec$, $59.2 \arcsec$] & [$15.7 \arcsec$, $30.7 \arcsec$] & [$32.2 \arcsec$, $59.1 \arcsec$]  }

\startdata
$\alpha $ & $-1.5$ & $-1.3$ & $-2.0$ & $-1.4$ & $-1.2$ & $-1.9$   \\ 
$\chi ^{2}$ & $0.33$ & $0.09$ & $0.04$ & $3.41$ & $0.30$ & $0.18$   \\ 
\hline
$\mathbf{p}$ & $\mathbf{2.0}$ & $\mathbf{1.8}$ & $\mathbf{2.6}$  &  $ \mathbf{1.9}$ & $\mathbf{1.7}$ & $\mathbf{2.4}$ \\ 
\hline
$\langle p_{Beuther } \rangle $ &  & $1.6\pm 0.5$ & $2.3\pm 0.7$ &  &  $1.6\pm 0.5$ & $2.3\pm 0.7$   \\ 
$\langle p_{Mueller } \rangle $ & $1.8\pm 0.4$ &  &  & $1.8\pm 0.4$ &  &   \\  
\enddata

\end{deluxetable}



\begin{thebibliography}{122}
\expandafter\ifx\csname natexlab\endcsname\relax\def\natexlab#1{#1}\fi

\bibitem[{{Adams}(1991)}]{Adams91}
{Adams}, F.~C. 1991, \apj, 382, 544

\bibitem[{{Allen}(1972)}]{Allen72}
{Allen}, D.~A. 1972, \apjl, 172, L55

\bibitem[{{Anandarao} {et~al.}(2004){Anandarao}, {Chakraborty}, {Ojha}, \&
  {Testi}}]{Anandarao04}
{Anandarao}, B.~G., {Chakraborty}, A., {Ojha}, D.~K., \& {Testi}, L. 2004,
  \aap, 421, 1045

\bibitem[{{Andr{\' e}} {et~al.}(1999){Andr{\' e}}, {Bacmann}, {Motte}, \&
  {Ward-Thompson}}]{AndreB99}
{Andr{\' e}}, P., {Bacmann}, A., {Motte}, F., \& {Ward-Thompson}, D. 1999, in
  The Physics and Chemistry of the Interstellar Medium, Proceedings of the 3rd
  Cologne-Zermatt Symposium, held in Zermatt, September 22-25, 1998, Eds.: V.
  Ossenkopf, J. Stutzki, and G. Winnewisser, GCA-Verlag, Herdecke, p. 241

\bibitem[{Andr\'e {et~al.}(1993)Andr\'e, Ward-Thompson, \& Barsony}]{Andre93}
Andr\'e, P., Ward-Thompson, D., \& Barsony, M. 1993, ApJ, 406, 122

\bibitem[{{Aspin}(1998)}]{Aspin98}
{Aspin}, C. 1998, \aap, 335, 1040

\bibitem[{{Bacmann} {et~al.}(2000){Bacmann}, {Andr{\' e}}, {Puget}, {Abergel},
  {Bontemps}, \& {Ward-Thompson}}]{Bacmann2000}
{Bacmann}, A., {Andr{\' e}}, P., {Puget}, J.-L., {Abergel}, A., {Bontemps}, S.,
  \& {Ward-Thompson}, D. 2000, \aap, 361, 555

\bibitem[{{Barsony} {et~al.}(1991){Barsony}, {Schombert}, \&
  {Kis-Halas}}]{Barsony91}
{Barsony}, M., {Schombert}, J.~M., \& {Kis-Halas}, K. 1991, \apj, 379, 221

\bibitem[{{Baudry} {et~al.}(1997){Baudry}, {Desmurs}, {Wilson}, \&
  {Cohen}}]{Baudry1997}
{Baudry}, A., {Desmurs}, J.~F., {Wilson}, T.~L., \& {Cohen}, R.~J. 1997, \aap,
  325, 255

\bibitem[{{Beuther} {et~al.}(2002){Beuther}, {Schilke}, {Menten}, {Motte},
  {Sridharan}, \& {Wyrowski}}]{Beuther2002II}
{Beuther}, H., {Schilke}, P., {Menten}, K.~M., {Motte}, F., {Sridharan}, T.~K.,
  \& {Wyrowski}, F. 2002, \apj, 566, 945

\bibitem[{{Bica} {et~al.}(2003{\natexlab{a}}){Bica}, {Dutra}, \&
  {Barbuy}}]{Bica03a}
{Bica}, E., {Dutra}, C.~M., \& {Barbuy}, B. 2003{\natexlab{a}}, \aap, 397, 177

\bibitem[{{Bica} {et~al.}(2003{\natexlab{b}}){Bica}, {Dutra}, {Soares}, \&
  {Barbuy}}]{Bica03b}
{Bica}, E., {Dutra}, C.~M., {Soares}, J., \& {Barbuy}, B. 2003{\natexlab{b}},
  \aap, 404, 223

\bibitem[{{Blitz} {et~al.}(1982){Blitz}, {Fich}, \& {Stark}}]{Blitz82}
{Blitz}, L., {Fich}, M., \& {Stark}, A.~A. 1982, \apjs, 49, 183

\bibitem[{{Bonnarel} {et~al.}(2000){Bonnarel}, {Fernique}, {Bienaym{\' e}},
  {Egret}, {Genova}, {Louys}, {Ochsenbein}, {Wenger}, \& {Bartlett}}]{Aladin}
{Bonnarel}, F., {Fernique}, P., {Bienaym{\' e}}, O., {Egret}, D., {Genova}, F.,
  {Louys}, M., {Ochsenbein}, F., {Wenger}, M., \& {Bartlett}, J.~G. 2000,
  \aaps, 143, 33

\bibitem[{{Bonnor}(1956)}]{Bonnor56}
{Bonnor}, W.~B. 1956, \mnras, 116, 351

\bibitem[{{Brand} \& {Blitz}(1993)}]{Brand93}
{Brand}, J. \& {Blitz}, L. 1993, \aap, 275, 67

\bibitem[{Brogui\`ere {et~al.}(1996)Brogui\`ere, Neri, \& Sievers}]{nic}
Brogui\`ere, D., Neri, R., \& Sievers, A. 1996, {NIC} Bolometer Users Guide,
  IRAM

\bibitem[{{Bronfman} {et~al.}(1996){Bronfman}, {Nyman}, \& {May}}]{Bronf96}
{Bronfman}, L., {Nyman}, L.-A., \& {May}, J. 1996, \aaps, 115, 81

\bibitem[{{Caplan} {et~al.}(2000){Caplan}, {Deharveng}, {Pe{\~ n}a}, {Costero},
  \& {Blondel}}]{Caplan00}
{Caplan}, J., {Deharveng}, L., {Pe{\~ n}a}, M., {Costero}, R., \& {Blondel}, C.
  2000, \mnras, 311, 317

\bibitem[{{Carpenter} {et~al.}(1990){Carpenter}, {Snell}, \&
  {Schloerb}}]{Carpenter90}
{Carpenter}, J.~M., {Snell}, R.~L., \& {Schloerb}, F.~P. 1990, \apj, 362, 147

\bibitem[{{Carpenter} {et~al.}(1995{\natexlab{a}}){Carpenter}, {Snell}, \&
  {Schloerb}}]{Carpenter95a}
---. 1995{\natexlab{a}}, \apj, 445, 246

\bibitem[{{Carpenter} {et~al.}(1995{\natexlab{b}}){Carpenter}, {Snell}, \&
  {Schloerb}}]{Carpenter95b}
---. 1995{\natexlab{b}}, \apj, 450, 201

\bibitem[{{Carpenter} {et~al.}(1993){Carpenter}, {Snell}, {Schloerb}, \&
  {Skrutskie}}]{Carpenter93}
{Carpenter}, J.~M., {Snell}, R.~L., {Schloerb}, F.~P., \& {Skrutskie}, M.~F.
  1993, \apj, 407, 657

\bibitem[{{Chan} {et~al.}(1996){Chan}, {Henning}, \& {Schreyer}}]{Chan1996}
{Chan}, S.~J., {Henning}, T., \& {Schreyer}, K. 1996, \aaps, 115, 285

\bibitem[{{Chandler} \& {Richer}(2000)}]{Chandler2000}
{Chandler}, C.~J. \& {Richer}, J.~S. 2000, \apj, 530, 851

\bibitem[{{Chini} {et~al.}(1997){Chini}, {Reipurth}, {Ward-Thompson}, {Bally},
  {Nyman}, {Sievers}, \& {Billawala}}]{Chini97}
{Chini}, R., {Reipurth}, B., {Ward-Thompson}, D., {Bally}, J., {Nyman}, L.-A.,
  {Sievers}, A., \& {Billawala}, Y. 1997, \apjl, 474, L135

\bibitem[{{Condon} {et~al.}(1998){Condon}, {Cotton}, {Greisen}, {Yin},
  {Perley}, {Taylor}, \& {Broderick}}]{NVSS}
{Condon}, J.~J., {Cotton}, W.~D., {Greisen}, E.~W., {Yin}, Q.~F., {Perley},
  R.~A., {Taylor}, G.~B., \& {Broderick}, J.~J. 1998, \aj, 115, 1693

\bibitem[{{Cox} {et~al.}(1990){Cox}, {Deharveng}, \& {Leene}}]{Cox90}
{Cox}, P., {Deharveng}, L., \& {Leene}, A. 1990, \aap, 230, 181

\bibitem[{{Crowther} \& {Conti}(2003)}]{Crowther03}
{Crowther}, P.~A. \& {Conti}, P.~S. 2003, \mnras, 343, 143

\bibitem[{{Deharveng} {et~al.}(1997){Deharveng}, {Zavagno}, {Cruz-Gonzalez},
  {Salas}, {Caplan}, \& {Carrasco}}]{Deharveng1997}
{Deharveng}, L., {Zavagno}, A., {Cruz-Gonzalez}, I., {Salas}, L., {Caplan}, J.,
  \& {Carrasco}, L. 1997, \aap, 317, 459

\bibitem[{{Diaz-Miller} {et~al.}(1998){Diaz-Miller}, {Franco}, \&
  {Shore}}]{Diaz-Miller98}
{Diaz-Miller}, R.~I., {Franco}, J., \& {Shore}, S.~N. 1998, \apj, 501, 192

\bibitem[{{Dorschner} \& {G{\"u}rtler}(1963)}]{Dorschner63}
{Dorschner}, J. \& {G{\"u}rtler}, J. 1963, Astronomische Nachrichten, 287, 257

\bibitem[{{Egan} {et~al.}(2003){Egan}, {Price}, {Kraemer}, {Mizuno}, {Carey},
  {Wright}, {Engelke}, {Cohen}, \& {Gugliotti}}]{MSXEgan2003}
{Egan}, M.~P., {Price}, S.~D., {Kraemer}, K.~E., {Mizuno}, D.~R., {Carey},
  S.~J., {Wright}, C.~O., {Engelke}, C.~W., {Cohen}, M., \& {Gugliotti}, M.~G.
  2003, Air Force Research Laboratory Technical Report, AFRL-VS-TR-2003-1589

\bibitem[{{Eiroa} {et~al.}(1994){Eiroa}, {Casali}, {Miranda}, \&
  {Ortiz}}]{Eiroa94}
{Eiroa}, C., {Casali}, M.~M., {Miranda}, L.~F., \& {Ortiz}, E. 1994, \aap, 290,
  599

\bibitem[{{Emerson}(1995)}]{EmersonII}
{Emerson}, D.~T. 1995, in ASP Conf. Ser. 75: Multi-Feed Systems for Radio
  Telescopes, p. 309

\bibitem[{{Emerson} {et~al.}(1979){Emerson}, {Klein}, \& {Haslam}}]{Emerson79}
{Emerson}, D.~T., {Klein}, U., \& {Haslam}, C.~G.~T. 1979, \aap, 76, 92

\bibitem[{{Evans} {et~al.}(1981){Evans}, {Beichman}, {Gatley}, {Harvey},
  {Nadeau}, \& {Sellgren}}]{Evans81b}
{Evans}, N.~J., {Beichman}, C., {Gatley}, I., {Harvey}, P., {Nadeau}, D., \&
  {Sellgren}, K. 1981, ApJ, 246, 409

\bibitem[{{Evans} \& {Blair}(1981)}]{Evans81a}
{Evans}, N.~J. \& {Blair}, G.~N. 1981, \apj, 246, 394

\bibitem[{{Evans} {et~al.}(2002){Evans}, {Shirley}, {Mueller}, \&
  {Knez}}]{Evans02}
{Evans}, N.~J., {Shirley}, Y.~L., {Mueller}, K.~E., \& {Knez}, C. 2002, in ASP
  Conf. Ser. 267: Hot Star Workshop III: The Earliest Phases of Massive Star
  Birth, 17

\bibitem[{{Fa{\'u}ndez} {et~al.}(2004){Fa{\'u}ndez}, {Bronfman}, {Garay},
  {Chini}, {Nyman}, \& {May}}]{Faundez04}
{Fa{\'u}ndez}, S., {Bronfman}, L., {Garay}, G., {Chini}, R., {Nyman},
  L.-{\AA}., \& {May}, J. 2004, \aap, 426, 97

\bibitem[{{Felli} {et~al.}(1977){Felli}, {Habing}, \& {Israel}}]{Felli77}
{Felli}, M., {Habing}, H.~J., \& {Israel}, F.~P. 1977, \aap, 59, 43

\bibitem[{{Felli} {et~al.}(1997){Felli}, {Testi}, {Valdettaro}, \&
  {Wang}}]{Felli97}
{Felli}, M., {Testi}, L., {Valdettaro}, R., \& {Wang}, J.-J. 1997, \aap, 320,
  594

\bibitem[{{Fich}(1993)}]{Fich93}
{Fich}, M. 1993, \apjs, 86, 475

\bibitem[{{Forbrich} {et~al.}(2004){Forbrich}, {Schreyer}, {Posselt}, {Klein},
  \& {Henning}}]{Forbrich04}
{Forbrich}, J., {Schreyer}, K., {Posselt}, B., {Klein}, R., \& {Henning}, T.
  2004, \apj, 602, 843

\bibitem[{{Franco} {et~al.}(1994){Franco}, {Shore}, \&
  {Tenorio-Tagle}}]{Franco94}
{Franco}, J., {Shore}, S.~N., \& {Tenorio-Tagle}, G. 1994, \apj, 436, 795

\bibitem[{{F\"urst} {et~al.}(1990){F\"urst}, {Reich}, {Reich}, \&
  {Reif}}]{Furst90}
{F\"urst}, E., {Reich}, W., {Reich}, P., \& {Reif}, K. 1990, \aaps, 85, 805

\bibitem[{Garay(2005)}]{GaraySicily05}
Garay, Guido. 2005, Massive and dense cores: the maternities of high mass
  stars, Talk at IAU Symposium 227, Acireale, Italy

\bibitem[{{Ghosh} {et~al.}(2000){Ghosh}, {Iyengar}, {Karnik}, {Rengarajan},
  {Tandon}, \& {Verma}}]{Ghosh00}
{Ghosh}, S.~K., {Iyengar}, K.~V.~K., {Karnik}, A.~D., {Rengarajan}, T.~N.,
  {Tandon}, S.~N., \& {Verma}, R.~P. 2000, Bulletin of the Astronomical Society
  of India, 28, 515

\bibitem[{{Hatchell} {et~al.}(2000){Hatchell}, {Fuller}, {Millar}, {Thompson},
  \& {Macdonald}}]{Hatchell2000}
{Hatchell}, J., {Fuller}, G.~A., {Millar}, T.~J., {Thompson}, M.~A., \&
  {Macdonald}, G.~H. 2000, \aap, 357, 637

\bibitem[{{Henning}(1990)}]{Henning90}
{Henning}, Th. 1990, Fundamentals of Cosmic Physics, 14, 321

\bibitem[{{Henning} {et~al.}(1992){Henning}, {Cesaroni}, {Walmsley}, \&
  {Pfau}}]{Henning92b}
{Henning}, Th., {Cesaroni}, R., {Walmsley}, M., \& {Pfau}, W. 1992, \aaps, 93,
  525

\bibitem[{{Henning} {et~al.}(2000){Henning}, {Klein}, {Launhardt}, {Schreyer},
  \& {Stecklum}}]{Henning00b}
{Henning}, Th., {Klein}, R., {Launhardt}, R., {Schreyer}, K., \& {Stecklum}, B.
  2000, in ISO Survey of a Dusty Universe, Proceedings of a Ringberg Workshop
  Held at Ringberg Castle, Tegernsee, Germany, 8-12 November 1999, Edited by D.
  Lemke, M. Stickel, and K. Wilke, Lecture Notes in Physics, vol. 548, p. 333

\bibitem[{{Henriksen} {et~al.}(1997){Henriksen}, {Andr\'{e}}, \&
  {Bontemps}}]{Henriksen97}
{Henriksen}, R., {Andr\'{e}}, P., \& {Bontemps}, S. 1997, \aap, 323, 549

\bibitem[{{Herbig}(1971)}]{Herbig71}
{Herbig}, G.~H. 1971, \apj, 169, 537

\bibitem[{{Hillenbrand} \& {Hartmann}(1998)}]{Hillenbrand98}
{Hillenbrand}, L.~A. \& {Hartmann}, L.~W. 1998, \apj, 492, 540

\bibitem[{{Holland} {et~al.}(1999){Holland}, {Robson}, {Gear}, {Cunningham},
  {Lightfoot}, {Jenness}, {Ivison}, {Stevens}, {Ade}, {Griffin}, {Duncan},
  {Murphy}, \& {Naylor}}]{scuba}
{Holland}, W.~S., {Robson}, E.~I., {Gear}, W.~K., {Cunningham}, C.~R.,
  {Lightfoot}, J.~F., {Jenness}, T., {Ivison}, R.~J., {Stevens}, J.~A., {Ade},
  P.~A.~R., {Griffin}, M.~J., {Duncan}, W.~D., {Murphy}, J.~A., \& {Naylor},
  D.~A. 1999, \mnras, 303, 659

\bibitem[{{Houk} \& {Smith-Moore}(1988)}]{Houk1988}
{Houk}, N. \& {Smith-Moore}, M. 1988, in Michigan Spectral Survey, Ann Arbor,
  Dept. of Astronomy, Univ. Michigan (Vol. 4) (1988), 0--+

\bibitem[{{Howard} {et~al.}(1997){Howard}, {Pipher}, \& {Forrest}}]{Howard97}
{Howard}, E.~M., {Pipher}, J.~L., \& {Forrest}, W.~J. 1997, \apj, 481, 327

\bibitem[{{Howard} {et~al.}(1998){Howard}, {Pipher}, \& {Forrest}}]{Howard98}
---. 1998, \apj, 509, 749

\bibitem[{{Hunter} {et~al.}(2000){Hunter}, {Churchwell}, {Watson}, {Cox},
  {Benford}, \& {Roelfsema}}]{Hunter2000}
{Hunter}, T.~R., {Churchwell}, E., {Watson}, C., {Cox}, P., {Benford}, D.~J.,
  \& {Roelfsema}, P.~R. 2000, \aj, 119, 2711

\bibitem[{{Ishii} {et~al.}(2002){Ishii}, {Hirao}, {Nagashima}, {Nagata},
  {Sato}, \& {Yao}}]{Ishii02}
{Ishii}, M., {Hirao}, T., {Nagashima}, C., {Nagata}, T., {Sato}, S., \& {Yao},
  Y. 2002, \aj, 124, 430

\bibitem[{{Itoh} {et~al.}(2001){Itoh}, {Tamura}, {Suto}, {Hayashi}, {Murakawa},
  {Oasa}, {Nakajima}, {Kaifu}, {Kosugi}, {Usuda}, \& {Doi}}]{Itoh2001}
{Itoh}, Y., {Tamura}, M., {Suto}, H., {Hayashi}, S.~S., {Murakawa}, K., {Oasa},
  Y., {Nakajima}, Y., {Kaifu}, N., {Kosugi}, G., {Usuda}, T., \& {Doi}, Y.
  2001, \pasj, 53, 495

\bibitem[{{Ivanov} {et~al.}(2002){Ivanov}, {Borissova}, {Pessev}, {Ivanov}, \&
  {Kurtev}}]{Ivanov02}
{Ivanov}, V.~D., {Borissova}, J., {Pessev}, P., {Ivanov}, G.~R., \& {Kurtev},
  R. 2002, \aap, 394, L1

\bibitem[{{Jenness} \& {Lightfoot}(1998)}]{surf}
{Jenness}, T. \& {Lightfoot}, J.~F. 1998, in ASP Conf. Ser. 145: Astronomical
  Data Analysis Software and Systems VII, Vol.~7, p. 216

\bibitem[{{Jiang} {et~al.}(2003){Jiang}, {Yao}, {Yang}, {Baba}, {Kato},
  {Kurita}, {Nagashima}, {Nagata}, {Nagayama}, {Nakajima}, {Ishii}, {Tamura},
  \& {Sugitani}}]{Jiang2003}
{Jiang}, Z., {Yao}, Y., {Yang}, J., {Baba}, D., {Kato}, D., {Kurita}, M.,
  {Nagashima}, C., {Nagata}, T., {Nagayama}, T., {Nakajima}, Y., {Ishii}, M.,
  {Tamura}, M., \& {Sugitani}, K. 2003, \apj, 596, 1064

\bibitem[{{K\"ompe} {et~al.}(1989){K\"ompe}, {Baudry}, {Joncas}, \&
  {Wouterloot}}]{Koempe89}
{K\"ompe}, C., {Baudry}, A., {Joncas}, G., \& {Wouterloot}, J.~G.~A. 1989,
  \aap, 221, 295

\bibitem[{{Kraemer} {et~al.}(2003){Kraemer}, {Shipman}, {Price}, {Mizuno},
  {Kuchar}, \& {Carey}}]{Kraemer03}
{Kraemer}, K.~E., {Shipman}, R.~F., {Price}, S.~D., {Mizuno}, D.~R., {Kuchar},
  T., \& {Carey}, S.~J. 2003, \aj, 126, 1423

\bibitem[{Kreysa {et~al.}(1998)}]{bolos}
Kreysa, E. {et~al.} 1998, in \procspie, Vol. 3357, 319

\bibitem[{{Kurtz} {et~al.}(1994){Kurtz}, {Churchwell}, \& {Wood}}]{Kurtz94}
{Kurtz}, S., {Churchwell}, E., \& {Wood}, D.~O.~S. 1994, \apjs, 91, 659

\bibitem[{{Lada} \& {Lada}(2003)}]{Lada03}
{Lada}, C.~J. \& {Lada}, E.~A. 2003, \araa, 41, 57

\bibitem[{{Mampaso} {et~al.}(1989){Mampaso}, {Vilches}, {Riera}, {Phillips}, \&
  {Pismis}}]{Mampaso89}
{Mampaso}, A., {Vilches}, J.~M., {Riera}, A., {Phillips}, J.~P., \& {Pismis},
  P. 1989, \aap, 220, 235

\bibitem[{{Margulis} {et~al.}(1988){Margulis}, {Lada}, \& {Snell}}]{Margulis88}
{Margulis}, M., {Lada}, C.~J., \& {Snell}, R.~L. 1988, \apj, 333, 316

\bibitem[{{McCutcheon} {et~al.}(2000){McCutcheon}, {Sandell}, {Matthews},
  {Kuiper}, {Sutton}, {Danchi}, \& {Sato}}]{McCutcheon2000}
{McCutcheon}, W.~H., {Sandell}, G., {Matthews}, H.~E., {Kuiper}, T.~B.~H.,
  {Sutton}, E.~C., {Danchi}, W.~C., \& {Sato}, T. 2000, \mnras, 316, 152

\bibitem[{{Mezger} {et~al.}(1988){Mezger}, {Chini}, {Kreysa}, {Wink}, \&
  {Salter}}]{Mezger88}
{Mezger}, P.~G., {Chini}, R., {Kreysa}, E., {Wink}, J.~E., \& {Salter}, C.~J.
  1988, \aap, 191, 44

\bibitem[{{Miller} \& {Scalo}(1979)}]{MillerScalo79}
{Miller}, G.~E. \& {Scalo}, J.~M. 1979, \apjs, 41, 513

\bibitem[{{Minier} {et~al.}(2004){Minier}, {Burton}, {Purcell}, {Hill},
  {Longmore}, \& {Walsh}}]{Minier04}
{Minier}, V., {Burton}, M., {Purcell}, C., {Hill}, T., {Longmore}, S., \&
  {Walsh}, A. 2004, in The Dusty and Molecular Universe: A Prelude to Herschel
  and ALMA, meeting held in Paris, France, October 27-29, 2004, Eds.: A.
  Wilson. ESA Conf. Ser., 159

\bibitem[{{Molinari} {et~al.}(2002){Molinari}, {Testi}, {Rodr{\' i}guez}, \&
  {Zhang}}]{Molinari02}
{Molinari}, S., {Testi}, L., {Rodr{\' i}guez}, L.~F., \& {Zhang}, Q. 2002,
  \apj, 570, 758

\bibitem[{{Mookerjea} {et~al.}(1999){Mookerjea}, {Ghosh}, {Karnik},
  {Rengarajan}, {Tandon}, \& {Verma}}]{Mook99}
{Mookerjea}, B., {Ghosh}, S.~K., {Karnik}, A.~D., {Rengarajan}, T.~N.,
  {Tandon}, S.~N., \& {Verma}, R.~P. 1999, \apj, 522, 285

\bibitem[{{Mookerjea} {et~al.}(2000){Mookerjea}, {Ghosh}, {Rengarajan},
  {Tandon}, \& {Verma}}]{Mookerjea2000}
{Mookerjea}, B., {Ghosh}, S.~K., {Rengarajan}, T.~N., {Tandon}, S.~N., \&
  {Verma}, R.~P. 2000, \apj, 539, 775

\bibitem[{{Motte} \& {Andr{\' e}}(2001)}]{Motte2001}
{Motte}, F. \& {Andr{\' e}}, P. 2001, \aap, 365, 440

\bibitem[{{Mueller} {et~al.}(2002){Mueller}, {Shirley}, {Evans}, \&
  {Jacobson}}]{Mueller2002}
{Mueller}, K.~E., {Shirley}, Y.~L., {Evans}, N.~J., \& {Jacobson}, H.~R. 2002,
  \apjs, 143, 469

\bibitem[{{Neckel} \& {Staude}(1984)}]{Neckel84}
{Neckel}, T. \& {Staude}, H.~J. 1984, \aap, 131, 200

\bibitem[{{Neckel} {et~al.}(1989){Neckel}, {Staude}, {Meisenheimer}, {Chini},
  \& {Guesten}}]{Neckel89}
{Neckel}, T., {Staude}, H.~J., {Meisenheimer}, K., {Chini}, R., \& {Guesten},
  R. 1989, \aap, 210, 378

\bibitem[{{Ogura} {et~al.}(2002){Ogura}, {Sugitani}, \& {Pickles}}]{Ogura2002}
{Ogura}, K., {Sugitani}, K., \& {Pickles}, A. 2002, \aj, 123, 2597

\bibitem[{{Ossenkopf} \& {Henning}(1994)}]{Ossenkopf94}
{Ossenkopf}, V. \& {Henning}, T. 1994, \aap, 291, 943

\bibitem[{{Osterloh} {et~al.}(1997){Osterloh}, {Henning}, \&
  {Launhardt}}]{Osterloh97}
{Osterloh}, M., {Henning}, T., \& {Launhardt}, R. 1997, \apjs, 110, 71

\bibitem[{{Palagi} {et~al.}(1993){Palagi}, {Cesaroni}, {Comoretto}, {Felli}, \&
  {Natale}}]{Palagi93}
{Palagi}, F., {Cesaroni}, R., {Comoretto}, G., {Felli}, M., \& {Natale}, V.
  1993, \aaps, 101, 153

\bibitem[{{Phelps} \& {Lada}(1997)}]{Phelps97}
{Phelps}, R.~L. \& {Lada}, E.~A. 1997, \apj, 477, 176

\bibitem[{{Pirogov}(1999)}]{Pirogov99}
{Pirogov}, L. 1999, \aap, 348, 600

\bibitem[{Posselt(2003)}]{PosseltDIP}
Posselt, B. 2003, Master's thesis, Universitity of Jena

\bibitem[{{Press} {et~al.}(1986){Press}, {Flannery}, \&
  {Teukolsky}}]{Numericalrecipes86}
{Press}, W.~H., {Flannery}, B.~P., \& {Teukolsky}, S.~A. 1986, {Numerical
  recipes. The art of scientific computing} (Cambridge: University Press, 1986)

\bibitem[{{Rodgers} {et~al.}(1960){Rodgers}, {Campbell}, \&
  {Whiteoak}}]{Rodgers60}
{Rodgers}, A.~W., {Campbell}, C.~T., \& {Whiteoak}, J.~B. 1960, \mnras, 121,
  103

\bibitem[{{Rodriguez-Gaspar} {et~al.}(1995){Rodriguez-Gaspar}, {Tenorio-Tagle},
  \& {Franco}}]{Rodriguez95}
{Rodriguez-Gaspar}, J.~A., {Tenorio-Tagle}, G., \& {Franco}, J. 1995, \apj,
  451, 210

\bibitem[{{Sandell}(2000)}]{Sandell2000}
{Sandell}, G. 2000, \aap, 358, 242

\bibitem[{{Schreyer}(1997)}]{KatharinaDOK}
{Schreyer}, K. 1997, PhD thesis, University of Jena

\bibitem[{{Schreyer} {et~al.}(1996){Schreyer}, {Henning}, {K\"ompe}, \&
  {Harjunpaeae}}]{Schreyer96}
{Schreyer}, K., {Henning}, Th., {K\"ompe}, C., \& {Harjunpaeae}, P. 1996, \aap,
  306, 267

\bibitem[{{Schreyer} {et~al.}(2002){Schreyer}, {Henning}, {van der Tak},
  {Boonman}, \& {van Dishoeck}}]{Schreyer2002}
{Schreyer}, K., {Henning}, Th., {van der Tak}, F.~F.~S., {Boonman}, A.~M.~S.,
  \& {van Dishoeck}, E.~F. 2002, \aap, 394, 561

\bibitem[{{Schreyer} {et~al.}(2003){Schreyer}, {Stecklum}, {Linz}, \&
  {Henning}}]{Schreyer2003a}
{Schreyer}, K., {Stecklum}, B., {Linz}, H., \& {Henning}, Th. 2003, \apj, 599,
  335

\bibitem[{{Sharpless}(1959)}]{Sharpless59}
{Sharpless}, S. 1959, \apjs, 4, 257

\bibitem[{{Shepherd} \& {Churchwell}(1996)}]{Shepherd96}
{Shepherd}, D.~S. \& {Churchwell}, E. 1996, \apj, 472, 225

\bibitem[{Shu {et~al.}(1987)Shu, Adams, \& Lizano}]{Shu}
Shu, F.~H., Adams, F.~C., \& Lizano, S. 1987, Ann. Rev. Astron. Astrophys., 26,
  23

\bibitem[{{Snell} {et~al.}(1990){Snell}, {Dickman}, \& {Huang}}]{Snell90}
{Snell}, R.~L., {Dickman}, R.~L., \& {Huang}, Y.-L. 1990, \apj, 352, 139

\bibitem[{{Snell} {et~al.}(1988){Snell}, {Huang}, {Dickman}, \&
  {Claussen}}]{Snell88}
{Snell}, R.~L., {Huang}, Y.-L., {Dickman}, R.~L., \& {Claussen}, M.~J. 1988,
  \apj, 325, 853

\bibitem[{{Sridharan} {et~al.}(2002){Sridharan}, {Beuther}, {Schilke},
  {Menten}, \& {Wyrowski}}]{Beuther2002I}
{Sridharan}, T.~K., {Beuther}, H., {Schilke}, P., {Menten}, K.~M., \&
  {Wyrowski}, F. 2002, \apj, 566, 931

\bibitem[{{Steinacker} {et~al.}(2004){Steinacker}, {Lang}, {Burkert},
  {Bacmann}, \& {Henning}}]{Steinacker04}
{Steinacker}, J., {Lang}, B., {Burkert}, A., {Bacmann}, A., \& {Henning}, T.
  2004, \apjl, 615, L157

\bibitem[{{Stine} \& {O'Neal}(1998)}]{Stine98}
{Stine}, P.~C. \& {O'Neal}, D. 1998, \aj, 116, 890

\bibitem[{{Szymczak} {et~al.}(2000){Szymczak}, {Hrynek}, \&
  {Kus}}]{Szymczak2000}
{Szymczak}, M., {Hrynek}, G., \& {Kus}, A.~J. 2000, \aaps, 143, 269

\bibitem[{{Townsley} {et~al.}(2003){Townsley}, {Feigelson}, {Montmerle},
  {Broos}, {Chu}, \& {Garmire}}]{Townsley2003}
{Townsley}, L.~K., {Feigelson}, E.~D., {Montmerle}, T., {Broos}, P.~S., {Chu},
  Y., \& {Garmire}, G.~P. 2003, \apj, 593, 874

\bibitem[{{Tuthill} {et~al.}(2002){Tuthill}, {Monnier}, {Danchi}, {Hale}, \&
  {Townes}}]{Tuthill02}
{Tuthill}, P.~G., {Monnier}, J.~D., {Danchi}, W.~C., {Hale}, D.~D.~S., \&
  {Townes}, C.~H. 2002, \apj, 577, 826

\bibitem[{{Valdettaro} {et~al.}(2001){Valdettaro}, {Palla}, {Brand},
  {Cesaroni}, {Comoretto}, {Di Franco}, {Felli}, {Natale}, {Palagi}, {Panella},
  \& {Tofani}}]{Valdettaro2001}
{Valdettaro}, R., {Palla}, F., {Brand}, J., {Cesaroni}, R., {Comoretto}, G.,
  {Di Franco}, S., {Felli}, M., {Natale}, E., {Palagi}, F., {Panella}, D., \&
  {Tofani}, G. 2001, \aap, 368, 845

\bibitem[{{van den Bergh}(1966)}]{Bergh1966}
{van den Bergh}, S. 1966, \aj, 71, 990

\bibitem[{{van der Tak} {et~al.}(2000){van der Tak}, {van Dishoeck}, {Evans},
  \& {Blake}}]{Tak2000}
{van der Tak}, F.~F.~S., {van Dishoeck}, E.~F., {Evans}, N.~J., \& {Blake},
  G.~A. 2000, \apj, 537, 283

\bibitem[{{Walsh} {et~al.}(2001){Walsh}, {Bertoldi}, {Burton}, \&
  {Nikola}}]{Walsh2001}
{Walsh}, A.~J., {Bertoldi}, F., {Burton}, M.~G., \& {Nikola}, T. 2001, \mnras,
  326, 36

\bibitem[{{Walsh} {et~al.}(1999){Walsh}, {Burton}, {Hyland}, \&
  {Robinson}}]{Walsh99}
{Walsh}, A.~J., {Burton}, M.~G., {Hyland}, A.~R., \& {Robinson}, G. 1999,
  \mnras, 309, 905

\bibitem[{{Ward-Thompson} {et~al.}(1994){Ward-Thompson}, {Scott}, {Hills}, \&
  {Andre}}]{Ward-Thompson94}
{Ward-Thompson}, D., {Scott}, P.~F., {Hills}, R.~E., \& {Andre}, P. 1994,
  \mnras, 268, 276

\bibitem[{{Ward-Thompson} {et~al.}(2000){Ward-Thompson}, {Zylka}, {Mezger}, \&
  {Sievers}}]{Wardthomson}
{Ward-Thompson}, D., {Zylka}, R., {Mezger}, P.~G., \& {Sievers}, A.~W. 2000,
  \aap, 355, 1122

\bibitem[{{Wolf-Chase} {et~al.}(2003){Wolf-Chase}, {Moriarty-Schieven}, {Fich},
  \& {Barsony}}]{Wolf2003}
{Wolf-Chase}, G., {Moriarty-Schieven}, G., {Fich}, M., \& {Barsony}, M. 2003,
  \mnras, 344, 809

\bibitem[{{Wouterloot} {et~al.}(1993){Wouterloot}, {Brand}, \&
  {Fiegle}}]{Wout93}
{Wouterloot}, J.~G.~A., {Brand}, J., \& {Fiegle}, K. 1993, \aaps, 98, 589

\bibitem[{{Wu} {et~al.}(1996{\natexlab{a}}){Wu}, {Huang}, \& {He}}]{Wu1996}
{Wu}, Y., {Huang}, M., \& {He}, J. 1996{\natexlab{a}}, \aaps, 115, 283

\bibitem[{{Wu} {et~al.}(1996{\natexlab{b}}){Wu}, {Huang}, \& {He}}]{Wu96}
---. 1996{\natexlab{b}}, VizieR Online Data Catalog, 411, 50283

\bibitem[{{Yun} {et~al.}(1999){Yun}, {Moreira}, {Afonso}, \& {Clemens}}]{Yun99}
{Yun}, J., {Moreira}, M.~C., {Afonso}, J., \& {Clemens}, D.~P. 1999, \aj, 118,
  990

\bibitem[{{Zapata} {et~al.}(2001){Zapata}, {Rodr{\'{\i}}guez}, \&
  {Kurtz}}]{Zapata2001}
{Zapata}, L.~A., {Rodr{\'{\i}}guez}, L.~F., \& {Kurtz}, S.~E. 2001, Revista
  Mexicana de Astronomia y Astrofisica, 37, 83

\end{thebibliography}
\end{document}